\documentclass[%
 reprint,
groupedaddress,
nofootinbib,
 amsmath,amssymb,
 aps,
prb,
]{revtex4-2}
\usepackage{stackrel}
\usepackage{graphicx}% Include figure files
\usepackage{dcolumn}% Align table columns on decimal point
\usepackage{bm}% bold math
\usepackage{dsfont}
\usepackage{amsmath}
\usepackage{braket}
\usepackage{courier}
\usepackage{appendix}
\usepackage[normalem]{ulem}
\usepackage{graphicx}
\usepackage{cancel}
\usepackage{printlen}
\usepackage[caption=false]{subfig}
\usepackage{xcolor}
\usepackage[normalem]{ulem}
\usepackage[colorlinks=true,
    linkcolor=blue,
    citecolor=red,
    filecolor=magenta,      
    urlcolor=cyan,
    pdftitle={Gauge},]{hyperref}
\usepackage{pgfplots}
\usepackage{tikz}
\usepackage{float}
\usepackage{comment}
\usepackage{manfnt}
\usepackage{mathrsfs}

\newcommand{\icol}[1]{% inline column vector
  \left(\begin{smallmatrix}#1\end{smallmatrix}\right)%
}

\makeatletter
\def\l@subsubsection#1#2{}
\makeatother

\begin{document}

\title{The Physics of (good) LDPC Codes II. Product constructions} 

\author{Tibor Rakovszky}
\author{Vedika Khemani}
\affiliation{%
 Department of Physics, Stanford University, Stanford, California 94305, USA
}%

\begin{abstract}

We continue the study of classical and quantum low-density parity check (LDPC) codes from a physical perspective. We focus on constructive approaches and formulate a general framework for systematically constructing codes with various features on generic Euclidean and non-Euclidean graphs. These codes can serve as fixed-point limits for phases of matter. To build our machinery,  we unpack various \emph{product constructions} from the coding literature in terms of physical principles such as symmetries and redundancies, introduce a new cubic product, and combine these products with the ideas of gauging and Higgsing introduced in Part I~\cite{LDPCGauge}.  We illustrate the usefulness of this approach in finite Euclidean dimensions by showing that using the one-dimensional Ising model as a starting point, we can systematically produce a very large zoo of classical and quantum phases of matter, including type I and type II fractons and SPT phases with generalized symmetries. We also use the  \emph{balanced product} to construct new Euclidean models, including one with topological order enriched by translation symmetry and another  exotic fracton model whose excitations are formed by combining those of a fractal spin liquid with those of a toric code, resulting in exotic mobility constraints. Moving beyond Euclidean models, we give a review of existing constructions of \emph{good qLDPC} codes and classical \emph{locally testable} codes and elaborate on the relationship between quantum code distance and classical energy barriers, discussed in Part I, from the perspective of product constructions. 

\end{abstract}

\maketitle

\section{Introduction}

Recent years have seen a number of breakthrough results in the theory of quantum error correction, culminating in the discovery of so-called \emph{good} quantum low-density parity check (qLDPC) codes~\cite{hastings2021fiber,panteleev2021degenerate,panteleev2021quantum,panteleev2022asymptotically,breuckmann2021balanced,leverrier2022quantum,dinur2023good,lin2022good}. These codes exhibit an optimal asymptotic scaling of some key metrics for error correction, allowing for the protection of quantum information robustly and at a low overhead\footnote{In good codes, both the code rate \emph{k} (the number of logical qubits encoded) and the code distance $d$ (the size of the smallest undetectable error) are proportional to $n$ (the number of physical qubits), which is the optimal scaling.}. A key ingredient in these constructions is to eschew geometric locality constraints in favor of a more general arrangement of qubits, represented by highly connected \emph{expander graphs}. However, the ``low density" property of LDPC codes enforces that every qubit still interacts only with finitely many others, so these codes still have a generalized notion of graph locality. While more work will be required to bring these theoretical ideas into contact with experimental reality, much progress has been made towards realizing the necessary ingredients~\cite{Kollar2019,Periwal2021,Bluvstein2022}, especially in the context of reconfigurable atom arrays~\cite{LukinLDPC}.

These advances present an exciting opportunity for quantum many-body physics. The close connection between quantum error correction and exotic phases of interacting quantum systems was realized early on in the work of Kitaev~\cite{kitaev2003fault} and, more recently, has played a central role in the study of fracton phases~\cite{chamon2005quantum,haah2011local,vijay2016fracton,nandkishore2019fractons}. In particular, commuting stabilizer models, the most commonly studied type of quantum error correcting codes~\cite{calderbank1996good,steane1996multiple,gottesman1997stabilizer},  can also describe the fixed point limits of some gapped phases of matter, exhibiting their universal properties in an exactly solvable setting\footnote{In a stabilizer code each individual term -- ``parity check" -- is a product of Pauli operators. The code subspace is spanned by the +1 eigenstates of all checks, which are the ground states of the corresponding stabilizer Hamiltonians.}. Thus, the discovery of good qLDPC codes (and some of the tantalizing properties they exhibit~\cite{anshu2023nlts}) prompt us to ask more general questions: what new quantum phases and phenomena can exist if we relax the constraints of spatial locality in a way that can accommodate LDPC codes? How to even properly define phases in this context? What kinds of quantum many-body states can appear as ground states of gapped Hamiltonians in this context? 

In Part I of the present paper~\cite{LDPCGauge} we embarked on establishing a theoretical framework aimed at addressing some of these questions. In particular, building on previous work on topological order and fracton phases~\cite{wegner1971duality,levin2012braiding,vijay2016fracton,williamson2016fractal,shirley2019foliated,kubica2018ungauging}, we formulated in detail how qLDPC codes arise within topologically ordered deconfined phases of \emph{generalized gauge theories}, which can be obtained from applying an appropriate ``gauging'' operation on \emph{classical} LDPC codes. As a side product of this construction, we also saw how a different set of stabilizer models, commonly exhibiting features associated with \emph{symmetry protected topological} (SPT) order, can be constructed systematically from classical codes and appear in the ``Higgs" phases of the same gauge theories.

In these duality transformations that map classical codes to quantum codes, a key role is played by the properties of the underlying classical codes from which the quantum models are obtained. One such crucial feature is given by the \emph{symmetries} of the classical model, which we identified with the logical operators of the classical code. It is these symmetries that are being ``gauged'' on the way to obtaining qLDPC codes. Another important feature we identified~\cite{LDPCGauge} is given by the structure of excitations (domain walls) in the classical model, which can be ``point-like'' or ``extended'': it is the latter which, upon gauging, give rise to non-trivial quantum codes (which are endowed with features reminiscent of \emph{higher-form symmetries}~\cite{gaiotto2015generalized,mcgreevy2023generalized}). The extended nature of domain walls is, in turn, related to the presence (and structure) of \emph{local redundancies}\footnote{By local redundancy, we mean a finite set of parity checks whose product equals the identity, implying that these checks are not mutually independent. These are also variously referred to as ``local relations" or ``meta-checks".} in the classical code~\cite{LDPCGauge}. Furthermore, we argued~\cite{LDPCGauge} that \emph{good} qLDPC codes in particular are associated with additional features in the corresponding classical codes to which they are gauge dual, notably a property called \emph{local testability}~\cite{goldreich2005short,dinur2023good}.

This discussion brings to the fore the centrality of the features (symmetries, redundancies, etc.) of the underlying classical codes. There remains a question of how to find codes with particular features. In the present paper, we turn to this problem by describing a framework of constructive approaches for obtaining such codes in a systematic manner. In this, a central role will be played by various \emph{product constructions} which can be used to build classical codes with desired properties from simpler, less structured ones. A whole zoo of such product constructions have been developed\footnote{Examples include the tensor product, hypergraph product, balanced product, check product, among others.} in the computer science literature~\cite{tillich2013quantum,bravyi2014homological,hastings2021fiber,panteleev2021quantum,breuckmann2021balanced,leverrier2022quantumxyz,cross2023quantum}. Here we unpack and analyze many of these from a \emph{physical} perspective, giving insight into how these constructions produce physically interesting properties and models.

A key principle underlying these various constructions is the fact, realized early on by Kitaev, that the properties of LDPC codes can be naturally formulated in terms of homology theory~\cite{kitaev2003fault,bombin2007homological}. More specifically, 
classical and quantum codes can be associated with algebraic structures called \emph{chain complexes}, and properties of the codes can be derived from topological features of the complexes.  The chain complexes associated with codes are defined by the structure and redundancies of the checks of the code; importantly, these complexes can have an effective dimensionality and define a local notion of ``geometry", which may be entirely distinct from the dimensionality and geometry of the \emph{physical} lattice or graph on which the qubits and checks live. Much of the literature on product constructions is framed in terms of obtaining chain complexes with desired properties from simpler inputs --- notably, building higher dimensional chain complexes from lower dimensional ones, similar to ideas in topology where higher dimensional manifolds can be constructed from lower dimensional ones.  We will explicate and use the physical properties of the chain complexes associated with the input and output codes of various product constructions as a central organizing principle in our discussion\footnote{Indeed, while the importance of properties such as the dimensionality and connectivity of the physical lattice is widely appreciated in condensed matter physics, a similarly widespread appreciation is lacking for analogous properties of the chain complexes associated with models (in cases where such associations can be made).}. 

We physically elucidate a number of different product constructions from the computer science~\cite{tillich2013quantum,breuckmann2021balanced,cross2023quantum} and physics~\cite{devakul2021fractalizing} literature, which we organize into two categories: those that create \emph{subsystem symmetries}\footnote{We say that a symmetry is ``global" is its support scales linearly with the number of bits, while ``subsystem" if it scales with some smaller power.} (familiar from the study of fracton phases) and those that produce the aforementioned local redundancies. We also introduce a novel product construction (which we call the \emph{cubic product}), which takes three classical codes as input and produces a code with both of the above features. Combining these product constructions with other ingredients, including the duality maps from Part I~\cite{LDPCGauge}, yields an entire machinery that can be used to systematically build classical and quantum models with increasingly intricate properties (Fig.~\ref{fig:Factory}) from  simple building blocks. From this perspective, known products for obtaining quantum codes, such as the \emph{hypergraph product}~\cite{tillich2013quantum}, are interpreted as a multi-step process which first builds a classical code with additional features out of simpler ones, and then gauges it to obtain a quantum code. Applying the same approach to our cubic product gives a family of quantum codes that generalizes the $X$-cube model~\cite{vijay2016fracton} to a large family of quantum LDPC codes. Indeed, a power of the machinery is its generality, which works for constructing models both in Euclidean space and in more general non-Euclidean  geometries. 

In the latter parts of the paper, we demonstrate the power of this ``code factory'', in a variety of ways. Firstly, we use it to recover a vast array of known gapped phases of matter and to elucidate a large web of connections between these different models. 
Taking only the simplest non-trivial stabilizer model, the one-dimensional classical Ising model, as a starting point, the machinery produces a variety of ordered phases including the topologically ordered toric code, type-I and type-II fracton phases, SPTs protected by higher form and subsystem symmetries, among others (Fig.~\ref{fig:MapOfPhases}). 
Turning to non-Euclidean geometries, we show how a recently discovered fracton model~\cite{gorantla2023gapped} (defined on a general graph) can also be understood in the language of product constructions.

Secondly, we use the machinery of products and gauging to construct two stabilizer models, in two and three spatial dimensions, which are, to the best our knowledge, novel. We argue that the first describes a non-trivial symmetry-enriched topological (SET) phase~\cite{essin2013classifying,mesaros2013classification,barkeshli2019symmetry}, where translations perform non-trivial operations on four logical qubits; this might have some practical relevance, given that translations are a natural operation e.g. in a Rydberg atom platform~\cite{Bluvstein2022}. The other new model we construct is a fracton phase whose excitations exhibit exotic mobility properties. In both of these constructions, an important role is played by the idea of ``modding out'' spatial (in this case, translation) symmetries, which is also at the heart of the \emph{balanced product}, which is used to obtain good qLDPC codes. We also give an overview of the known varieties of such good codes, their properties and their relationships, which we believe will be of use to the physics audience.  

Finally, we return to the question of the relationship between locally testable classical codes and good quantum codes, which we discussed previously in Part I~\cite{LDPCGauge}. There, we observed that gauging a good classical locally testable code (LTC) generically tends to give rise to a qLDPC code with a good distance to one type of (i.e., either $X$ or $Z$) error (genuinely good qLDPC codes are then dual to \emph{a pair} of LTCs). Here, having reviewed the features that ensure the goodness of qLDPC codes, we are able to pinpoint why they also give rise to local testability. Moreover, we discuss a general argument that relates quantum code distance to the \emph{energy barriers} (which enter the definition of local testability) of the corresponding classical model for a family of codes that arise as hypergraph product codes. We also discuss how the argument might generalize to the balanced product construction that gives rise to good qLDPC codes.

The remainder of the paper is organized as follows. In Sec.~\ref{Sec:Definitions} we give an overview of some general definitions pertaining to classical and quantum stabilizer codes within the framework of chain complexes, which we will rely on throughout.  In Sec.~\ref{sec:codefactory} we give an overview of the ingredients of the ``code factory''; the details of these ingredients are then laid out in Sec.~\ref{sec:BuildingBlocks}, which describes prescriptions for defining classical codes on generic graphs and basic transformations between them; these codes can then serve as the building blocks of more complicated classical codes obtained through various product constructions discussed in Sec.~\ref{sec:products}. In Sec.~\ref{sec:GaugeAndHiggs}, we describe dualities that turn classical codes into quantum stabilizer models and we describe how the properties of the quantum codes are inherited from the classical codes that enter some of the product constructions of Sec.~\ref{sec:products}. In Sec.~\ref{sec:Examples} we discuss how many gapped phases on Euclidean lattices arise from our machinery, including known and new examples.  In Sec.~\ref{sec:NonEuclidean} we turn to models on non-Euclidean lattices and, after reviewing some examples from recent physics literature, we give an exposition to the construction of good qLDPC codes and their properties. Finally, in Sec.~\ref{sec:EnergyBarriers}, we discuss the relationship between good qLDPC codes and classical LTCs in the context of product constructions.

\tableofcontents

\section{Codes and Chain Complexes}\label{Sec:Definitions}

In this section, we summarize some of the important definitions and notation relating to classical and quantum stabilizer codes, which we will use throughout the paper. For a more detailed exposition, we refer the reader to Part I~\cite{LDPCGauge} (see Sec. III in particular). Motivated by the discussion there, we put the notion of chain complexes front and center; this language will also be important when we discuss product constructions in Sec.~\ref{sec:products} below. 
 
\subsection{Definitions}

\subsubsection{Chain Complexes}

A \emph{chain complex} is defined by a sequence of linear maps $\{\delta_p\}$ between vector spaces $\{V_p\}$:
\begin{equation}
V_{\mathscr{D}_c} \rightarrow  \cdots V_p \xrightarrow{\delta_p} V_{p-1} \cdots \underset{\text{``plaquettes"}}{V_2}  \xrightarrow{\delta_2} \underset{\text{``edges"}}{V_1} \xrightarrow{\delta_1} \underset{\text{``vertices"}}{V_0}
\label{eq:chaincomplex}
\end{equation}
with the defining property that the composition of two successive maps is the zero map, 
\begin{equation}
 \delta_p \delta_{p+1}=0.   
 \label{eq:ccboundarymap}
\end{equation}
These maps are called ``boundary operators" and the condition of two successive maps being zero can be colloquially stated as ``the boundary of a boundary is zero". 
A useful example of chain complexes to keep in mind are the cellulations of manifolds, where the successive vector spaces (``levels") represent cells of increasing dimension (i.e. $V_0$ is associated with vertices, $V_1$ with one-dimensional edges,  $V_2$ with two-dimensional plaquettes and so on, as indicated in Eq.~\eqref{eq:chaincomplex}). Generalizing from this intuition, the map $\delta_{2}$ can  be visualized as a map from a subset of generalized plaquette-like objects to a subset of generalized edge-like objects defining the collective boundary of the plaquettes.
In this way, a chain complex defines a local notion of geometry, and has an effective dimension, $\mathscr{D}_c$, defined by the length of the chain complex\footnote{The word ``dimension'' is not conventional for this. We use it here to emphasize the physical intuition behind it.}. Note, however that the generalized edges in question can involve more than two vertices and are thus more like the \emph{hyper-edges} of a hypergraph. Similarly, two plaquettes can share more than two edges and so on.
It is also customary to refer to a $\mathscr{D}_c$ dimensional chain complex as a ``level-($\mathscr{D}_c+1$) chain-complex". 
We can also talk of the \emph{dual chain complex} with  transposed maps $\{\delta_p^T\}$:
\begin{equation}\label{eq:DualComplex}
    V_{\mathscr{D}_c} \leftarrow \cdots V_p \xleftarrow{\delta^T_p} V_{p-1} \cdots V_2 \xleftarrow{\delta^T_2} V_1 \xleftarrow{\delta^T_1} V_0,
\end{equation} 
which satisfy 
\begin{equation}
 \delta^T_p \delta^T_{p-1}=0.   
 \label{eq:dualccboundarymap}
\end{equation}

In the correspondence between CSS codes and chain complexes, (qu)bits and checks define vector spaces associated with cells of different dimensions (i.e. with different $V_r$'s), as we now discuss. 

\subsubsection{Classical Codes}\label{sec:ClassicalReview}

A \emph{classical code} is defined by a set of parity checks acting on a set of bits. It is specified through a function $\delta$ which defines which bits are part of each check. The classical bits/spins are denoted as $\sigma_i = \pm 1$ with $i=1,\ldots,n$\footnote{From the coding perspective, it is more conventional to use bits that can take a value or $0,1$, but we use classical spin variables which can be $\pm1$ to connect to the statistical mechanical perspective.}. The parity checks are labeled by $a=1,\ldots,m$, and the check with label $a$ corresponds to a product of spins within a subset of sites $i \in \delta(a)$, $C_a \equiv \prod_{i\in \delta(a)} \sigma_i$. The codewords are defined as the set of configurations where all $m$ checks are satisfied (i.e. equal to +1). These are the ground states of a classical Hamiltonian, $H_{\rm cl}~=~-\sum_a C_a = - \sum_{a=1}^m \prod_{i\in \delta(a)} \sigma_i$.  

Each spin configuration is represented as a vector in an $n$ dimensional vector space over the binary field $\mathbb{Z}_2$\footnote{Each $i$ is assigned a basis vector $|i\rangle$, and each spin configuration is a vector $\sum_i \alpha_i |i\rangle$ with $\alpha_i = 0(1)$ if $\sigma_i = +1(-1)$ in the configuration. Vector addition is modulo 2. Equivalently, one can view each vector as specifying a subset of spins, where spin $i$ is included in the subset if $\alpha_i=1$ and vector addition is the symmetric difference of subsets.}. Likewise, each check takes a value $\pm 1$, and configurations of checks are also represented as vectors in an $m$ dimensional vector space over the binary field $\mathbb{Z}_2$. 
The map $\delta$ is defined from subsets of checks to the subset of spins that their product acts on. This is a linear map over two $\mathbb{Z}_2$ vectors spaces, and can thus be represented as an $n\times m$ binary matrix, $\delta_{ia} =1 \; {\rm iff}\; i \in \delta(a)$. Its transpose $\delta^T$ maps subsets of spins onto the set of checks that change sign if the spins in question are flipped. Throughout this paper, we will focus on LDPC codes. The low-density property amounts to the condition that the number of non-zero elements in each row of $\delta$ and $\delta^T$ is finite (i.e. does not scale with $n$ as $n$ is increased). Colloquially, ``every bit talks to finitely many bits''.

A useful representation of this structure is through the so-called \emph{Tanner graph} of the code. This is a bi-partite graphs with two sets of vertices, $V_1$ and $V_2$, which are in one-to-one correspondence with the bits and checks of the code, respectively. One then draws an edge between $i \in V_1$ and $a \in V_2$ iff $\delta_{ia} = 1$ (See Part I~\cite{LDPCGauge}, figure 5 for an illustration). The LDPC property then amounts to the requirement that the degree (number of neighbors) of any vertex in the Tanner graph is bounded from above by some $n$-independent constant. 

There is a correspondence between the properties, specifically the ground state degeneracy and the symmetries, of $H_{\rm cl}$ and the structure of logical information of the classical code. A ground state degeneracy of $2^k$ means that the code has $2^k$ distinct codewords, and is said to encode $k$ bits of logical information.   The `all-up' state, $\sigma_i=1 \; \forall \; i$ is always a codeword. 
Each of the other codewords is associated with a logical operator which flips the sign of the bits which are $-1$ in that codeword, thereby mapping between the codeword and the `all-up' state. Each such logical is a symmetry of $H_{\rm cl}$. Thus, each of the degenerate codewords represent different symmetry broken states, and $H_{\rm cl}$ exhibits spontaneous symmetry breaking. 
Finally, $d$, the code distance, is the smallest number of spins that can be flipped to go from any codeword to any other.  The triplet of numbers $[n, k,d]$ is commonly used to characterize codes. 

In the following, it will be sometimes useful to use a quantum language, even for classical codes. To do so, we can associate a qubit for each $i$ and equate $\sigma$ with the Pauli $z$ component of this qubit, so that the (diagonal) Hamiltonian reads 
\begin{equation}\label{eq:QuantizedHam}
    {H}_{\rm cl} = - \sum_{a=1}^m \prod_{i\in \delta(a)} \sigma_i^z.
\end{equation}
The logical operators (symmetries of $H_{\rm cl}$) are then products of Pauli $X$ matrices, e.g.\footnote{Here, and later, we abuse notation slightly by using $\lambda$ as both a label for the logical and as the subset of spins on which it acts.} $\mathcal{X}_\lambda = \prod_{i\in\lambda} \sigma_i^x$. As a quantum model, we can also define the corresponding ``logical $Z$'' operators, $\mathcal{Z}$, one for each logical $X$, in such a way that the eigenvalues $\pm 1$ of the $k$ different $Z$ logicals uniquely label the $2^k$ codewords. One can always choose a basis of logicals where these classical $Z$ logicals each act on a single qubit. For example, the simplest classical code, the repetition code, can be associated with the 1D Ising model with nearest-neighbor Ising checks, $\sigma_i^z \sigma_{i+1}^z$ on $n$ qubits. The codewords are the two degenerate ground states (`all-up' and `all-down') so that one logical bit is encoded ($k=1$). The logical $X$ operator which flips the state of the logical bit is the Ising symmetry operator, $\mathcal{X} = \prod_{i=1}^n \sigma_i^x$ acting on all spins, so that $d=n$. There is a single logical $Z$ which can be chosen as $\mathcal{Z} = \sigma_i^z$ on any site $i$. 

To connect to chain complexes, we note that a classical code is simply a sequence of two vector spaces, $V_0$ and $V_1$, associated with the bits and checks respectively, with maps $\delta, \delta^T$ between these. This is nothing but a one dimensional chain complex as defined in Eq.~\eqref{eq:chaincomplex}:
$$
\underset{\text{checks}}{V_1} \stackrel[\delta^T]{\delta}{\rightleftarrows}  \underset{\text{bits}}{V_0},
$$
where the bits are associated with vertex-like objects, and the checks are associated with edge-like objects acting on the bits. That is, a classical code  has a geometrical interpretation as a \emph{hypergraph}, where bits define vertices and each check $a$ corresponds to a hyperedge containing the vertices in $\delta(a)$.

A classical code may also be embedded within a higher dimensional chain complex, for example by only using the lowest two vector spaces (``levels") for bits and checks\footnote{Note that if we used some intermediate levels instead, then the level below that of bits would correspond to logical operators of the code because of Eq.~\eqref{eq:dualccboundarymap}. This would imply a code with a small, $O(1)$ code distance, assuming that all the boundary maps $\delta_p$ satisfy the LDPC property.}. In this case, the presence of the additional levels in the chain complex enforces constraints on the classical checks, with important physical consequences. We will return to this point in the next subsection. 

\subsubsection{Quantum Codes}

A quantum code is defined on a set of $n$ qubits, labeled by $a = 1, \cdots, n$, with Pauli operators $X_a,Y_a,Z_a$ acting on them.  We  focus on CSS stabilizer codes~\cite{calderbank1996good, steane1996multiple} defined by $X$ and $Z$ type checks (i.e. checks involving exclusively $X_a$ and $Z_a$ operators, respectively), which we label by $i=1\ldots,m_X$ and $p=1,\ldots,m_Z$. These individually define classical codes through  functions $\delta_{X}$ and  $\delta_Z$, which again define the set of spins that a given check acts on i.e., $X$-checks take the form $A_i \equiv \prod_{a \in \delta_X(i)} X_a$ and  $Z$-checks takes the form $B_p \equiv \prod_{a \in \delta_Z(p)} Z_a$. Note that, taken separately, the two checks define two classical codes which we denote $\mathscr{C}_{X}$ and $\mathscr{C}_{Z}$. Once again, the LDPC condition enforces that only finitely many qubits interact with each other via checks of either type. We will refer to such quantum codes as qLDPC codes, to distinguish them from their classical counterparts which we denote as cLDPC codes.

All checks commute in a stabilizer CSS code, $[A_i,B_p] = 0$ for any pair $(i,p)$. With this condition, the codespace is formed by the common (+1) eigenstates of all checks, $A_i\ket{\psi} = B_p \ket{\psi} = \ket{\psi} \, \forall i,p$, and has dimension $2^k$. Just as in the classical case, it is possible to combine all the checks into a Hamiltonian
\begin{equation}\label{eq:H_quantum}
    H_\text{q} = -\sum_i A_i - \sum_p B_p = -\sum_{i} \prod_{a \in \delta_X(i)} X_a - \sum_{p} \prod_{a\in \delta_Z(p)} Z_a,
\end{equation}
which has the code subspace as its ground state subspace. The logical operators of the quantum code leave the code subspace invariant, but act on it non-trivially. There are $k$ independent logicals of both $X$ and $Z$ type which must commute with each of the $A_i$ and $B_p$ checks (in order to leave the code subspace invariant) and be orthogonal to the subspace spanned by the checks (in order to act on the code subspace non-trivially). Thus, each quantum logical commutes with $H_{\rm q}$ and is therefore a symmetry operator. However, unlike the symmetries of the classical code, the quantum logicals are ``deformable", since multiplying an $X/Z$-type logical with any of the $X/Z$ checks yields an equivalent logical operator. In the physics literature, such deformable symmetry operators are associated with \emph{higher-form symmetries}~\cite{gaiotto2015generalized,mcgreevy2023generalized}, and the presence of a degenerate ground state subspace in $H_{\rm q}$ can be interpreted as  spontaneously breaking of these symmetries; in a more conventional language, this corresponds to topological order.  The $X$ and $Z$ type logicals can be grouped into anticommuting pairs with the same algebra as Pauli operators, e.g. $\mathcal{X}_\lambda$ and $\mathcal{Z}_{\lambda'}$ anti-commute if $\lambda=\lambda'$ and commute otherwise. We can define the $X$- and $Z$-distances of the code, $d_X$ and $d_Z$, as the \emph{smallest} Pauli weight (i.e., number of qubits being acted upon) that a logical operator of each type can have. The overall code distance is the smaller of the two: $d \equiv \min(d_X,d_Z)$. The canonical example of a quantum CSS code is the 2-dimensional toric code  with $X$-checks, qubits and $Z$-checks associated with the sites, edges and plaquettes of a 2D lattice.

To connect to chain complexes, we can again associate the $X$-checks, qubits, and $Z$-checks to vector spaces over the binary field $\mathbb{Z}_2$, which we suggestively label by $V_0, V_1, V_2$  respectively.
 The maps $\delta_X$ and $\delta_Z$ which specify the support of the checks map between these spaces: $\delta_Z: V_2 \to V_1$ and $\delta_X: V_0 \to V_1$.  So far, this just looks like two separate classical codes acting on the same spins (viewed in different bases). However, the two are related by the condition that all checks in the quantum code must commute. 
The  commutativity of the $X$ and $Z$ checks means that they must always overlap on an even number of qubits which is equivalent to the condition $    \delta_X^T \delta_Z = 0,$ where multiplication is defined modulo 2. In words, this says that if one flips all the qubits that are part of a $Z$ check (or a product of $Z$ checks), this does not flip the sign of any $X$ checks, which means that the $X$ and $Z$ checks commute.  If we associate $\delta_2 \equiv \delta_Z$ and $\delta_1^T \equiv \delta_X$, then the vector spaces $V_0, V_1, V_2$ and the maps $\delta_{1}, \delta_2$ together define a \emph{two dimensional chain complex}, cf. Eqs.~\eqref{eq:chaincomplex}, \eqref{eq:ccboundarymap}:
\begin{equation}
\underset{Z-\text{checks}}{V_2} \xrightarrow{\delta_2 \equiv \delta_Z}
\underset{\text{qubits}}{V_1} \xrightarrow{\delta_1 \equiv \delta_X^T} \underset{X-\text{checks}}{V_0},
\label{eq:quantumcc}
\end{equation}

where the qubits are  associated with edge-like objects, the $X$-checks with vertex-like objects, and the $Z$-checks with plaquette-like objects. We emphasize again that, in general, the edges are hyperedges that may involve more than two vertices, plaquettes are hyper-plaquettes etc. 
The commutativity of the $X$ and $Z$ checks is ensured by the condition in Eq.~\eqref{eq:ccboundarymap} satisfied by the chain-complex maps. The LDPC condition again enforces that the maps $\delta_{1,2}$ are sparse. 

The logical operators of the CSS code are defined by the topological properties of the chain complex~\cite{kitaev2003fault,bombin2007homological,bravyi2014homological}.  The notion of deforming logical operators by multiplication with checks has a natural topological interpretation in terms of the chain complex. In particular, the support of the quantum logicals correspond to \emph{non-contractible loops} on the geometry defined by the chain complex, and the equivalence classes of logicals correspond to the homology and cohomology classes of the chain complex. 

Finally, we note that while a quantum CSS code minimally requires a two dimensional chain complex, it can be embedded within a higher dimensional chain complex where the presence of the additional levels will again impose constrains on the checks, with important consequences for physical properties, as we will discuss below.

\subsection{Physical role of graph vs. chain complex dimensionality}
\label{subsec:physicalandccgeometry}

As we have reviewed in the previous subsection, classical stabilizer codes are naturally associated with one-dimensional chain complexes, while quantum CSS codes are associated with two dimensional chain complexes. Indeed, one central goal of the product constructions we discuss below is to build higher dimensional chain complexes out of lower dimensional ones, which allows one to build quantum codes using classical codes as inputs~\cite{tillich2013quantum,bravyi2014homological}. 

At the same time, given a higher-dimensional chain complex, we can also associate \emph{classical} codes to it, by taking two subsequent levels of the chain complex to represent the bits and checks of a classical code (a relationship between some of these classical and quantum CSS codes, living on the same chain complex, is provided by the  generalized gauge dualities~\cite{LDPCGauge,kubica2018ungauging}, which we review in Sec.~\ref{sec:GaugeAndHiggs} below). In this view, which is the one we will take in Sec.~\ref{sec:products}, product constructions have the effect of creating new classical codes with features absent in their inputs.

What features should be associated to classical codes that correspond to higher dimensional chain complexes with $\mathscr{D}_c \geq 2$? Consider the case, as we often will below, of a two-dimensional chain complex whose lower two levels $V_0$ and $V_1$ are associated with the bits and checks of a classical code, respectively. In the latter case, the presence of the additional third level imposes extra structure on the code. In particular, the basis elements of the vector space $V_2$ are \emph{local redundancies} between the checks of the classical code, which force products of some finitely many of said checks to be equal to the identity i.e. $\prod_{a \in R} C_a = +1$. In equations
\begin{equation}
\underset{\text{local redundacies}}{V_2} \xrightarrow{\delta_2} \underset{\text{checks}}{V_1} \xrightarrow{\delta_1}, \underset{\text{bits}}{V_0},  
\label{eq:classical_2dcc}
\end{equation}
where the condition $\delta_1 \delta_2=0$ precisely enforces the redundancy. In this work, we will always consider maps $\delta$ with satisfy the LDPC property of sparseness (i.e. of having rows and columns with finitely many non-zero elements). This, in turn, enforces that the redundancies are \emph{local} i.e. small weight ($|R|$ is finite, independent of $n$) and that each check is involved in finitely many redundancies. We could generalize this to even higher dimensional chain complexes (still associating bits and checks to the lowest two levels). For example, elements of $V_3$ correspond to ``meta-redundancies'' (linear relationships between local redundancies) and so on. We will refer to the dimension of the chain complex, $\mathscr{D}_c$ as the ``code dimensionality''\footnote{Here, we take the point of view of starting from a predefined chain complex and associating a classical code to it. This is natural from the perspective of product constructions, where the chain complex is built systematically. More generally, given a classical code in terms of its bits and checks alone, one could ask what is the largest dimensional chain complex into which it can be embedded in a non-trivial way, while maintaining the LDPC property of all boundary maps, although a rigorous way of formulating this question is not obvious.}. The simplest models to keep in mind to visualize classical codes with increasing $\mathscr{D}_c$ are Ising models with bits living on the sites of hypercubic Euclidean lattices of dimension $D$ and nearest-neighbor Ising checks on the edges of the lattice\footnote{Thus, the hyperedges defining the checks coincide with the physical edges of the lattice.}. The 1D Ising model has no local redundancies, and hence has $\mathscr{D}_c=1$; the 2D Ising model has local redundancies and hence has $\mathscr{D}_c=2$ (the product of four Ising checks on plaquettes of the 2D square lattice is the identity); the 3D Ising model has local redundancies on plaquettes, and meta-redundancies on cubes so $\mathscr{D}_c=3$ etc.  

In a similar vein, while defining a \emph{quantum} code minimally requires a two dimensional chain complex, it can be embedded within a higher dimensional chain complex with $\mathscr{D}_c \geq 3$. If we associate the lowest three levels with $X$-checks, qubits and $Z$-checks as in Eq.~\eqref{eq:quantumcc}, then the elements of $V_3$ correspond to local redundancies between the $Z$-checks, which follows from $\delta_2\delta_3=0$:
\begin{equation}
\underset{\textrm{\shortstack{local redundancies\\ of $Z-$checks}}}{V_3} \xrightarrow{\delta_3}\underset{Z-\text{checks}}{V_2} \xrightarrow{\delta_2 \equiv \delta_Z}
\underset{\text{qubits}}{V_1} \xrightarrow{\delta_1 \equiv \delta_X^T} \underset{X-\text{checks}}{V_0}.
\label{eq:3dtc}
\end{equation}
An example of such a code is the toric code in three dimensions. Additional levels in the chain complex enforce yet more constraints. For example, if $\mathscr{D}_c=4$, then the elements of $V_4$ would correspond to meta-reduncies of $Z-$checks. Alternatively, given a four-dimensional chain complex, one could choose to populate the \emph{middle} three levels with $X-$checks, qubits and $Z-$checks, in which case the elemnts of $V_0$ correspond to local-redundancies of $X-$checks, while elements of $V_4$ correspond to local redundancies of $Z-$checks, which follows from $\delta_2^T \delta_1^T=0$ and $\delta_3\delta_4=0$ respectively: 
\begin{align}
\underset{\textrm{\shortstack{local\\ redundancies\\ of $Z-$checks}}}{V_4} \xrightarrow{\delta_4}\underset{\textrm{\shortstack{\\$Z-$\\\text{checks}}}}{V_3} \xrightarrow[\delta_Z]{ \delta_3}
\underset{\text{qubits}}{V_2} \xrightarrow[\delta_X^T]{ \delta_2} \underset{\textrm{\shortstack{\\$X-$\\\text{checks}}}}{V_1}\xrightarrow{\delta_1}\underset{\textrm{\shortstack{local \\redundancies\\ of $X-$checks}}}{V_0}.
\label{eq:4dtc}
\end{align}

The presence of additional relations between checks (in both classical and quantum codes) has important physical consequences, which we discuss below. In the examples discussed thus far (Ising models and toric codes defined on Euclidean lattices of spatial dimension $D$), the dimension of the \emph{physical} lattice on which the degrees of freedom live ($D$) and the code-dimension ($\mathscr{D}_c$) coincide. The importance of the dimensionality of the lattice is well understood in condensed matter physics and leads to constraints on possible ordered phases at zero and finite temperatures, for example through Peierls-Hohenberg-Mermin-Wagner type theorems. However, the geometry of the physical lattice and the ``code-geometry" (defined by the chain complex associated with checks as hyperedges) need not coincide in general. 

We emphasize this crucial point: even when the checks of a code are local in some Euclidean lattice or \emph{graph} of dimension $D$, this dimension might be distinct from the dimension of the \emph{code} defined on this graph, $\mathscr{D}_c \neq D$. One way of getting a handle on this difference is by considering closed loops or cycles in both the graph and code geometries. In the graph, these are defined in the obvious way, while in the code they correspond  to redundancies between checks. Local redundancies or ``short loops" are kept track of through additional levels in the chain complex. 

Thus, if classical codes defined on Euclidean graphs do not have any local redundancies, they are associated to a one-dimensional chain complex, $\mathscr{D}_c=1$, even if the graph dimension is $D \geq 2$.  Examples of this are provided by certain classical codes that exhibit \emph{subsystem symmetries} and are classical analogues of fracton phases. Two examples of this in $D=2$ (that we will discuss frequently in this paper) are provided by the plaquette Ising~\cite{vijay2016fracton} and Newman-Moore (NM)~\cite{newman1999glassy} models, both of which have bits arranged on the sites of 2D lattices with periodic boundaries, with four-bit and three-bit checks respectively living on the plaquettes of the lattice (see Fig.~\ref{fig:CodesLocalRed} as well as Fig. 4 in Part I~\cite{LDPCGauge}). Note that these are hyper-edges involving four- and three- bits respectively, and are distinct from the actual edges of the physical lattice. 
The symmetries (logicals) of the NM and plaquette Ising models are subsystem, scaling with a non-trivial power of $n$ that is smaller than 1: in the plaquette-Ising model, these are rigid ``line-like" symmetries corresponding to flipping all spins along any horizontal or vertical line, while the NM model has fractal symmetries corresponding to flipping spins along Sierpinski tetrahedra. Neither of these models feature any local redundancies, i.e. there are no finite, system-size independent number of checks whose product is trivial, and so their code dimension is $\mathscr{D}_c = 1$, despite the fact that they are defined on a physical two-dimensional graph\footnote{Both models, however, have \emph{global} redundancies which involve a diverging number of checks. These are isomorphic to, and in 1:1 correspondence with, the logicals in both cases, being line-like for the plaquette Ising and fractal for NM.}. Likewise, the 3D plaquette Ising model has bits on the sites of a 3D lattice with four-bit checks on the plaquettes~\cite{vijay2016fracton} (see Fig.~\ref{fig:CodesLocalRed}). This model has local redundancies, given by the product of four plaquettes oriented along two perpendicular directions on every cube), but there are no local meta-redundancies. Thus, $\mathscr{D}_c=2$ while $D=3$. In the quantum setting, fractonic models such as the 3D $X-$cube model~\cite{vijay2016fracton} and the 3D Haah code~\cite{haah2011local} are two-dimensional chain-complexes ($\mathscr{D}_c=2$) despite living in three spatial dimensions, because neither the $X$ nor the $Z$ checks have local redundancies\footnote{In fact, the classical 3D plaquette Ising model and the quantum $X-$cube model live on the same two-dimensional chain complex and are gauge dual~\cite{vijay2016fracton}.}.

Another context where the notions of spatial and code dimensionality diverge is in the cases when codes are associated to non-Euclidean graphs, so that $D$ is not even well-defined or is nominally infinite. For example, expander graphs, which are crucial in the construction of good codes (see Sec.~\ref{sec:BuildingBlocks} for a definition), can be thought of as infinite dimensional (in the sense that ``volume'' grows exponentially with ``radius''). Nevertheless, a classical code defined on such a graph might not have any local redundancies and may thus still have $\mathscr{D}_c = 1$. We will see examples of this in Sec.~\ref{sec:BuildingBlocks} below. 

We now discuss the physical relevance of these redundancies, meta-redundancies etc. In Part I~\cite{LDPCGauge}, we argued that they allow one to talk about the dimensionality of domain walls (excitations formed by violations of the classical checks) in a way that is independent of any notion of an underlying \emph{spatial} geometry. In particular, we associate codes with $\mathscr{D}_c \geq 2$ with \emph{extended} domain walls, as opposed to the \emph{point-like} domain walls of codes without local redundancies(See, for example, Fig. 3 in ~\cite{LDPCGauge}); one can then further sub-divide extended domain walls into loop-like ($\mathscr{D}_c = 2$), surface-like ($\mathscr{D}_c = 3$) etc. Considering Ising models in various dimensions (for which $D=\mathscr{D}_c$), we indeed see that the 1D Ising models has point-like domain-walls, the 2D Ising model has loop-like domain walls (enforced by the plaquette redundancy of checks), the 3D Ising model has surface-like domain walls (enforced by the plaquette redundancies and the cubic meta-redundancies) etc. Turning to cases where $D \neq \mathscr{D}_c$, it is $\mathscr{D}_c$ that determines the dimensionality of excitations. This is illustrated by the classical ``fractonic" Newman-Moore and plaquette-Ising models mentioned above. Both live in two spatial dimensions but have $\mathscr{D}_c=1$ and feature point-like excitations: in the plaquette-Ising model, flipping spins in a rectangular domain violates four checks at the four corners, while for NM, flipping spins along the shape of a Sierpinski triangle creates three excitations at the three corners of the triangle. In this sense, these models resemble the \emph{one-dimensional} Ising model, despite existing on a 2D lattice. The connection between the absence of local redundancies and the point-like nature of excitations applies more generally to translation-invariant models in finite dimensions. One can also make a connection between the existence of redundancies and the nature of excitations for arbitrary LDPC codes, although a similarly general statement is missing in that context---see Part I~\cite{LDPCGauge} for a discussion of this issue. 

The point-like vs. extended nature of domain walls has important physical consequences, which are illustrated by the comparison of the 1D and 2D Ising models. One notable feature of the latter is its stability at finite temperature, a feature that arises precisely because of the extended domain walls that provide a macroscopic energy (and free energy) barrier between the two symmetry broken states. In contrast to this, the aforementioned ``fractonic'' codes, with their point-like excitations, fail to be thermally stable despite existing in two spatial dimensions\footnote{In the NM model, due to the fractal nature of its symmetries, there is an energy cost that grows logarithmically with the number of flipped spins. While this results in slow, glassy dynamics at low temperatures, it is insufficient for thermal stability~\cite{newman1999glassy,prem2017glassy}.}~\cite{newman1999glassy,prem2017glassy}. For non translation invariant Euclidean models, large energy barriers are possible even in the absence of local redundancies, but they generically seem to be insufficient to provide true thermal stability due to large entropic contributions to the free energy~\cite{siva2017topological}. The case of non-Euclidean models, however, is less clear, with e.g. random expander codes exhibiting an $O(n)$ energy barrier and no redundancies~\cite{ben2010locally}; whether this suffices for thermal stability or this too can be destabilized by entropic factors is, to the best of our knowledge, unknown.

Similar considerations apply to CSS codes when defined on chain complexes with $\mathscr{D}_c \geq 3$. For example, in the 3D toric code (cf. Eq.~\eqref{eq:3dtc}), the ``vortex" excitations of the $Z-$checks form closed loops due to local redundancies of $Z$ checks (while $X$ checks are still point-like), and hence the code is thermally stable in the presence of perturbations that create $Z$ excitations but not $X$ excitations. In contrast, the so-called $(2,2)$ toric code in 4D (cf. \eqref{eq:4dtc}) has loop like $X$ and $Z$ excitations due to local redundancies for both types of checks and is thermally stable~\cite{dennis2002topological}. Also in the case of CSS codes, it is $\mathscr{D}_c$ that is relevant and not $D$; for example, the $X-$cube model and Haah's code have point-like $X$ and $Z$ excitations despite living in three spatial dimensions.  

Finally, we note that another difference between $\mathscr{D}_c = 1$ and $\mathscr{D}_c \geq 2$ classical codes, more pertinent to our purposes here, is in the behavior of the gauge theories obtained from gauging their symmetries (given by the classical logicals of the code). As discussed in detail in Part I~\cite{LDPCGauge}, it is classical codes with $\mathscr{D}_c \geq 2$ (and hence with extended domain walls) that gives rise to non-trivial gauge theories that can exhibit a deconfined topologically ordered phase. This phase is characterized, in its fixed point limit, by a quantum CSS code that is precisely the CSS code associated to the two-dimensional chain complex of the classical code with local redundancies. 

\section{The Code/Model Factory}
\label{sec:codefactory}

\begin{figure*}
    \centering
    \includegraphics[trim={1cm 0 0 0},width = 0.8\linewidth]{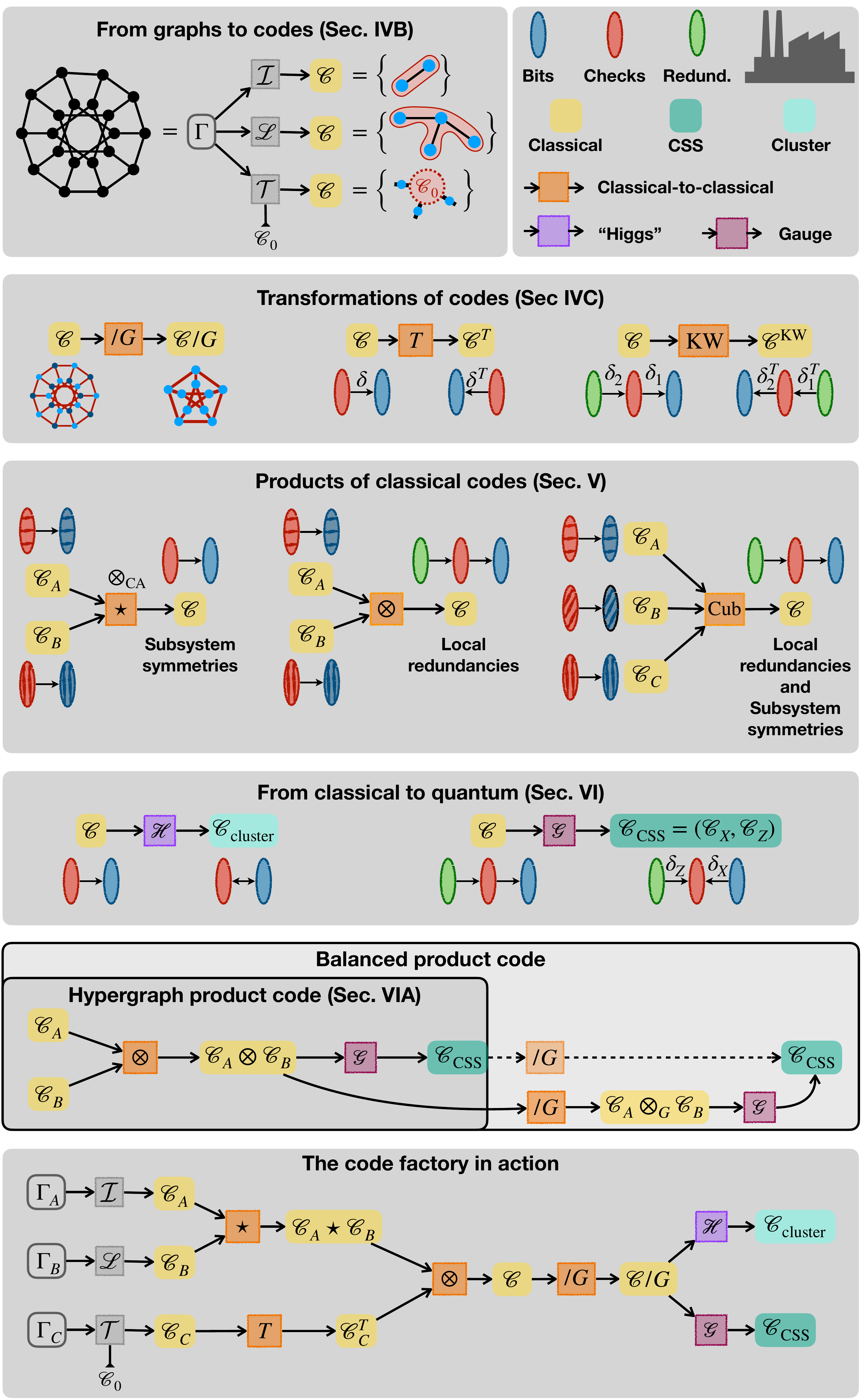}
    \caption{\textbf{The code factory.} See Sec.~\ref{sec:codefactory} for a description.}
    \label{fig:Factory}
\end{figure*}

\begin{figure*} 
    \centering
    \includegraphics[trim={1cm 0 0 0},width = 0.8\linewidth]{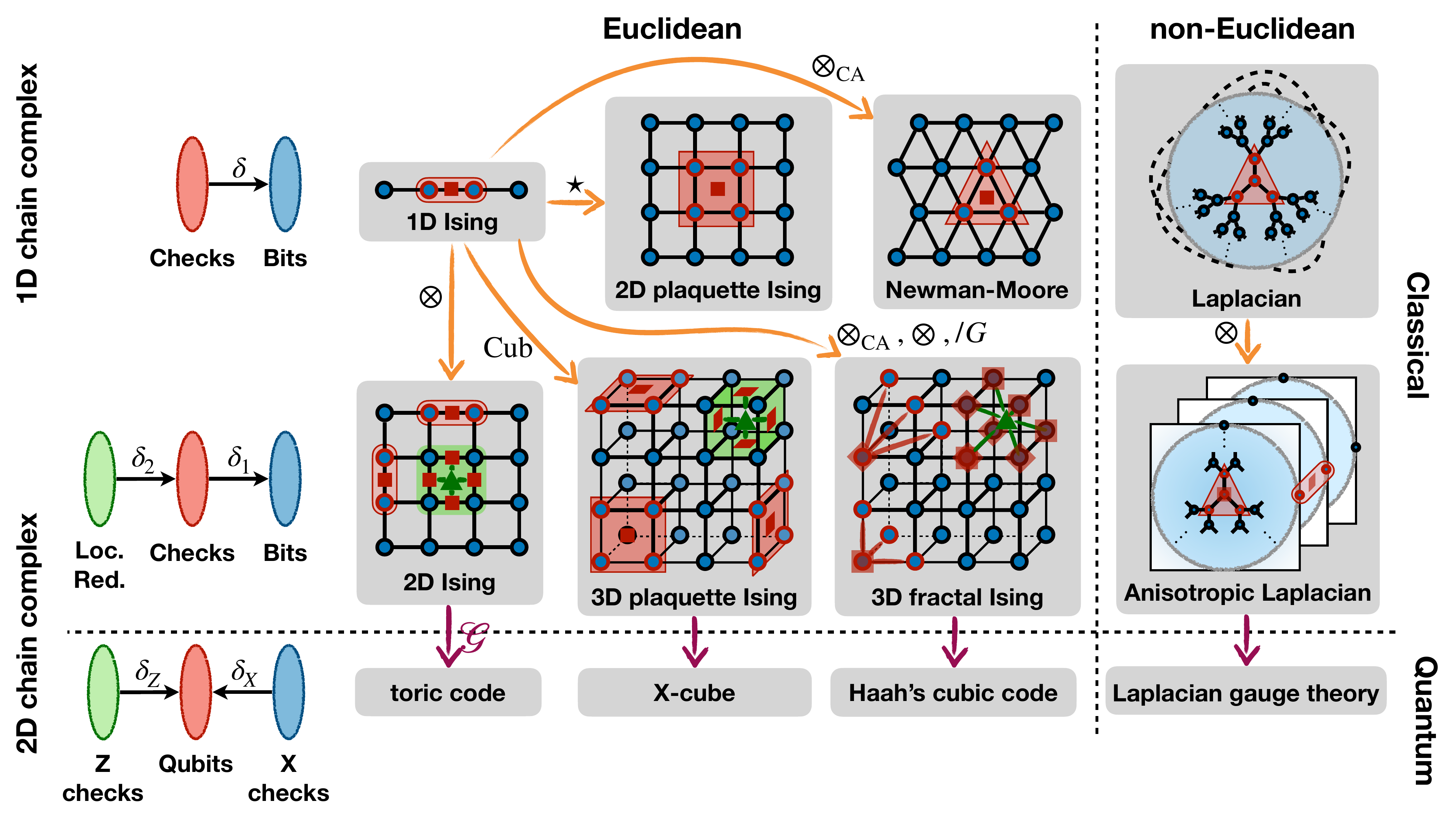}
    \caption{\textbf{Examples of classical and quantum codes and their interrelations through product constructions and gauge dualities.} Many spatially local (``Euclidean'') codes can be obtained from  the code factory with the 1D Ising model as input. Some products ($*$ and $\otimes_{\rm CA}$) produce local redundancies but leave the dimensionality of the chain complex unchanged.  Other products ($\otimes$, $\rm{Cub}$) produce local redundancies (i.e. 2D chain complexes) and hence can be gauged ($\mathscr{G}$) to give quantum CSS codes, including fracton models (see also Fig.~\ref{fig:MapOfPhases}). Similar constructions can be applied for codes on non-Euclidean graphs, an example of which involves the Laplacian gauge theory of Ref. \onlinecite{gorantla2023gapped}.}
    \label{fig:CodesLocalRed}
\end{figure*}

In this section we briefly outline the machinery that is to be described in Sections~\ref{sec:BuildingBlocks}-\ref{sec:GaugeAndHiggs}, also illustrated graphically in Fig.~\ref{fig:Factory} and Fig.~\ref{fig:CodesLocalRed}. We will use the 1D Ising model and its transformations as a concrete example in the discussion below, see also Section~\ref{subsec:IsingUniverse} and  Fig.~\ref{fig:MapOfPhases}.

\begin{enumerate}
    \item \emph{Basic building blocks (Sec.~\ref{sec:BuildingBlocks}).}
    \begin{enumerate}
        \item \emph{Codes on graphs (Sec.~\ref{subsec:CodesOnGraphs}).} We begin by outlining a few different ways of associating classical codes to some underlying graph in a local way. These can serve as the initial ``building-block" codes which are inputs to the constructions that follow. The simplest example would be to produce a 1D Ising model given a 1D lattice, but more general checks on general graphs (Euclidean and non-Euclidean) can be constructed to generate different families of classical codes in a systematic way.  
        \item \emph{Transformations of Codes (Sec.~\ref{subsec:OldToNew}).} We review some ways in which a given classical code can be mapped to a new classical code with potentially new properties. This includes operations for associating different classical codes to the same chain complex (say by taking a ``transpose" ($T$) which interchanges bits and checks or, more generally, by Kramers-Wannier (KW) dualitues) or operations in which new codes are generated by ``modding out"
 spatial symmetries of the input code ($/G$). 
    \end{enumerate}
    \item \emph{Products of classical codes (Sec.~\ref{sec:products}).}
    \begin{enumerate}
        \item  \emph{Products that create local redundancies (Sec.~\ref{subsec:LocRedProd}).} We discuss the \emph{tensor product} ($\otimes$) that takes two classical codes with no local redundancies ($\mathscr{D}_c=1$) and produces a classical code that has such redundancies ($\mathscr{D}_c = 2$)\footnote{More generally, the tensor product of two classical codes $\mathscr{C}_A$, $\mathscr{C}_B$ with code-dimensionality $\mathscr{D}_A$, $\mathscr{D}_B$ yields a yields a classical code with code-dimension $\mathscr{D}_A + \mathscr{D}_B$.}. For example, the tensor product of two 1D Ising models yields the 2D Ising model. 
        
        Combining it with the modding out of symmetries yields the \emph{balanced product} ($\otimes_G$).
        
        \item \emph{Products that create subsystem symmetries (Sec.~\ref{subsec:SubSysProd}).} We discuss \emph{check} ($*$) and \emph{cellular automaton} ($\otimes_{\rm CA}$) products, which leave $\mathscr{D}_c$ invariant but change the structure of symmetries (and may increase $D$); for example, these products can turn global symmetries that act on a finite fraction of bits to subsystem symmetries that act only on some parameterically smaller subset. As an example, the check product of two 1D Ising models generates the 2D plaquette Ising model which has line-like subsystem symmetries, while the CA product of two 1D Ising models generates the Newman-Moore model with fractal subsystem symmetries corresponding to flipping spins on Sierpinski triangles. Such subsystem symmetries play an important role in construction of fractonic models in which excitations have restricted mobility.

        \item \emph{Cubic product (Sec.~\ref{subsec:CubicProd}).} We introduce a new product that creates both local redundancies (i.e. increases $\mathscr{D}_c$) and creates subsystem symmetries: it takes three $\mathscr{D}_c = 1$ classical codes as its input and creates a classical code with $\mathscr{D}_c = 2$ which also has (``planar'') subsystem symmetries. For example, the cubic product of three 1D Ising models yields the 3D plaquette Ising model. 
        
    \end{enumerate}
    \item \emph{From classical to quantum (Sec.~\ref{sec:GaugeAndHiggs}).} 
    \begin{enumerate}
        \item \emph{Gauging and Higgsing.} We discuss two mappings that turn classical codes into quantum stabilizer models: ``gauging" ($\mathscr{G}$) results in non-trivial CSS codes while ``Higgsing" ($\mathscr{H}$) produces cluster states corresponding to SPT phases.
        \item \emph{Hypergraph product codes (Sec.~\ref{subsec:HGP}).} Combining tensor products with gauging yields hypergraph product codes, whose properties are inherited from the two classical codes that are fed into the tensor product. Example: the toric code is the hypergraph product of two 1D Ising models, and can be obtained by gauging a 2D Ising model. 
        
        \item \emph{Generalized $X$-cube models (Sec.~\ref{subsec:GXC}).} Combining the cubic product with gauging yields a new family of quantum CSS codes, whose properties are derived from a triple of classical codes and generalize those of the $X$-cube model (which is obtained by gauging the 3D plaquette Ising model). 
    \end{enumerate}
    \item \emph{Examples (Secs.~\ref{sec:Examples} and~\ref{sec:NonEuclidean})}. 
    
    Combining these ingredients leads to a combinatorial explosion of systematic ways to generate new models with desired features.  
    \begin{enumerate}
        \item \emph{Euclidean models (Sec.~\ref{sec:Examples}).} The power of the code factory is demonstrated first by recovering a large family of known phases from the 1D Ising model (see Figs.~\ref{fig:CodesLocalRed} and~\ref{fig:MapOfPhases}) and then by constructing new examples (Sec.~\ref{subsec:NewCodes}).
        \item \emph{non-Euclidean models (Sec.~\ref{sec:NonEuclidean})}. We discuss product codes on non-Euclidean graphs, first identifying recent examples that have appeared in the physics literature (Sec.~\ref{subsec:LaplacianArboreal}; see also Fig.~\ref{fig:CodesLocalRed}) and then reviewing constructions of good qLDPC codes (Sec.~\ref{subsec:GoodCodes}).
    \end{enumerate}
\end{enumerate}

\section{Building blocks: classical codes on graphs}\label{sec:BuildingBlocks}

In this section, we review a few different constructions to obtain classical codes either on Euclidean lattices or more general graphs. These can serve as building blocks to construct classical codes with increasing structure via various product constructions, discussed in Sec.~\ref{sec:products}, as well as starting points for constructing quantum stabilizer models as discussed in Sec.~\ref{sec:GaugeAndHiggs}.

\subsection{Graphs of interest}

We are interested in LDPC codes which are not necessarily spatially local in Euclidean space. A way to obtain such codes---one that is natural from a physical perspective wherein one usually begins with some underlying geometry on which the physical degrees of freedom are arranged---is to begin with some graph $\Gamma$ of bounded degree, and define codes that are local on this underlying graph geometry. 

To make this notion more precise, we need to consider a \emph{family} of graphs, $\Gamma_l$, labeled by some integer $l$, with the number of vertices going to infinity as $l$ is increased, so that we can speak of some notion akin to that of a ``thermodynamic limit''. At the same time, we want the degree (maximal number of neighbors that a vertex can have) of $\Gamma_l$ to have a constant upper bound, independent of $l$, to have some meaningful notion of locality. In particular, we can define the distance between any two vertices as the number of edges in the shortest path connecting them\footnote{Note that, in contrast with our discussion of general chain complexes, here we mean edges in the usual sense, connecting exactly two vertices.}, and we can consider a ``ball of radius $r$'' in the graph as all vertices with distance at most $r$ from a given vertex. From a physical perspective, to have a well-defined thermodynamic limit, we would also want to require that the members of this family of graphs all look ``locally similar'' in some sense\footnote{One possible way of making this more precise is as follows. Let us assume the graphs $\Gamma_l$ are \emph{vertex transitive}, meaning that there is a graph automorphism mapping any vertex to any other. In that cast, we can speak of a ball of radius $r$, $B_l(r)$ without having to specify its origin. We can then require that for any $r$, there exists an $l(r)$ such that whenever $l > l(r)$ the ball $B_l(r) = B(r)$ becomes independent of $l$.}$^{,}$\footnote{We could relax this notion to allow for families of random graphs which look \emph{statistically} similar locally.}.

We want to consider models which are local with respect to $\Gamma_l$. For example, if we imagine placing degrees of freedom (bits, qubits) on the vertices of $\Gamma$, we would want the interactions between them (e.g., the parity checks of some stabilizer code) to only involve vertices that are within a finite graph distance from one another. 

For example, we can place degrees of freedom (bits, qubits) on the vertices of $\Gamma$ and require the interactions between them (e.g., the parity checks of some stabilizer code) to only involve vertices that are within a finite graph distance from one another. In other words, the Hamiltonian associated with such a code is a sum of terms such that each term only contains vertices within some ball radius $r$ where $r$ can be chosen independent of $l$. This, along with the bounded degree, ensures that the energy is extensive. The bounded degree also gives rise to a meaningful notion of a Lieb-Robinson bound on these graphs~\cite{hastings2006spectral}, which plays an important role in the theory of stable (gapped) phases~\cite{bravyi2011short}. 

An obvious set of graphs that satisfy our criteria are finite-dimensional \emph{Euclidean lattices}, which arise as cellulations of $D$-dimensional Euclidean space. Such lattices are an obvious setting for studying phases of matter in condensed matter physics and they are also a natural setting from the perspective of error correction, where they correspond to the intrinsic layout of various qubit architectures. However, they come with limitations: as shown by Ref. \onlinecite{bravyi2010tradeoffs}, quantum stabilizer codes that are local in Euclidean lattices satisfy the bound $k d^{\frac{2}{D-1}} \leq O(n)$, signalling a fundamental tradeoff between the amount of information encoded $k$ and its robustness, as measured by the code distance $d$. This motivates considering codes on more general graphs, which can evade this bound. More generally, one can ask about the ground states of local gapped Hamiltonians on such graphs: unlike their Euclidean counterparts~\cite{hastings2006spectral,hastings2007area}, we know little about the limitations of many-body quantum states that can arise in this context.

Of particular importance for constructing stabilizer codes are \emph{expander graphs}~\cite{hoory2006expander,sipser1996expander,breuckmann2021quantum}. These have a number of different definitions. One, which we will refer to below is \emph{vertex expansion}: a graph with vertices $V$ is a $(\gamma,\alpha)$ vertex expander if for any set of vertices $A \subset V$, such that $|A| \leq \gamma |V|$, we have $|N(A)| \geq \alpha |A|$, where $N(A)$ is the set of vertices neighboring $A$. Another one is that of \emph{spectral expansion}, which is measured by the gap between the two largest eigenvalues of the adjacency matrix of $\Gamma$. This is connected to vertex expansion by Cheeger's inequality~\cite{hoory2006expander}, which upper bounds the constant $\alpha$ in terms of the spectral gap. 

A useful way of obtaining interesting graphs, which can include expanders, is as \emph{Cayley graphs} of some discrete group $G$. To define these, one needs to choose some generating set $S$, i.e. a subset of $G$ such that any element of $G$ can be written as a product of elements from $S$. The Cayley graph $\Gamma(G,S)$ is then constructed by assigning a vertex to every element of $G$, and drawing an edge between two elements if they are related by multiplication with an element of $S$, i.e. there is an edge $(g,gs)$ for every $s \in S$; the degree of the graph is thus set by the size of the set $S$. An advantage of this construction is that the resulting graph is highly symmetric, i.e. it is invariant under mapping vertex $g$ to $gh$, which ensures that all vertices are equivalent. It is known that appropriate choices of $G$ and $S$ can yield expander graphs. In particular, the group $G=\text{PSL}(2,\mathbb{F}_q)$, given by $2\times 2$ matrices with elements from the field $\mathbb{F}_q$ (for $q$ a prime that is equal to $1$ modulo $4$) up to an overall multiplication by scalars, has Cayley graphs that are optimal spectral expanders, known as \emph{Ramanujan graphs}~\cite{lubotzky1988ramanujan}. These graphs are used in the construction of some of the good qLDPC codes we discus in Sec.~\ref{subsec:GoodCodes}.

\subsection{Classial codes on a generic graph $\Gamma$}\label{subsec:CodesOnGraphs}

Given a graph $\Gamma$ that sets the underlying geometry, there are many ways of defining classical codes (and hence Hamiltonians) that are local on this graph. We now discuss a few ways of achieving this, which will play a role in what follows. These are also depicted pictorially in the top-left panel of Fig.~\ref{fig:Factory}. As we will see in the rest of the paper, these codes can then be used as ingredients to obtain other, more elaborate models using the various constructions we discuss below. Throughout this section, $\Gamma=(V,E)$ stands for a graph with vertices $V$ and edges $E$. 

\subsubsection{Ising model}

Arguably the simplest example of a code local on $\Gamma$ is the Ising model, $\mathcal{I}(\Gamma)$, defined by assigning a spin $\sigma_i$ to every vertex $i\in V$ and a two-spin check $\sigma_i\sigma_j$ to all pairs of vertices $i,j\in V$ that are connected by an edge $e=(i,j)\in E$. This corresponds to a code with parameters $[n,k,d] = [|V|,1,|V|]$, encoding a single bit of logical information into the ``all up'' and ``all down'' spin configurations. In the following we will often encounter the case where $\Gamma$ is either the cycle graph (i.e., a closed 1D chain) or a 2D square lattice; we will denote these by $\mathcal{I}_\text{1D}$ and $\mathcal{I}_\text{2D}$, and refer to them as the 1D and 2D Ising models, respectively. 

While the code rate and distance are thus independent of the structure of the graph, other features of the Ising model can depend on the geometry in crucial ways as discussed in Sec.~\ref{subsec:physicalandccgeometry}. This is illustrated by contrasting the cases of the Ising model on one- and two-dimensional lattices, only the latter of which exhibits thermal stability. Another example was pointed out by Freedman and Hastings~\cite{freedman2013quantum}, who showed that the Ising model on an expander graph can have features that are not realizable on any finite dimensional lattice. In particular, they may display the property that any state (i.e., probability distribution over spin configurations) with low energy density with respect to the Ising Hamiltonian must have long-range spin correlations, providing a classical version of the so-called ``no low-energy trivial states'' (NLTS) property.

In the Ising model, there is a direct relationship between the underlying graph $\Gamma$ and the code considered from the chain complex perspective discussed in Sec.~\ref{sec:ClassicalReview}, with a one-to-one correspondence between bits (checks) and vertices (edges) in $\Gamma$. This also implies that redundancies of the code correspond to loops in the graph. The local redundancies that would form the third level of a chain complex are ``short loops'' containing a finite number of edges. On the other hand, if the size of the smallest loop\footnote{Known as the \emph{girth} of the graph.} diverges in the thermodynamic limit then the code has no local redundancies. Indeed, the presence of short loops is responsible for the aforementioned thermal stability of the 2D Ising model.

\subsubsection{Laplacian model}

Moving on the examples where $k$ and $d$ have a non-trivial dependence on the graph $\Gamma$, we consider a set of models inspired by Ref. \cite{gorantla2023gapped}. Let us again place bits on the vertices of $\Gamma$. We will now also assign a check to each vertex in the following way:
\begin{equation}\label{eq:Laplacian}
    C_i = \prod_{e=(i,j)} \sigma_i \sigma_j.
\end{equation}
In other words, $C_i$ is the product of all the Ising checks along edges that emanate from vertex $i$. Equivalently, the matrix $\delta$ for this code is the \emph{Laplacian matrix}\footnote{The graph Laplacian is defined as $L = A - D$ where $A$ is the adjacency matrix of the graph and $D$ is the degree matrix, i.e. a diagonal matrix where the element $D_{ii}$ is equal to the degree (number of neighbors) of vertex $i$.} of $\Gamma$ with each matrix element taken modulo 2. For this reason, we will refer to it as the \emph{Laplacian model} on graph $\Gamma$ and denote it by $\mathscr{L}(\Gamma)$. 

By construction, $\mathscr{L}(\Gamma)$ is Ising symmetric, i.e., invariant under flipping all the spins, which means that it encodes at least one bit of logical information. However, it might have many more codewords depending on the choice of the graph $\Gamma$. An expression of $k$ and a choice of basis for codewords can be constructed from the so-called \emph{Smith decompositon} of the graph Laplacian, as was shown in Ref. \cite{gorantla2023gapped}\footnote{As we will discuss below in Sec.~\ref{sec:NonEuclidean}, the quantum codes of Ref. \onlinecite{gorantla2023gapped} are \emph{hypergraph products} between the Laplacian model and the 1D Ising model. The logicals are inherited from these classical codes, as we discuss in Sec.~\ref{subsec:HGP}, so that the expressions for ground state degeneracy (Eq. 3.7) and symmetry operators (Eqs. 3.4-3.5) in Ref. \cite{gorantla2023gapped} can be re-interpreted as properties of the Laplacian model.}. Reading off $k$ and $d$ of the Laplacian model $\mathscr{L}(\Gamma)$ from the graph $\Gamma$ is thus not straightforward but as Ref. \onlinecite{gorantla2023gapped} shows, one can make both of these scale non-trivially with the number of bits. The motivation of constructing the models of Ref. \onlinecite{gorantla2023gapped} was inspired by the study of fracton phases and indeed, they show that excitations associated to the Laplacian model tend to be immobile and thus fracton-like. 

It can be shown that redundancies of $\mathscr{L}(\Gamma)$ are in one-to-one correspondence with its logicals. As a consequence, if $\Gamma$ is such that $\mathscr{L}(\Gamma)$ has a non-trivial code distance (i.e., one that diverges in the thermodynamic limit), then there are no local redundancies present. 

\subsubsection{Tanner codes} 

Another construction that associates a family of classical codes to a given graph $\Gamma$, and has played an important role in the development of classical and quantum LDPC codes, is the so-called \emph{Tanner code}~\cite{tanner1981recursive,sipser1996expander}\footnote{Not to be confused with the \emph{Tanner graph} defined in Sec.~\ref{Sec:Definitions}. The Tanner graph is a representation of an arbitrary classical code, while Tanner codes provide a specific construction of codes on a given graph. The two are related, however; see below.}. In this case, bits are placed on the edges, rather than vertices of $\Gamma$. Considering a vertex $i$, one can define the ``local view'' of the code, consisting of bits on edges adjacent to $i$ (see Fig.~\ref{fig:Factory}). For a vertex of degree $n_i$, this local view includes $n_i$ bits. On these one can define a ``small code'' $\mathscr{C}_i$, which is some $[n_i,k_i,d_i]$ code consisting of checks supported on the local view of $i$. The Tanner code $\mathcal{T}(\Gamma,\{\mathscr{C}_i\})$ is defined by assigning such a small code to every vertex and combining all their checks to define a code\footnote{Note that the code defined by this prescription is local, involving only nearest neighbor interactions, on the \emph{line graph} of $\Gamma$.}. If $\Gamma$ is a regular graph, such that $n_i = n_0$ is the same for all vertices, then we can choose the small code to be fixed, $\mathscr{C}_i = \mathscr{C}_0$, for all $i$. In that case the Tanner code $\mathcal{T}(\Gamma,\mathscr{C}_0)$ is defined only by the graphs structure and a finite amount of local data\footnote{This also involves some choice of labeling of the edges at each vertex to match them with the bits of $\mathscr{C}_0$.}.

Tanner codes play a prominent role in coding theory thanks to the fact that one can derive general bounds on their code distance based on properties of $\Gamma$ and $\mathscr{C}_0$. In particular, one can prove that if $\Gamma$ is a sufficiently good spectral expander and $\mathscr{C}_0$ has a suffciently large relative distance $d_0 / n_0$, then $\mathcal{T}(\Gamma,\mathscr{C}_0)$ itself will have a relative distance $d/n$ that can be lower bounded by a constant~\cite{sipser1996expander}. One can understand this at an intuitive level, by imagining that we try to construct a non-trivial codeword step-by-step, starting some vertex $v$. Flipping spins to create a non-trivial logical of $\mathscr{C}_v$ creates excitations at neighboring vertices; to remove these, further spins need to be flipped. The combination of large $d_0/n_0$, along with graph expansion, ensure that this ``wave'' of flipped spins keeps spreading until it covers some finite fracton of all edges. At each step one also has multiple choices of which spins to flip, resulting in a finite rate $k/n$ that can be lower bounded by a simple counting argument. This way, Tanner codes on expander graphs can lead to good cLDPC codes that satisfy $k,d \propto n$. One particular construction uses the Ramanujan graphs mentioned in the previous section~\cite{sipser1996expander}. 

Tanner codes involve some familiar examples. For example, the Tanner code on a cycle graph with the obvious non-trivial choice of local code that involves both edges meeting at a vertex is equivalent to the one-dimensional Ising model. The two-dimensional Ising model can also be written in the Tanner form by grouping four checks together into a small code (see Fig.~\ref{fig:goodQLDPC}(a),(b) below). More generally, we point out that any cLDPC code can be embedded into a Tanner code at the cost of increasing the number of bits by an $O(1)$ multiplicative factor as follows. Given a code $\mathscr{C}$, let us take as $\Gamma$ its Tanner graph, defined in Sec.~\ref{Sec:Definitions}, whose two sets of vertices correspond to bits and check of $\mathscr{C}$ while edges represent the adjacency relations between them. We can define a Tanner code on the Tanner graph as follows: for the vertices that originally represented checks, we take the small code to contain a single check, involving the product of all its adjacent edges. For the other set of vertices, labeled by the original spin indices $i$, we take the small code to be formed by Ising-like two-spin checks between all pairs of edges adjacent to the same vertex\footnote{If $|\delta^T(i)| > 2$ then some of these will be redundant and could be omitted at the cost of making the construction look less symmetric.}. These Ising interactions have the effect of forcing these set of spins to agree with each other, which ensures that the codewords of this Tanner code coincide with those of $\mathscr{C}$. We thus obtain a Tanner code that has the same $k$ as $\mathscr{C}$ but with a larger number of bits $n = \prod_i |\delta^T(i)|$\footnote{The code distance will also be increased as a codeword supported on some set $\lambda$ of spins will turn into a codeword involving $\prod_{i\in\lambda} |\delta^T(i)|$ spins. If we assume that every $i$ is involved in exactly $w$ checks in $\mathscr{C}$ than the ``Tannerized'' code has parameters $[wn,k,wd]$.}. In more physical terms, we can think of this as splitting every site $i$ into a small cluster, containing as many sites as the number of checks that involve $i$, and adding Ising interactions between the members of the cluster that prefer them to align. If we associate a Hamiltonian to the code, we could imagine multiplying these Ising terms by a large coupling constant, such that at low energies the Tanner code model reduces to the original code $\mathscr{C}$. 
 
\subsection{Transformations of codes}\label{subsec:OldToNew}

Given a classical code, there are various ways of transforming it into a different code in systematic ways. We review some of these mappings here. They will also play a role in the product constructions discussed in the next section.

\subsubsection{Transpose code} 

A simple way of turning a cLDPC code $\mathscr{C}$ into another is by taking its \emph{transpose}, which we denote $\mathscr{C}^T$. This amounts to the map $\delta \to \delta^T$, which has the effect of exchanging bits and checks of the code. By construction, the transpose also exchanges codewords i.e. symmetries (elements of $\text{Ker}(\delta^T)$) with redundancies (elements of $\text{Ker}(\delta)$).

Some examples of classical codes and their transposes are shown in Fig. 4 of Part I~\cite{LDPCGauge}. In particular, for some of the examples mentioned in Sec.~\ref{Sec:Definitions}, such as the 1D Ising, 2D plaquette Ising and 2D Newman-Moore models, one finds that the transpose code is isomorphic to the original\footnote{More generally, this is true of translation invariant codes in Euclidean space with a single bit and a single check per unit cell; see also App.~\ref{app:Polynom}.}. The same is true of the Laplacian model, which is invariant under transposition for any choice of $\Gamma$. Since transposition switches the role of redundancies and symmetries, these must also be isomorphic to each other. In the examples mentioned, $\mathscr{D}_c=1$ so there are no local redundancies and both symmetries and redundancies scale non-trivially with $n$.  

On the other hand, the tranpose code is \emph{not} isomorphic to the original for the Ising model on graphs other than the cycle graph; instead $\mathcal{I}(\Gamma)^T$, is a code that has a bit on every edge and a check on each vertex that corresponds to the product of all edge-bits incident on that vertex\footnote{Note that this is also a simple example of a Tanner code.}. In particular, on a 2D lattice, this transpose code has small logical operators (symmetries), corresponding to the local redundancies of the Ising model. This is true more generally of codes that correspond to chain complexes of dimension $\mathscr{D}_c \geq 2$.

\subsubsection{Dual chain complex and Kramers-Wannier dualities}

For codes associated to $\mathscr{D}_c \geq 2$ chain complexes, we can define generalized versions of the transposition operation. For example, one can always map a chain complex to its dual complex, obtained by inverting the order of the various objects, as in Eq.~\eqref{eq:DualComplex}. We can then associate a new classical code to the lowest two levels of this dual complex. When $\mathscr{D}_c=2$, this is analogous to the classical Kramers-Wannier duality that maps the Ising model on a given 2D lattice to the Ising model on the dual lattice. Motivated by this, we will refer to this as the (classical) Kramers-Wannier dual code and denote it by $\mathscr{C}^\text{KW}$. Since $\mathscr{C}$ and $\mathscr{C}^\text{KW}$ are associated to the same two-dimensional chain complex, they both give rise to the same qLDPC code upon gauging~\cite{LDPCGauge}.  

When $\mathscr{D}_c \geq 3$, we have multiple options. As mentioned in Sec.~\ref{Sec:Definitions}, we can take the dual of the entire chain complex (for example, mapping a $D$-dimensional Ising model with bits on sites onto a $D$-dimensional Ising model on the dual lattice). This is the most natural in the sense that, by using the lowest two levels of the dual complex (highest two levels of the original), we still end up with classical codes with a diverging code distance. However, when we come to turning these classical codes into quantum models in Sec.~\ref{sec:GaugeAndHiggs} and \ref{sec:Examples}, it will be useful also to consider classical codes that correspond to populating some intermediate levels of the chain complex, which leads to small logical operators and hence a small code distance for the classical model\footnote{The rough idea is that in the quantum model, we can include the small logicals of these codes as additional stabilizers, thus keeping only a smaller set of ``global'' logicals which are not generated by these. The (2,2) toric code in 4D is obtained via such a construction.}. One could obtain such codes by performing a ``partial Kramers-Wannier duality'': i.e., we could take the chain complex defined by the lowest three levels of a $\mathscr{D}_c \geq 2$ chain complex, and take the dual of this sub-complex to define a new code. 

\subsubsection{Dual code}

Another concept that often appears in the literature of error correcting codes is that of the \emph{dual code} $\mathscr{C}^\perp$ (not to be confused with the dual complexes just discussed). The notation is motivated by the fact that the code subspace of $\mathscr{C}^\perp$ is the orthogonal complement of that of $\mathscr{C}$. This can be achieved by defining the checks of $\mathscr{C}^\perp$ to have the same support as the logical operators of $\mathscr{C}$\footnote{This involves choosing some basis for the logicals first and then mapping each into a check in the dual code. Different basis choices lead to equivalent dual codes, in the sense that they will share the same set of logicals and thus the same $k$ and $d$.}. Clearly, if $\mathscr{C}$ has a large code distance, then $\mathscr{C}^\perp$ will not be LDPC but will instead have checks with large supports that grow with $n$. Nevertheless, the dual code is often useful in analyzing codes and it can be used as an ingredient in certain product constructions that do map LDPC codes to LDPC codes (as we discuss below in Sec.~\ref{sec:products}).

\subsubsection{Modding out symmetries} 

Another important concept is that of the \emph{symmetries} or automorphisms of the code $\mathscr{C}$. Here, we mean ``spatial symmetries'', i.e., permutations of the bits that preserve the structure of the code. More precisely, let $\pi$ be a permutation of $n$ elements. We say that it is a symmetry of the code if, for every check $C_a = \prod_{i\in\delta(a)} \sigma_i$, there is another check $C_b$ such that $\prod_{i\in\delta_a} \sigma_{\pi(i)} = C_b$; for example, the Ising model on a regular Euclidean lattice will be symmetric under lattice translations, and Tanner codes defined in the previous subsection can be made invariant under the action of the group $G$ if the graph $\Gamma$ is chosen to be the Cayley graph of $G$. 

The symmetries of $\mathscr{C}$ clearly form a group. Taking $G$ to be some subgroup of this symmetry group, we can define a new code by \emph{modding out} $G$, which we denote by $\mathscr{C} / G$. This amounts to identifying all the bits that are related to each other by a symmetry action (i.e., are in the same \emph{orbit} of $G$). This induces a corresponding identification of the checks of the code, defining a ``quotient code'' $\mathscr{C} / G$. i.e., let $C_a$ be a check in $\mathscr{C}$ supported on a subset of bits $\delta(a)$, which we can divide into equivalence classes with respect to the action of $G$: the equivalence classes that contain an odd number of bits will constitute the support of a new check in $\mathscr{C} / G$\footnote{This is perhaps easier to see in the linear algebraic language. The support of $C_a$ corresponds to a vector $\sum_{i\in\delta(a)} \ket{i}$. After modding out $G$, each $\ket{i}$ maps to one of the new basis vectors that correspond to equivalence classes of bits. Equivalence classes that occur an even number of times drop out, since we are considering a vector space over $\mathbb{Z}_2$.}. 

One can similarly define symmetries of higher dimensional chain complexes, as a permutation of the vertices such that edges get mapped to edges, plaquettes to plaquettes, etc. In other words, the symmetry is a simultaneous permutation of the different levels that leaves the overall incidence structure (which vertex is part of which edge, which is part of which plaquette, etc.) invariant. We also note that when the symmetry is non-Abelian (as is the case of the Cayley graph-based constructions that enter the examples discussed in Sec.~\ref{subsec:GoodCodes}), one has to specify whether the symmetry is acting from left or right\footnote{Denoting by $g(i)$ the symmetry action of group element $g \in G$ on site $i$, acting from the \emph{left} means that $(gh)(i) = g(h(i))$, while acting from the \emph{right} means that $(gh)(i) = h(g(i))$. The former is naturally thought of as ``multiplying by $g$ from the left'' and the latter as ``multiplying from the right''.}. 

This operation of modding out symmetries plays an important role in the \emph{balanced product} construction described in Sec.~\ref{sec:products}, which was a key step in the construction of good qLDPC codes as discussed in Sec.~\ref{subsec:GoodCodes}.

\section{Combining the building blocks: product constructions}\label{sec:products}

In the preceding section, we described some simple ways of defining cLDPC codes on various graphs. In this section, we turn to various constructions that take multiple such classical codes as inputs and use these to build other interesting classical codes in a systematic way. Such ``product constructions'' can be used to build codes with additional structure (symmetries, redundancies etc.) that may be absent in the original building blocks. 

One important application of this idea is to increase $\mathscr{D}_c$, the effective dimensionality of the chain complex, for example by turning input classical codes without local redundancies ($\mathscr{D}_c=1$)  into a new code that has such redundancies ($\mathscr{D}_c=2$).  This not only produces classical codes with new features (see e.g. our discussion of 1D vs 2D Ising model in Sec.~\ref{Sec:Definitions} above) but also  
plays an important role in obtaining non-trivial \emph{quantum codes} by gauging (as detailed in Part I~\cite{LDPCGauge}).  Converting one-dimensional chain complexes into two dimensional ones has indeed been the original use of a product construction~\cite{tillich2013quantum}. Other products we discuss naturally leave $\mathscr{D}_c$, but change the structure of logical operators / symmetries in interesting ways, in particular by creating \emph{subsystem symmetries}, which have played an important part in the theory of fracton phases~\cite{vijay2016fracton,williamson2016fractal,shirley2019foliated}. Finally, we introduce a novel construction, named ``cubic product'' which takes three classical codes as input and induces \emph{both} local redundancies \emph{and} subsystem symmetries in the output. In Sec.~\ref{sec:Examples}, we will describe how many known interesting models arise from simpler ones via these product constructions and also illustrate their use by obtaining new models in $D=2, 3$ spatial dimensions.

While we will describe most of the product constructions in a general way which applies to arbitrary input codes (and this is indeed where some of their power lies), it will be useful throughout to also specifically consider  models that exhibit translation invariance on a finite dimensional (hypercubic) lattice.  For these, we can make use of the polynomial representation developed in~\cite{vijay2016fracton}, which review in App.~\ref{app:Polynom}. The various products we consider have a simple representation in this language, which we will discuss along with the more general definitions. We will also rely on this formalism to simplify some calculations when we consider specific examples in Sec.~\ref{sec:Examples}.

\subsection{Products that create local redundancies}\label{subsec:LocRedProd}

Here, we discuss a number of constructions which can be used to construct classical codes with local redundancies ($\mathscr{D}_c \geq 2$) from codes that may possess no such redundancies. 

\begin{figure*} 
    \centering
    \includegraphics[trim={1cm 0 0 0},width = 1.\linewidth]{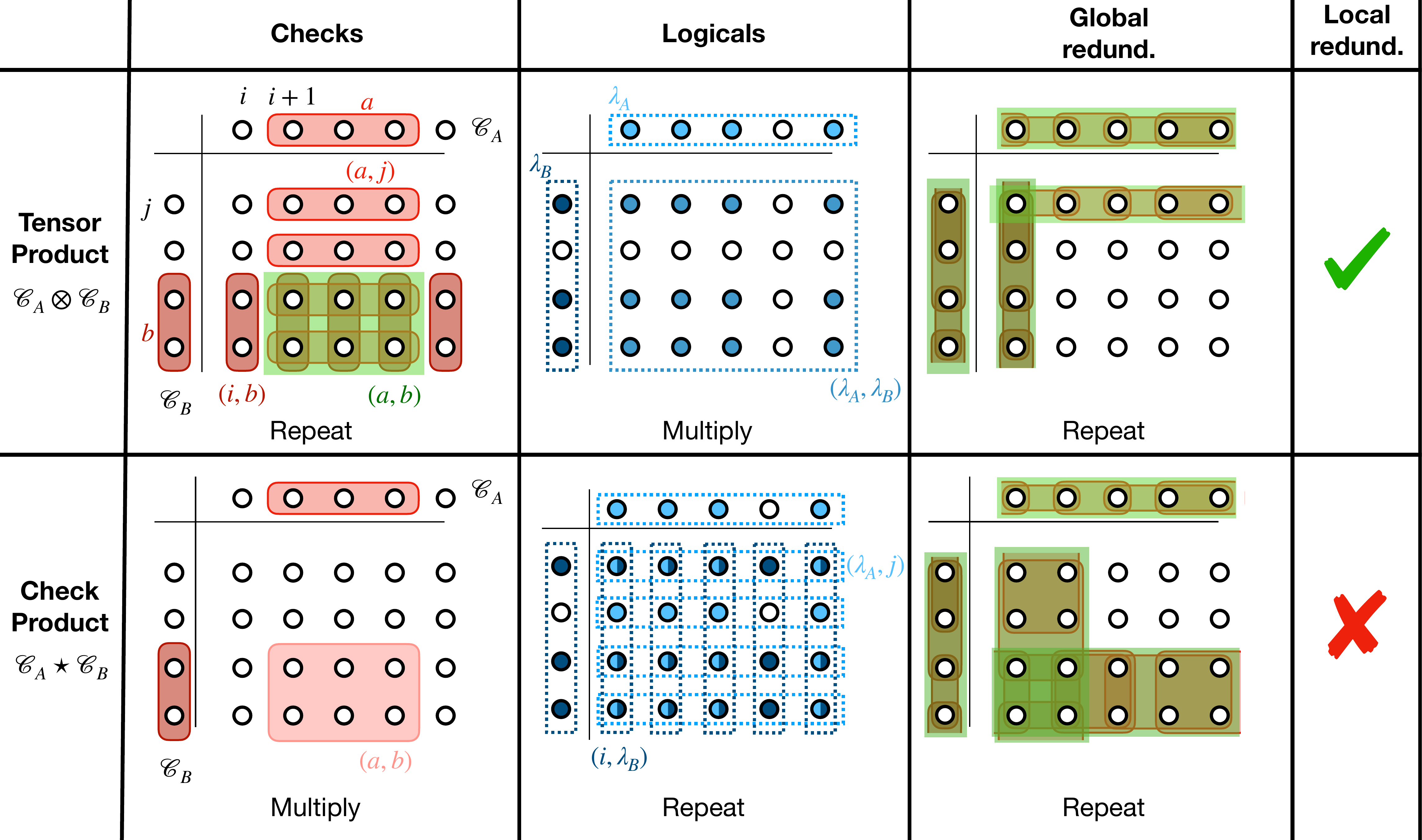}
    \caption{\textbf{Tensor and check products.} Both products take two classical codes, $\mathscr{C}_A$ and $\mathscr{C}_B$. The resulting codes have bits labeled by a pair of indices, $(i,j)$, and can be visualized on a 2D grid. In the tensor product, checks of the input codes (which need not be local on the grid) get repeated along every ``horizontal row'' and ``vertical column'' of the product, while logicals multiply together to form ``rectangles''. For the check product the situation is reversed: logicals get repeated and act along rows or columns, thereby forming ``line-like" subsystem symmetries, while checks multiply.  The tensor product naturally produces local redundancies, one for every pair of input checks $(a,b)$, e.g. the five checks in the green shaded region on the top left multiply to $+1$. In contrast, the check product does not produce local redundancies. Both the tensor and check products inherit the redundancies of their inputs, which is shown for the case of global redundancies in the third column: both  product codes have global redundancies repeated along each row and each column (we only show one row and column in the grid for clarity), with the redundancy of the check product needing a slight modification to account for the plaquette like nature of the checks (see main text). 
    }
    \label{fig:TensorCheckProd}
\end{figure*}

\subsubsection{Tensor product $(\otimes)$}

The simplest construction~\cite{macwilliams1977theory,tillich2013quantum} is the so-called tensor (or direct) product of two codes. A simple example of this is $\mathcal{I}_{\rm 2D}$, the Ising model on a 2D square lattice, which can be understood as a tensor product of two $\mathcal{I}_{\rm 1D}$ Ising models in one dimension. The checks of $\mathcal{I}_{\rm 2D}$ come in two flavors, corresponding to horizontal and vertical edges of the 2D lattice. If we restrict to only one flavor,  $\mathcal{I}_{\rm 2D}$ reduces to multiple decoupled copies of 1D Ising models. We thus view each flavor of check as coming from one of two $\mathcal{I}_{\rm 1D}$ input codes. By combining both horizontal and vertical types of edges/checks together, we end up with local redundancies given by the product of four checks (two vertical, two horizontal) that form the sides of a square plaquette. We write this as $\mathcal{I}_\text{1D} \otimes \mathcal{I}_\text{1D} = \mathcal{I}_\text{2D}$. 

More generally, we can define the tensor product $\mathscr{C} = \mathscr{C}_A \otimes \mathscr{C}_B$ of two arbitrary classical codes $\mathscr{C}_A$ and $\mathscr{C}_B$ as follows.

\paragraph{Checks.}  Let $\mathscr{C}_A$ be a code on bits labeled by $i=1,\ldots,n_A$ and checks labeled by $a=1,\ldots,m_A$, each acting on a subset $\delta_A(a)$ of bits, with $\delta_A$ being a linear map defining the code $\mathscr{C}_A$. Similarly, $\mathscr{C}_B$ is defined on bits $j=1,\ldots,n_B$ through checks labeled by subsets $b$ and defined via $\delta_B(b)$. The tensor product code, $\mathscr{C}_A \otimes \mathscr{C}_B$, is a code acting on $n=n_A n_B$ bits, labeled by the pairs $(i,j)$. The checks of the product code come in two flavors: $\text{\textbf{A}}$-type checks are labeled by the pair $(a,j)$, i.e. a check from the $\mathscr{C}_A$ code and a bit from $\mathscr{C}_B$, while $\text{\textbf{B}}$-type checks are labeled by pairs $(i,b)$, i.e. a bit from $\mathscr{C}_A$ and a check from $\mathscr{C}_B$. The two types of checks are defined as
\begin{align}
    C_{a,j}^\text{\textbf{A}} = \prod_{i\in \delta_A(a)} \sigma_{ij}, & & C_{i,b}^\text{\textbf{B}} = \prod_{j\in\delta_B(b)}\sigma_{ij}.
\label{eq:tensorprodchecks}
\end{align}
The combination of all of these checks defines the tensor product code. 
    
There is a geometrical way to visualize the this construction, shown in Fig.~\ref{fig:TensorCheckProd}. We will make use of this representation repeatedly in the following. First, imagine that the bits of the input code $\mathscr{C}_A$ are arranged along a one-dimensional horizontal line with sites labeled by $i$. Similarly $\mathscr{C}_B$ has sites laid out in a one-dimensional vertical line with sites labeled $j$. Note that we can always lay out the bits of each input code on such a linear geometry, but we are \emph{not} assuming that checks are local along the line. In the product code, the bits are then arranged on a two-dimensional grid, with $i$ and $j$ labeling columns and rows, respectively; the checks of $\mathscr{C}_A$ are repeated along every row, resulting in the checks labeled by the pair $(a,j)$. Since these correspond to edges of a chain complex (see Sec.~\ref{Sec:Definitions}), we will refer to them as ``horizontal edges'' or as ``edges pointing in the $\text{\textbf{A}}$ direction''\footnote{We remind the reader that these are hyperedges which can involve multiple bits.}. Similarly, the checks of $\mathscr{C}_B$ are repeated along every column, resulting in ``vertical edges'', pointing in the $\text{\textbf{B}}$ direction, labeled by the pair $(i,b)$.  Physically, this is a kind of `coupled wire' construction: one takes multiple copies of $\mathscr{C}_A$ (envisaged as a 1D system) and couple them using terms from $\mathscr{C}_B$ (or vice versa).

\paragraph{Logicals.} By construction, the logicals of the product code are products of the logicals of the input codes. In particular, if $\mathscr{C}_A$ has a logical that flips spins in a subset $\lambda_A$ and $\mathscr{C}_B$ has a logical that flips spins in a subset $\lambda_B$, then flipping all spins in the set $\{(i,j)| i \in \lambda_A, j \in \lambda_B\}$ will be a logical operation of $\mathscr{C}_A \otimes \mathscr{C}_B$, as illustrated in Fig.~\ref{fig:TensorCheckProd}. Using the quantum notation, this correspond to an operator $$\mathcal{X}^\text{\textbf{AB}}_{\lambda_A,\lambda_B} = \prod_{i\in \lambda_A, j\in \lambda_B} \sigma_{ij}^x,$$  where the superscript denotes that the logical operator lives in the $\text{\textbf{AB}}$ plane. We therefore have that upon taking the tensor product, both the number and size of logical operators multiplies, so that the product code has 
\begin{align}
    k=k_Ak_B, & & d=d_Ad_B.
\end{align}

\paragraph{Redundancies.} Importantly, a tensor product  construction generates local redundancies even if the input codes did not have any local redundancies\footnote{We remind the reader that  a ``local" redundancy is a low-weight redundancy which satisfies the LDPC property. We are not assuming spatial locality here.}. In particular, there is a redundancy associated to any pair of checks, $(a,b)$. This is due to the fact that 
$$\prod_{j \in \delta_B(b)} C_{a,j}^\text{\textbf{A}} = \prod_{i \in \delta_A(a)}C_{i,b}^\text{\textbf{B}} = \prod_{i \in \delta_A(a)} \prod_{j\in\delta_B(b)}\sigma_{ij}$$ 
which can be seen by writing out the checks explicitly in terms of individual spins using Eq.~\eqref{eq:tensorprodchecks}. This 
implies that we can define a local redundancy $$\prod_{j \in \delta_B(b)} C_{a,j}^\text{\textbf{A}} \prod_{i \in \delta_A(a)}C_{i,b}^\text{\textbf{B}} = 1$$ for every pair of checks $(a,b)$. 
Visually, in Fig.~\ref{fig:TensorCheckProd}, this amounts to either taking a product of rows or columns, resulting in the same rectangular shape. A big advantage of this is that one immediately has a description of all the local redundancies\footnote{This assumes that all the local redundancies are a consequence of the product construction, and the input codes $\mathscr{C}_{A,B}$ themselves have no local redundancies. We will return to this point below}. One can include this explicitly in the description as a third level of a $2$-dimensional chain complex. In this language, the product construction generates a two-dimensional chain complex from two one-dimensional ones, in analogy with the notion of Cartesian product of manifolds. 

Apart from these local redundancies created by taking the product, the tensor product code also inherits the existing redundancies of its inputs. Of particular importance are inputs that may have \emph{global} redundancies (that involve a number of checks that diverges with $n$) even if they lack local redundancies. For example, the $\mathscr{D}_c=1$ codes discussed previously ($\mathcal{I}_{\rm 1D}$, Newman-Moore and 2D plaquette Ising) all have global redundancies even though they lack local ones.  More precisely, if $\rho_A$ denotes the support of a redundancy in $\mathscr{C}_A$ i.e. it denotes a collection of checks $C_a$ in $\mathscr{C}_A$ such that $\prod_{a \in \rho_A}C_a = +1$, then $\prod_{a \in \rho_A} C^{\text{\textbf{A}}}_{a,j} = +1$ for any $j$ and similarly for redundancies of $\mathscr{C}_B$ (see Fig.~\ref{fig:TensorCheckProd}).

\paragraph{Generalizations.} From the chain complex perspective, one can naturally generalize this construction to take tensor products of two higher dimensional chain complexes. The product of two complexes, $\mathscr{C}_A$ and $\mathscr{C}_B$, with dimensions $\mathscr{D}_A$ and $\mathscr{D}_B$ is a new chain complex $\mathscr{C}_A \otimes \mathscr{C}_B$ with dimension $\mathscr{D}_A + \mathscr{D}_B$\footnote{For example, we can write a basis of the linear space $V_p$ in the product $\mathscr{C}_A \otimes \mathscr{C}_B$ as $\ket{v_A}\ket{v_B}$ where $\ket{v_A}$ ($\ket{v_B}$) is a vector from some vector space $V_{p'}^A$ ($V_{p''}^B$) in $\mathscr{C}_A$ ($\mathscr{C}_B$) with $p'+p''=p$. Schematically, the boundary map then acts on this vector as $\delta_p \ket{v_A}\ket{v_B} = (\delta_{p'}^A \ket{v_A})\ket{v_B} + \ket{v_A} (\delta_{p''}^B\ket{v_B})$.}. This also allows one to take repeated tensor products, combining multiple codes into one. For example, by taking repeated tensor product of the 1D Ising model, we can build up the Ising model in arbitrary dimensions\footnote{Note that, if we are concerned only with the classical codes, defined in terms of their bits and checks, then we could just apply the tensor product as we defined it above. However, if we want to correctly keep track of all the local redundancies (and meta-redundancies, etc.) then we need to consider products directly in terms of chain complexes.}.

\paragraph{Polynomial representation.} We now briefly discuss how the tensor product construction works for translationally invariant codes. In the polynomial representation\footnote{See App.~\ref{app:Polynom} for a summary of the polynomial formalism for translation invariant stabilizer codes.}, $\mathscr{C}_{A}$ and $\mathscr{C}_{B}$ are both represented by a matrix of polynomials $S^{A,B}$, of sizes $N_A \times M_A$ and $N_B \times M_B$, respectively, over variables $x_1,\ldots,x_{D_A}$ and $y_1,\ldots,y_{D_B}$, with $D_{A,B}$ the spatial dimensions of the two codes\footnote{Not to be confused with the dimension of the corresponding chain complex, which might be different, as discussed in Sec.~\ref{Sec:Definitions}.} and $N_{A,B}$ ($M_{A,B}$) denoting the number of bits (checks) per unit cell. Their product, $\mathscr{C}_A \otimes \mathscr{C}_B$, lives in $D_A+D_B$ spatial dimensions, and has a stabilizer matrix $S$ made out of polynomials over $D_A + D_B$ variables, and has size $N_A N_B \times (N_A M_B + M_A N_B)$. It naturally divides into two sub-matrices, of size $N_A N_B \times N_A M_B$ and $N_A N_B \times M_A N_B$, which correspond to vertical and horizontal checks. For the vertical ones, we have $S^V_{(IJ)(I'b)} = \delta_{II'} S^B_{Jb}$ and for the horizontal ones $S^H_{(IJ)(J'a)} = \delta_{JJ'} S^A_{Ia}$, where $I (J) = 1,2, \cdots N_{A (B)}$, and $a (b) = 1, 2, \cdots M_{A (B)}$, label the unit cells and checks of both codes. 

To gain some insight, consider the case when $N_A = N_B= 1$, so that there is just one site per unit cell. Then $\mathscr{C}_{A,B}$ are described by a set of polynomials, one for each check per unit cell, and $\mathscr{C}_A \otimes \mathscr{C}_B$ is described by combination of both sets of polynomials, the first set acting on the first variables $x_1,\ldots,x_{D_A}$ while the second set on $x_{D_A + 1},\ldots x_{D_A+D_B}$. For example, the $1D$ Ising model is described by a single polynomial $1+x$, which, upon taking a product with itself, turns into a pair of polynomials, $(1+x,1+y)$ which correspond to Ising checks along horizontal and vertical edges of a $2D$ square lattice. The fact that checks along a square plaquette form a redundancy can be incorporated into the equation $(1+x,1+y) \icol{1+y\\1+x} = 0$ which follows from the fact that we are working with binary variables.

\subsubsection{Balanced product $(\otimes_G)$}

The balanced product, introduced in Ref. \onlinecite{breuckmann2021balanced} is a generalization of the tensor product construction which makes use of the idea of modding out symmetries introduced in Sec.~\ref{subsec:OldToNew}\footnote{For other related constructions see Refs. \onlinecite{panteleev2022asymptotically,hastings2021fiber}. For a review see Ref. \onlinecite{breuckmann2021quantum}.}. As such, it creates codes that still have the local redundancies of tensor product codes, but allows for more general codes that can evade the limitations of tensor products. This will be especially important when we turn these classical codes into quantum ones (see Sec.~\ref{sec:GaugeAndHiggs} below), where the reduction in the number of qubits induced by modding out a symmetry can be used to ``boost" the relative code distance $d/n$. This has played a central role in various recent breakthroughs, including the construction of \emph{good} qLDPC codes~\cite{panteleev2022asymptotically,lin2022good}, as we will review below in Sec.~\ref{subsec:GoodCodes}. The corresponding classical codes also have interesting features that cannot be realized within the confines of the simple tensor product construction, the most important of which is their \emph{locally testability}, which we discuss in Sec.~\ref{sec:EnergyBarriers}.

\begin{figure} 
    \centering
    \includegraphics[trim={1cm 0 0 0},width = 0.85\linewidth]{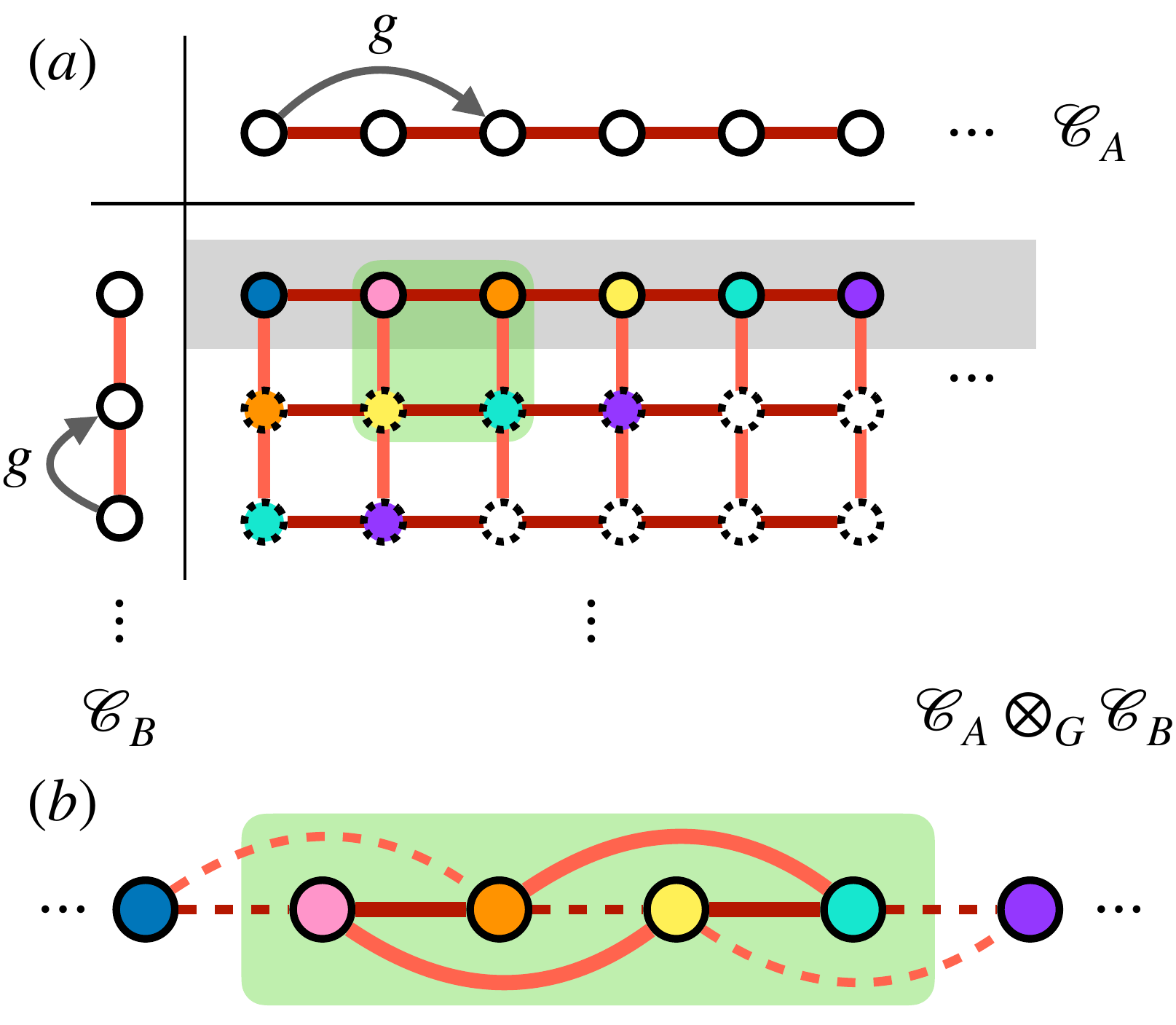}
    \caption{\textbf{Balanced product.} (a) Example of balanced product with the input $\mathscr{C}_A$ ($\mathscr{C}_B$) a 1D Ising model of length $2L$ ($L$). The symmetry group $G$ acts as translation by $2$ sites on $\mathscr{C}_A$ and translation by $1$ site on $\mathscr{C}_B$. Sites related by simultaneous action of both (represented as having the same color) are identified in the product. We can use this to restrict to a single row (shaded gray). (b) The local redundancy, associated to a plaquette in (a), turns into a local redundancy involving nearest and next nearest neighbor checks (thick solid lines) in the balanced product.
    }
    \label{fig:BalancedProduct}
\end{figure}

A somewhat trivial, but illustrative example of the balanced product can be defined using the tensor product of of two 1D Ising models of lengths $2L$ and $L$, respectively. As we saw, this give rise to a 2D Ising model on a $2L \times L$ square lattice. To turn this into a balanced product we can now mod out a symmetry of the product, which translates the first code by two sites and the second one by one site (See Fig.~\ref{fig:BalancedProduct}(a)). As discussed in Sec.~\ref{sec:BuildingBlocks}, modding out the symmetry ``glues together'' sites related by the simultaneous action of the symmetry on both input codes. As a result, the modded out code acts only on $2L$ bits which can be placed on a 1D line again. For this new model,  vertical checks of the product code turn into next-nearest neighbor checks of the modded out model. As a result, the new code maintains the local redundancies of the 2D Ising model, which are now small loops consisting of two nearest neighbor and two next nearest neighbor terms, as shown by Fig.~\ref{fig:BalancedProduct}(b). 

To define the balanced product more generally, one requires a pair of input codes, $\mathscr{C}_A$ and $\mathscr{C}_B$, that are both symmetric under some action of the same symmetry group $G$. When $G$ is non-Abelian (as it is in the case of the good code constructions discussed in Sec.~\ref{subsec:GoodCodes}), one takes it to act from the left on $\mathscr{C}_A$ and from the right on $\mathscr{C}_B$. The balanced product is then defined by first taking the tensor product $\mathscr{C}_A \otimes \mathscr{C}_B$ and then modding out the simultaneous (diagonal) action of the symmetry on both sides of the product:
\begin{equation}
    \mathscr{C}_A \otimes_G \mathscr{C}_B \equiv (\mathscr{C}_A \otimes \mathscr{C}_B) / G.
\end{equation}
Thus, the balanced product combines the tensor product with the idea of modding out symmetries to create a new code out of the two codes $\mathscr{C}_{A,B}$. In doing so, it reduces the number of bits: for example, if we assume, as will be the case for relevant examples, that $G$ acts freely, i.e. no bits are left in place by any symmetry operator (other than the trivial one), then $\mathscr{C}_A \otimes_G \mathscr{C}_B$ will be a code on $n_A n_B / |G|$ bits. The other structures, such as checks, logicals and redundancies, are also inherited from the product code after applying the appropriate identifications. 

In the polynomial language for translationally inviariant codes (reviewed in Appendix~\ref{app:Polynom}), a natural set of symmetries to consider is given by the spatial translations in various directions of the lattice. The result has a simple representation. Let $T_{l}$ represent translations in direction $l=1,\ldots,D$ and let us mod out by the generic translation $T_1^{a_1} \ldots T_{D}^{a_D}$. This has the effect of introducing a relation between the different variables, making $x_1^{a_1} \ldots x_D^{a_D} = 1$. For example, modding out the diagonal translation $T_x T_y$ gives $xy = 1$. This can be used to eliminate $y$ in favor of $x$ in all polynomials, yielding a model in one lower spatial dimension.

\subsection{Products that create subsystem symmetries}\label{subsec:SubSysProd}

Previous studies of gauging maps and stabilizer models have distinguished between usual $\mathbb{Z}_2$ gauge theories (e.g., the toric code) that result from gauging a \emph{global} $\mathbb{Z}_2$ symmetry, and fracton models, which originate from the gauging of \emph{subsystem symmetries}~\cite{vijay2015new,shirley2019foliated}. One can make a similar distinction even for general codes that are not defined on a Euclidean lattice: one might say that a symmetry is global if its support scales linearly with the number of bits, while subsystem if it scales with some smaller power. Here we discuss how certain subsystem symmetries can arise from taking products of codes that only have global symmetries. 

\subsubsection{Check product $(*)$}

One of the simplest models exhibiting subsystem symmetries~\cite{you2018subsystem} is the two-dimensional plaquette Ising model, which we introduced earlier in Fig.~\ref{fig:CodesLocalRed}, which has bits on sites of a square lattice, parity checks acting on the four bits at the corners of a plaquette, and logicals flipping the bits along any row or column. Similar to the 2D Ising model, we can understand this as a product of two one-dimensional Ising models, except now instead of placing the checks of the 1D Ising model on the 2D lattice directly, we combine a pair of two such checks into the 4-body plaquette check of the plaquette Ising model. The subsystem symmetries of this model are then inherited from the logicals of the 1D Ising model. In equations, we write $\mathcal{I}_\text{1D} \star \mathcal{I}_\text{1D} = \mathscr{C}_\text{2DPI}$ where $\mathscr{C}_\text{2DPI}$ denotes the plaquette Ising model in two dimensions. We now describe a construction, the \emph{check product}~\cite{cross2023quantum}, that generalizes this to a product of two arbitrary classical codes. 

\paragraph{Checks.} Just like the tensor product, the \emph{check product} $\mathscr{C}_A \star \mathscr{C}_B$ again acts on $n_A n_B$ bits, labeled by pairs $(i,j)$. Now for every pair of checks $(a,b)$ from the two input codes, we define a new check 
\begin{equation}
C^\text{\textbf{AB}}_{a,b} = \prod_{i \in \delta_A(a), j\in \delta_B(b)} \sigma_{ij}.
\label{eq:checkprodchecks}
\end{equation}
The combination of all of these checks defines the check product. 
Each check is visualized in Fig.~\ref{fig:TensorCheckProd} as a ``plaquette" in the $\text{\textbf{AB}}$ plane using a similar pictorial notation of laying out the two input codes along the ``horizontal" and ``vertical" directions prior to taking the product. We remind the reader that this is merely a matter of visual convenience: the checks are not local along the two directions. 

Formally, we can write this in a compact form by making use of the notion of the dual codes, introduced in Sec.~\ref{sec:BuildingBlocks}. We can then write the check product as
\begin{equation}\label{eq:CheckProdDef}
    \mathscr{C}_A \star \mathscr{C}_B \equiv (\mathscr{C}_A^\perp \otimes \mathscr{C}_B^\perp)^\perp. 
\end{equation}
As noted earlier, the dual of an LDPC code is not necessarily itself LDPC. However, by taking the dual twice in the Eq.~\eqref{eq:CheckProdDef}, one ensures that the resulting code is still LDPC if the inputs were. 

\paragraph{Logicals.} By construction, the check product has logical operators that live along individual horizontal rows and vertical columns in the $\text{\textbf{AB}}$ plane; the former are inherited from the logicals of $\mathscr{C}_A$, while the latter are inherited from $\mathscr{C}_B$. Using the quantum language, if the subset $\lambda_A (\lambda_B)$ defines a logical of $\mathscr{C}_A (\mathscr{C}_B)$, then $$\mathcal{X}^\text{\textbf{A}}_{\lambda_A,j} \equiv \prod_{i \in \lambda_A} \sigma_{ij}^x, \qquad \mathcal{X}^\text{\textbf{B}}_{i, \lambda_B} \equiv \prod_{j \in \lambda_B} \sigma_{ij}^x 
$$ 
 commute with all the checks of $\mathscr{C}_A \star \mathscr{C}_B$ and define ``line-like" logical operators pointing in the $\text{\textbf{A}}$ and $\text{\textbf{B}}$ directions respectively (see Fig.~\ref{fig:TensorCheckProd}). Since the logicals are stretched only along the ``lines" (subdimensional manifolds) of the ``2D $\text{\textbf{AB}}$ grid", the output code exhibits subsystem symmetries. Finally, we note that not all of the logicals are independent, as $\prod_{j \in \lambda_B}\mathcal{X}^\text{\textbf{A}}_{\lambda_A,j} = \prod_{i\in\lambda_A} \mathcal{X}^\text{\textbf{B}}_{i,\lambda_B}$. This implies that the check product code has
\begin{align}\label{eq:CheckProd_k_and_d}
    k= k_A n_B + n_A k_B -k_A k_B, & & d=\text{min}(d_A,d_B)
\end{align}

\paragraph{Redundancies.} In a sense, we can think of the the check product as being dual to the tensor product construction introduced in the previous subsection. In the tensor product, checks are repeated along rows/columns while logicals multiply together; in the check product, the reverse is true: checks are multiplied to form ``plaquette-like'' terms, while logicals get repeated along rows and columns. One consequence of this is that the check product does not induce local redundancies. However, it can be combined with other types of products (such as the tensor product) to construct interesting models which do have such redundancies. We will introduce another way to achieve this in Sec.~\ref{subsec:CubicProd} and discuss some other examples in Sec.~\ref{sec:Examples}. 

On the other hand, the check product does inherit the existing redundancies of its inputs, with a notable case again being the one where the input codes only have global redundancies. For example, if $\rho_A$ is the support of a redundancy in $\mathscr{C}_A$, then we have $\prod_{a \in \rho_A}C^\text{\textbf{AB}}_{a,b}=+1$ for any $b$ and similarly for redundancies of $\mathscr{C}_B$ (see Fig.~\ref{fig:TensorCheckProd}). 

\paragraph{Polynomial representation.} In the polynomial formalism, the stabilizer matrix of $\mathscr{C}_A \star \mathscr{C}_B$ has size $N_A N_B \times M_A M_B$ with polynomials $S_{(IJ)(ab)} = S^A_{Ia} S^B_{Jb}$. Again, this is simplest when there is a single site per unit cell, in which case the two input codes $\mathscr{C}_{A,B}$ are described by a set of polynomials $f^A_a(\mathbf{x})$ and $f^B_b(\mathbf{y})$, while $\mathscr{C}_A \star \mathscr{C}_B$ is defined by their products $f^A_a(\mathbf{x}) f^B_b(\mathbf{y})$. For example, from two $1D$ Ising models we get $(1+x)(1+y) = 1 + x + y + xy$, which corresponds to a 4-spin plaquette check.  

\subsubsection{Cellular automaton product $(\otimes_{\rm CA})$} 

Finally, we discuss a construction originally described in Ref. \onlinecite{devakul2021fractalizing} (see also Refs. \onlinecite{newman1999glassy,bravyi2010tradeoffs,yoshida2013exotic}). While this is not originally formulated as a product, it can be recast in a form that resembles a product construction, at least in some cases, as we explain here. We will refer to this as a \emph{cellular automaton} (CA) product, as the original formulation in Ref. \onlinecite{devakul2021fractalizing} was in terms of a cellular automaton (CA) that is used to extend a code into an additional spatial dimension. The consequence of this is that, for appropriate choices of CA, the resulting code has \emph{fractal} subsystem symmetries. In the translation invariant case, CA and codes are both represented as polynomials, so that we can also interpret this construction as a product of two classical codes. Consequently, in this section we only consider translation invariant codes on a hypercubic lattice, while noting that generalizing this construction to arbitrary codes is an interesting challenge. 

We can again illustrate the idea behind the CA product on the example where both of its inputs are taken to be 1D Ising models. Since we are dealing with translation invariant codes, we formulate  the CA product directly in terms of the polynomial representation\footnote{For a more general formulation, still restricted to Euclidean lattices, see Ref. \onlinecite{devakul2021fractalizing}.}. As we have seen, we can represent one Ising model by the polynomial $f^A(x) = 1+x$ and the other by $f^B(y) = 1+y$. We now define their CA product as a 2D model, with a check associated to the polynomial 
\begin{equation}
f(x,y) = f^A(f^B(y)x) = 1 + x(1+y) = 1+x + xy. 
\end{equation}
This describes a 3-spin check acting on a triangle (see Fig.~\ref{fig:CodesLocalRed}.), corresponding to the Newman-Moore model~\cite{newman1999glassy,devakul2021fractalizing,LDPCGauge} (see Fig.~\ref{fig:CodesLocalRed}). This model has logical operators that correspond to flipping spins along subsets that form Sierpinski triangles. This is indeed the feature that motivates considering the CA construction: it naturally gives rise to a variety of fractal symmetries~\cite{devakul2021fractalizing}, which both result in interesting classical codes, and can serve as the basis of building interesting fracton phases~\cite{vijay2016fracton,williamson2016fractal,shirley2019foliated}.

We can easily generalize the above construction to the case when $\mathscr{C}_A$ is an arbitrary one-dimensional code and $\mathscr{C}_B$ is a code in spatial dimension $D_B$, both described by a single polynomial $f^A(x)$ and $f^B(\mathbf{y})$ (i.e., they have a single site and check per unit cell). Their CA product will be then defined by its polynomial on $D_B+1$ variables via
\begin{equation}\label{eq:CA_1D}
f(x,\mathbf{y}) = f^A(f^B(\mathbf{y})x). 
\end{equation}
The term inside the parenthesis describes the checks of $\mathscr{C}_B$, defined in $D_B$ spatial dimensions, translated by one in the extra dimension. If we write $f^A = x^{p_1} + x^{p_2} + \ldots$, for some set of integers $p_1,p_2,\ldots$, we find that Eq.~\eqref{eq:CA_1D} describes a check where on column $p_r$ in the $(D_B+1)$-th direction we put the check $(f^B)^{p_r}$. 

Finally, we can define the CA product for two codes in dimensions $D_A$ and $D_B$ possibly with multiple checks and sites per unit cell. We represent the two codes by stabilizer matrices $S^A$ and $S^B$, whose columns correspond to polynomials that represent the checks defining each code. To define the product, we will need the two to satisfy a compatibility condition, $M_B = D_A$, i.e., $S^B$ should have one check for each direction in $S^A$. Then we define a new stabilizer matrix as
\begin{equation}\label{eq:CAProd}
    S_{Ia}(\mathbf{x},\mathbf{y}) = S^A_{Ia}\left(\left\{\sum_{J=1}^{N_B} S^B_{Jb}(\mathbf{y}) x_b^J\right\}\right),
\end{equation}
where the expression in brackets is to be interpreted as a series of coordinates $\{x_B\}_{b=1}^{D_A}$ that are fed into the polynomials $S^A_{Ia}$.  This can give rise to a variety of fractal sybsystem symmetries, as was detailed in Ref. \onlinecite{devakul2021fractalizing}.

A notable feature of the CA product is that it is non-commutative: $\mathscr{C}_A \otimes_\text{CA} \mathscr{C}_B \neq \mathscr{C}_B \otimes_\text{CA} \mathscr{C}_A$. As we mentioned, Eq.~\eqref{eq:CAProd} was interpreted originally not as a product, but as a way of using cellular automata to propagate a $D_B$ dimensional code into some extra ``time'' directions (this is clearest when $D_A = 1$ so there is a single cellular automaton and a single time-like direction)~\cite{devakul2019fractal}. Here, we rewrote it as a product by noting that the representation used there for a cellular automaton can equally well be used to define a one-dimensional classical code. The asymmetry arises because in taking the product one of the codes effectively `acts' on the other as a cellular automaton. This feature is reminiscent to the construction of \emph{fibre bundle codes} in Ref. \onlinecite{hastings2021fiber}. It would be interesting to explore further whether there is a closer connection between these two concepts. This could also pave the way towards defining generalizations of the CA product for non-Euclidean geometries. 

\subsection{The cubic product}\label{subsec:CubicProd}

\begin{figure} 
    \centering
    \includegraphics[trim={1cm 0 0 0},width = 0.9\linewidth]{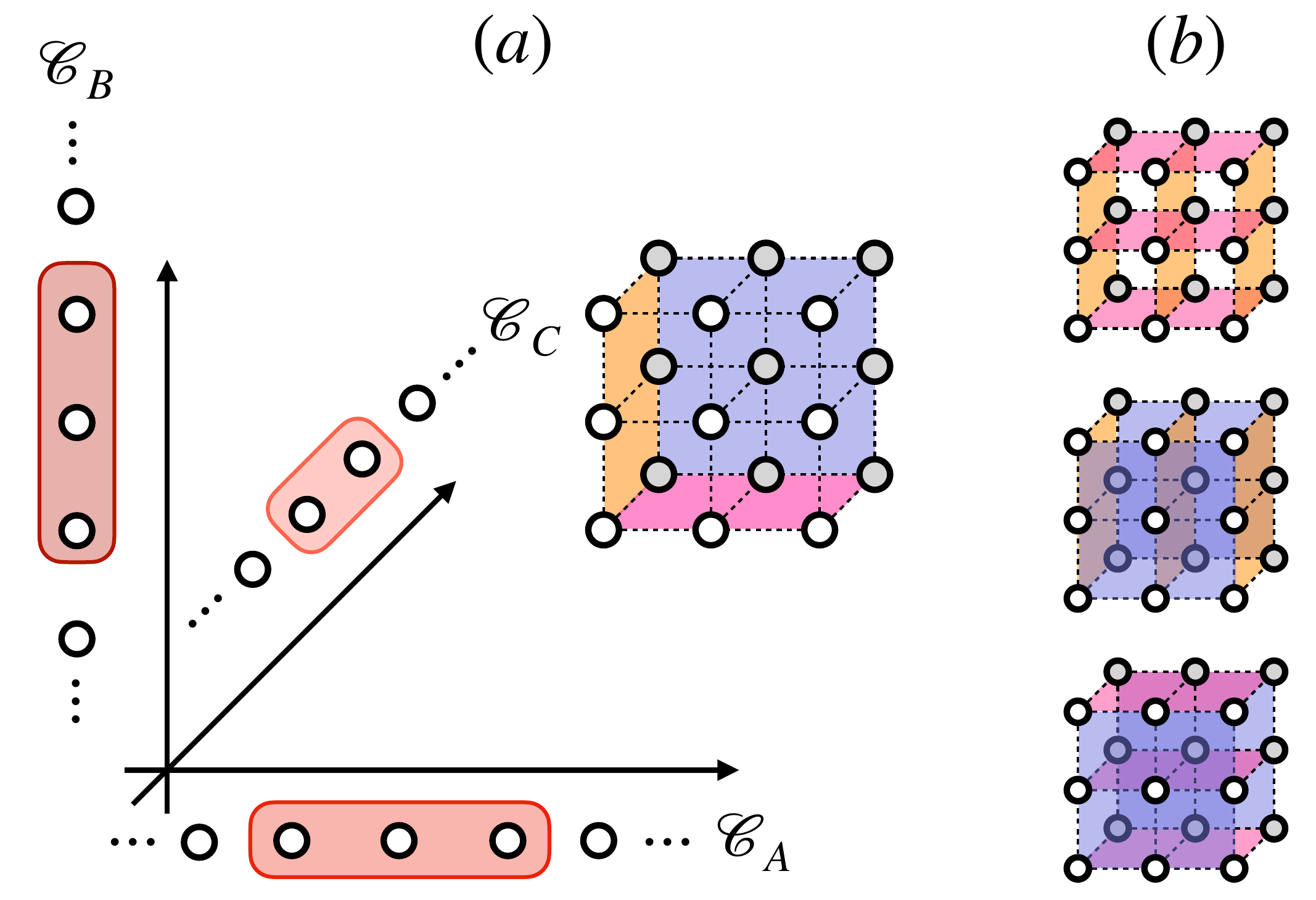}
    \caption{(a) \textbf{The cubic product construction.} Classical bits are placed on a three dimensional grid and copies of three different check products, constructed out of the three input codes $\mathscr{C}_{A,B,C}$ are placed on the three differently oriented planes. (b) Structure of redundancies of the cubic product (see Eq.~\eqref{eq:CubicRedundancies}). A ``stack'' of checks from the $\text{\textbf{AB}}$ plane repeated in the $\text{\textbf{C}}$ direction equals a stack of checks from the $\text{\textbf{BC}}$ plane repeated in the $\text{\textbf{A}}$ direction and also a stack of checks from the $\text{\textbf{AC}}$ plane repeated in the $\text{\textbf{B}}$ direction.}
    \label{fig:CubicProduct}
\end{figure}

The product constructions discussed so far can be used to create \emph{either} local redundancies or subsystem symmetries. A combination of both of these features can lead to exotic physics, as exemplified by the existence of fracton phases~\cite{vijay2016fracton}. Here, we introduce a construction that achieves this. In particular, we describe a kind of ``triple product'' which takes as input three classical codes and output a classical code that has both local redundancies and subsystem-like symmetries; we will call this the \emph{cubic product} construction.

Our construction is inspired by the three dimensional plaquette Ising model, which known to be gauge dual to the $X$-cube fracton model~\cite{vijay2016fracton}. This is a higher dimensional generalization of the 2D plaquette Ising model we have already encountered: it involves placing spins on the sites of a 3D cubic lattice and adding 4-spin interactions on all the faces of the lattice. From our earlier discussion, we can see this as a combination of check products of 1D Ising models, assigned to each two-dimensional plane of the 3D lattice. The resulting model has \emph{planar} subsystem symmetries, corresponding to flipping the spins along any of the 2D planes~\cite{vijay2016fracton} and also has local redundancies (Fig.~\ref{fig:CodesLocalRed}).  We now generalize this in a way that can be applied to arbitrary input codes, including those on non-Euclidean geometries, and therefore has the potential to create various types of interesting chain complexes with corresponding quantum CSS codes. 

\paragraph{Checks.} Let us consider three arbitrary classical codes, $\mathscr{C}_{A,B,C}$, defined on $n_{A,B,C}$ bits, respectively. The cubic product of these three codes will be a classical code with redundancies, defined on $n_A n_B n_C$ bits and we will denote it by $\text{Cub}(\mathscr{C}_A,\mathscr{C}_B,\mathscr{C}_C)$. The bits of this code can be labeled by the triples $(ijk)$, with $i=1,\ldots,n_A$ etc. Generalizing our discussion of the tensor and check product, we will now imagine the bits being arranged on a \emph{three-dimensional} grid with its exes labeled $\text{\textbf{A}}$, $\text{\textbf{B}}$ and \text{\textbf{C}}, after the three codes $\mathscr{C}_{A/B/C}$ (see Fig.~\ref{fig:CubicProduct}). 

If we fix a coordinate $k$, it defines a 2D plane on this grid ($\text{\textbf{AB}}$) with $n_A n_B$ bits. On this 2D grid, we place all the checks of the check product code $\mathscr{C}_A \star \mathscr{C}_B$ in the obvious way. We can label these checks as $C_{a,b,k}^{\text{\textbf{AB}}}$, where $a = 1,\ldots,m_A$ ($b=1,\ldots,m_B$) label checks of $\mathscr{C}_A$ ($\mathscr{C}_B$) and the upper index $\text{\textbf{AB}}$ is used to denote the fact that these checks are defined on the $\text{\textbf{AB}}$ planes of the 3D grid\footnote{Note that, just as in our previous description of tensor and check products, the grid is only used as convenient way of visualizing and labeling the bits and checks and we are not assuming that the codes are spatially local with respect to the distance on the grid.}. We can similarly define checks $C_{a,j,c}^{\text{\textbf{AC}}}$ which are checks of $\mathscr{C}_A \star \mathscr{C}_C$ (with $c=1,\ldots,m_C$) defined on fixed-$j$ planes, and checks $C_{i,b,c}^{\text{\textbf{BC}}}$ on fixed-$i$ planes. The combination of all these checks ($m_A m_B n_C + m_A n_B m_C + n_A m_B m_C$ in number) defines the cubic product of $\mathscr{C}_{A,B,C}$. This construction is illustrated in Fig.~\ref{fig:CubicProduct}(a). 

\paragraph{Logicals.} The cubic product inherits its logical operators from those of the three input codes $\mathscr{C}_{A,B,C}$. For example, if $\lambda_A$ ($\lambda_B$) labels a logical operator of $\mathscr{C}_A$ ($\mathscr{C}_B$), then the operator $\mathcal{X}^{\text{\textbf{AB}}}_{\lambda_A,\lambda_B,k} = \prod_{i\in\lambda_A}\prod_{j\in\lambda_B} \sigma_{ijk}^x$ will be a logical of the cubic product for any choice of $k$. In the 3D grid representation, these logicals live along two dimensional planes. There are two other similar sets of logicals along the $\text{\textbf{AC}}$ and $\text{\textbf{BC}}$ planes. Because these logicals (symmetries) live along planes, the cubic product code has ``planar" subsystem symmetry.

There are also relations between these logicals, for example $\prod_{k \in \lambda_C} \mathcal{X}^{\text{\textbf{AB}}}_{\lambda_A,\lambda_B,k} = \prod_{i \in \lambda_A} \mathcal{X}^{\text{\textbf{BC}}}_{i,\lambda_B,\lambda_C}$ etc. Taking these into account, the overall number of independent logicals is
\begin{align}
    k = k_Ak_Bn_C + k_An_Bk_C + n_Ak_Bk_C - \nonumber \\
    -k_Ak_B - k_Ak_C - k_Bk_C + k_Ak_Bk_C.
\end{align}
The code distance of the cubic product code is 
\begin{equation}
    d = \text{min}(d_Ad_B,d_Ad_C,d_Bd_C).
\end{equation}

\paragraph{Redundancies.} The cubic code construction also naturally gives rise to local redundancies. Taking a check $C_{a,b,k}^{\text{\textbf{AB}}}$ and repeating it along each plane specified by $k \in \delta_C(c)$ gives an operator acting on a three-dimensional volume specified by $a,b,c$. There are three different ways of getting the same volume, obtained by permuting the roles of the three checks (see Fig.~\ref{fig:CubicProduct}(b) for an illustration). In equations, this gives rise to the following set of relations between the checks of the cubic product code:
\begin{equation}\label{eq:CubicRedundancies}
    \prod_{i \in \delta_A(a)} C_{i,b,c}^{\text{\textbf{BC}}} = \prod_{j \in \delta_B(b)} C_{a,j,c}^{\text{\textbf{AC}}} = \prod_{k \in \delta_C(c)} C_{a,b,k}^{\text{\textbf{AB}}}.
\end{equation}
This gives three independent redundancies for each triple $(a,b,c)$, making the total number of local redundancies $3 m_A m_B m_C$\footnote{Note that these are not all linearly independent. however, we include all of them, in order to make the construction symmetric.}. We can now define a 2-dimensional chain complex $\text{Cub}(\mathscr{C}_A,\mathscr{C}_B,\mathscr{C}_C)$ with $V_0$,$V_1$,$V_2$ corresponding to the bits, checks and local redundancies of the code thus constructed. 

This prescription describes a new type of product, which takes a triple of classical codes / $1$-complexes and constructs a $2$-complex $\text{Cub}(\mathscr{C}_A,\mathscr{C}_B,\mathscr{C}_C)$ out of them. We can then assign a CSS stabilizer code to this complex (i.e., by ``gauging'' the classical code with redundancies), which provides a kind of generalization of the $X$-cube model in the same sense that hypergraph product codes are a generalization of the toric code, as we will discuss in the next section. There, we will also explain how the properties of these codes can be related to those of the three hypergraph product codes that can be constructed out of the inputs $\mathscr{C}_{A,B,C}$. 

Apart from the local redundancies that arise from its definition, the cubic product code also inherits the redundancies of the inputs $\mathscr{C}_{A,B,C}$, which turn into redundancies of the pairwise checks products as already discussed above. 

\paragraph{Polynomial representation.} In the polynomial language, we consider translationally invariant codes in Euclidean space, and let us imagine that each of the three input codes is specified by a single polynomial $f^A(\mathbf{x})$ etc. Then the cubic product will have three checks per unit cell, given by the pairwise check products
\begin{align}
    f_1 = f^A(\mathbf{x})f^B(\mathbf{y}), & & f_2 = f^A(\mathbf{x})f^C(\mathbf{z}), & & f_3 = f^B(\mathbf{y})f^C(\mathbf{z}).
\end{align}
The local redundancies correspond to the equation 
\begin{equation}
    f_1(\mathbf{x},\mathbf{y}) f^C(\mathbf{z}) = f_2(\mathbf{x},\mathbf{z}) f^B(\mathbf{y}) = f_3(\mathbf{y},\mathbf{z}) f^A(\mathbf{x}).
\end{equation}
The generalization to more general inputs, with multiple checks per unit cell, is straightforward using the polynomial description of the check product we summarized in the previous section. 

\section{From classical to quantum: Gauging and ``Higgsing''}\label{sec:GaugeAndHiggs}

Our discussion so far has focused on the construction of various classical codes. We can further extend the set of models we can obtain by mapping these classical codes into quantum Hamiltonians, using the ideas expounded in Part I~\cite{LDPCGauge} (building on earlier literature~\cite{wegner1971duality,levin2012braiding,vijay2016fracton,williamson2016fractal,kubica2018ungauging,shirley2019foliated,devakul2019fractal,verresen2022higgs}) which we review here. We will then describe two constructions, hypergraph product codes~\cite{tillich2013quantum} and generalized $X$-cube models, which build on the product code ideas introduced in Sec.~\ref{sec:products} to construct quantum CSS codes. In both cases, we describe how the logical operators and other properties of these quantum models derive from their lower-dimensional classical inputs. 

\subsubsection{Gauging}
\label{sec:Gauging}

One mapping we consider, which we will refer to as \emph{gauging} is aimed at turning a classical code with local redundancies into a quantum CSS code. We take $\mathscr{C}$ to be a classical code with local redundancies, represented as a 2-dimensional chain complex with boundary maps $\delta_{1,2}$ (Eq.~\eqref{eq:classical_2dcc}). We can then define a CSS code as
\begin{align}\label{eq:H_Gauge}
    H_\text{Gauge}(\mathscr{C}) = -\sum_i A_i -\sum_p B_p \\ A_i; \equiv \prod_{a \in \delta_1^T(i)} X_a, & & B_p \equiv \prod_{a\in\delta_2(p)} Z_a.    
\end{align}
Here, $A_i$ ($B_p$) are the $X$- ($Z$-) checks of the CSS code, assigned to sites (plaquettes) of the chain complex, and we combined them into a code Hamiltonian $H_\text{CSS}$. This corresponds to taking the 2-dimensional chain complex associated to a classical code and reinterpreting it as a quantum CSS code as follows (cf. Eqs.~\eqref{eq:quantumcc}, \eqref{eq:classical_2dcc}):
\begin{align}
    \text{classical bits} &\to X\text{-checks} \nonumber \\
    \text{classical checks} &\to \text{qubits} \nonumber \\
    \text{local redundancies} &\to Z\text{-checks}, \nonumber
\end{align}
In other words, this corresponds to the choice 
\begin{align}\label{eq:Gauge_XZ}
    \mathscr{C}_X = \mathscr{C}^T, & & \mathscr{C}_Z = \left(\mathscr{C}^\text{KW}\right)^T,
\end{align}
for the two codes that define the CSS code. By construction, this ensures the commutativity of the quantum checks. We will denote the CSS code obtained by gauging $\mathscr{C}$ as $\mathscr{G}[\mathscr{C}]$.

Properties of this CSS code are inherited from $\mathscr{C}$. For example, redundancies of $\mathscr{C}$ turn into ``Wilson loops'', i.e. products of Pauli $Z$ operators that commute with $H_\text{Gauge}$. As mentioned in Sec.~\ref{Sec:Definitions}, these can be understood as a kind of higher-form symmetry. The Wilson loops can be divided into two kinds, contractible and non-contractible, defined in terms of the underlying chain complex, with the latter corresponding to non-trivial logical $Z$ operators of the CSS code\footnote{In terms of the classical code $\mathscr{C}$, these are \emph{global redundancies}, not generated by the local ones.}. There is a similar picture for logical $X$ operators in terms of the redundancies of the dual code $\mathscr{C}^\text{KW}$. As discussed in Part I~\cite{LDPCGauge}, the relationship between the quantum version of $\mathscr{C}$ (i.e., Eq.~\eqref{eq:QuantizedHam}) and the $X$-checks of $H_\text{Gauge}$ can be understood as a generalized \emph{quantum} Kramers-Wannier transformation. This is distinct from the \emph{classical} KW dualities discussed in Sec.~\ref{sec:BuildingBlocks}; we will discuss a relationship between the two, via a quantum-classical mapping, and also making use of tensor product codes, in App.~\ref{app:QuantumToClassical}. For more details on gauging, see Part I~\cite{LDPCGauge} as well as earlier literature~\cite{wegner1971duality,levin2012braiding,vijay2016fracton,williamson2016fractal,shirley2019foliated,kubica2018ungauging}

\subsubsection{Higgsing}

Another mapping is what we, for lack of a better term, will refer to as \emph{Higgsing}. In Ref.~\cite{LDPCGauge}, motivated by earlier examples~\cite{devakul2019fractal,verresen2022higgs} (see also Ref. \onlinecite{lu2022measurement}), it was described in terms of the Higgs phase of a gauge theory associated to the classical code. More simply, we can describe it as mapping a classical code $\mathscr{C}$ onto the cluster model\footnote{In general, the cluster model on a generic graph $\Gamma=(V,E)$ may be defined as $H=-\sum_{v} X_v \prod_{v'\in N(v)}Z_{v'}$, where $N(v)$ is the set of neighbors of vertex $v$. The definition in Eq.~\eqref{eq:H_Higgs} differs from this by applying a Hadamard transformation, $X \leftrightarrow Z$ on the qubits labeled by $a$. The terms in the cluster model define a unique pure stabilizer state for any $\Gamma$, known as a cluster state or \emph{graph state}~\cite{briegel2001persistent}.} defined on its Tanner graph, namely
\begin{equation}\label{eq:H_Higgs}
    H_\text{Higgs}(\mathscr{C}) = -\sum_a Z_a \prod_{i\in\delta(a)} \tilde{Z}_i - \sum_i \tilde{X}_i \prod_{a \in \delta^T(i)} X_a,
\end{equation}
which is defined on $n+m$ qubits labeled by $i$ and $a$ with Pauli matrices $Z_a$,$X_a$ and $\tilde{Z}_i$,$\tilde{X}_i$, respectively. We will also denote the model $H_\text{Higgs}(\mathscr{C})$ as $\mathscr{H}[\mathscr{C}]$.

The Hamiltonian~\eqref{eq:H_Higgs} has two sets of degrees of freedom, associated to the bits and checks of $\mathscr{C}$. By construction, it is symmetric under both the logical operators of $\mathscr{C}$ and those of its transpose $\mathscr{C}^T$ (with the latter acting as products of $Z$ operators $\mathcal{Z}_\lambda^T = \prod_{a \in \lambda} Z_a$) and thus inherits the nature of these symmetries: for example, if $\mathscr{C}$ had subsystem symmetries, so will $H_\text{Higgs}$. On the other hand, if $\mathscr{C}$ has local redundancies, forming a higher dimensional chain complex, then $\mathscr{C}^T$ has local logicals and we can consider these as generating higher-form symmetries of the same type that would appear in the gauge theory~\eqref{eq:H_Gauge}\footnote{This can be made more explicit by adding the terms $B_p$ to $H_\text{Higgs}$, which would not change its ground states since these are generated by products of the terms already present in Eq.~\eqref{eq:H_Higgs}.}. As we argued in Part I~\cite{LDPCGauge}, the Hamiltonians $H_\text{Higgs}$ generically have features associated with symmetry protected topological quantum phases protected by these set of symmetries. Indeed, they reproduce a number of known examples and we will review some such cases when we discuss examples below. 

\subsection{Hypergraph product (HGP) codes}\label{subsec:HGP}

The first appearance of product constructions in the literature of quantum error correcting codes was in the \emph{hypergraph product} (HGP) code construction of Tillich and Zemor~\cite{tillich2013quantum}. It corresponds to taking taking a tensor product of two classical codes and and then applying the gauging duality to the resulting code\footnote{In fact, the original definition of Ref. \onlinecite{tillich2013quantum} is using the code $\mathscr{C}_A \otimes \mathscr{C}_B^\text{T}$. The reason to take a transpose in one of the input codes is that while is less natural for the resulting classical code, the CSS code obtained after gauging looks more symmetrical in this case, i.e. the $X$ and $Z$ codes defining the CSS code are $\mathscr{C}_X = (\mathscr{C}_A \otimes \mathscr{C}_B^\text{T})^\text{T}$ and $\mathscr{C}_Z = (\mathscr{C}_A^\text{T} \otimes \mathscr{C}_B)^\text{T}$. With this prescription $k_{q} \neq 0$ whenever $k_A,k_B \neq 0$, see Eq.~\eqref{eq:k_hypergraph_prod} below. Nevertheless, we stick to the physically more transparent definition, without taking the transpose of the second term in the product.}$^{,}$\footnote{An important caveat, already mentioned above, is that, in the definition of a hypergraph product code, during the gauging step we only use local redundancies that are generated by the product construction, which might not include all local redundancies, depending on whether $\mathscr{C}_{A,B}$ have any local redundancies themselves. In particular, if they do have such redundancies, it will result in a small quantum code distance}. In equations:
\begin{equation}
    \text{HGP}(\mathscr{C}_A,\mathscr{C}_B) = \mathscr{G}[\mathscr{C}_A \otimes \mathscr{C}_B].
\end{equation}
Combining this definition with Eq.~\eqref{eq:Gauge_XZ}, and using the fact that $(\mathscr{C}_A \otimes \mathscr{C}_B)^\text{KW} = \mathscr{C}_A^T \otimes \mathscr{C}_B^T$, we find that 
\begin{align}
    \mathscr{C}_X = (\mathscr{C}_A \otimes \mathscr{C}_B)^T & & \mathscr{C}_Z = (\mathscr{C}_A^\text{T} \otimes \mathscr{C}_B^T)^T.
\end{align}
so that the $X$ and $Z$ properties (checks, logicals) of the HGP code are related to each other by swapping the input codes $\mathscr{C}_{A/B}$ with their transposes. In this section, we review how properties of these quantum codes, in particular their logical operators, are inherited from the classical codes $\mathscr{C}_A$, $\mathscr{C}_B$, $\mathscr{C}_A^T$ and $\mathscr{C}_B^T$, emphasizing the physical intuition and analogies with the toric code.

\begin{figure} 
    \centering
    \includegraphics[trim={1cm 0 0 0},width = 0.6\linewidth]{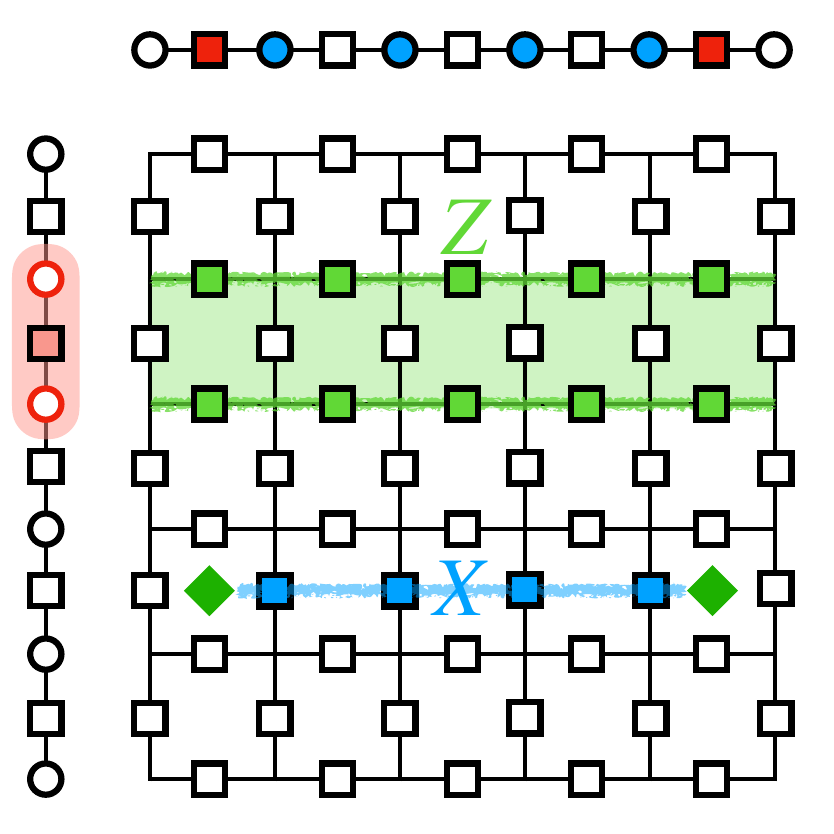}
    \caption{\textbf{The toric code as a hypergraph product.} The $Z$ logicals are inherited from the redundancy of the 1D Ising model and can be shifted by multiplying with Ising checks in the transverse direction. $X$ logicals are inherited from the symmetry of the 1D Ising model and the plaquette excitations created by a truncated logical correspond to a pair of Ising domain walls. 
    }
    \label{fig:HGP}
\end{figure}

\paragraph{The toric code as HGP.} Before discussing the general case, let us discuss the simplest example, where $\mathscr{C}_{A,B}$ are both given by the 1D Ising model $\mathcal{I}_\text{1D}$; as we shall see, much of the general structure generalizes easily to more general HGP codes. In this case, the HGP code is simply the usual 2D toric code but considering it as a hypergraph product gives a useful perspective on its properties (see Fig.~\ref{fig:HGP}). For example, one can decompose the ($X$ or $Z$) checks of the toric code into two parts, acting only on horizontal and vertical edges and notice that taken separately, these have the form of the checks of the 1D Ising model. Relatedly, the logical $X$ and $Z$ operators take the form of the global redundancy and the logical operator of the 1D Ising model, repeated along a particular row or column of the 2D square lattice, with edges oriented either perpendicular (for $X$ logicals) or perpendicar (for Z logicals) to the direction in which the logical is extended. One also needs to consider the equivalence relations induced between these by the checks of the code. For example, the $Z$ logical on row $i$, $\mathcal{Z}_i$, can be multiplied by a set of toric code plaquettes to turn it into a different representation $\mathcal{Z}_{i+1}$ acting on the subsequent row, as we illustrate in Fig.~\ref{fig:HGP}. This feature is directly inherited from the fact that in the 1D Ising model we can transform $\sigma_i^z$ into $\sigma_{i+1}^z$ by multiplying it with the check $\sigma_i^z \sigma_{i+1}^z$. Thus, the rules determining how the logicals can be deformed are also inherited from the classical inputs. 

Closely related to the properties of logicals, properties of the local excitations of the toric code can also be understood from this perspective. Let us, for example, consider a truncated version of a logical $X$ operator, supported on a finite horizontal line with two endpoints (see Fig.~\ref{fig:HGP}). This operator creates two plaquette excitations at its endpoints. These excitations can be separated without any additional energy cost by applying additional $X$ operators. This property is again directly inherited from the 1D Ising model, where a truncated version of the global symmetry operator creates a pair of local domain wall excitations that can be moved around at no additional energy cost. Thus the mobility of excitations and the shape of the energy landscape encountered in interpolating between different ground states is inherited from those of the lower-dimensional classical codes used in the HGP construction. 

Let us now turn to the case of generic HGP codes. As we shall see, using the abstract 2D grid representation of the tensor product we introduced in Sec.~\ref{sec:products}, much of the structure of the toric code generalizes in a straightforward manner.

\paragraph{Checks.} The hypergraph product is associated to a 2D chain complex. We again visualize this on a 2D grid generated by the two input codes, as in our discussion of products in Sec.~\ref{sec:products} (see Fig.~\ref{fig:TensorCheckProd}). In particular, sites are labeled by $(i,j)$, horizontal $\textbf{A}$-type edges by $(a,j)$, vertical $\textbf{B}$-type edges by $(i,b)$ and faces by $(a,b)$ ; these shall play roles analogous to the similar objects in the toric code. We once again remind the reader that this 2D grid is defined by the input codes and their checks (with the output HGP code corresponding to a $\mathscr{D}_c=2$ chain complex). The grid \emph{not} the physical lattice on which the output code lives, which could have any dimension $D\geq 2$; likwise, the checks are not local on this 2D grid, but satisfy the LDPC property. 

Concretely, each edge $(a,j), (i,b)$ hosts a qubit and the checks of the HGP code are defined as
\begin{align}\label{eq:HGP_checks}
    A_{i,j} &= \prod_{a \in \delta_A^T(i)} X_{a,j} \prod_{b \in \delta_B^T(j)} X_{i,b}, \nonumber \\ 
    B_{a,b} &= \prod_{i \in \delta_A(a)} Z_{i,b} \prod_{j \in \delta_B(b)} Z_{a,j}.
\end{align}
We note that if we restrict to one type of edge (horizontal or vertical), then the $Z$-checks simply become those of the input codes $\mathscr{C}_A$ and $\mathscr{C}_B$ alongs rows and columns, while the $X$-checks take the form of the checks in the transpose codes $\mathscr{C}_A^T$ and $\mathscr{C}_B^T$. 

\paragraph{Logicals.} Much like in the toric code, logicals will correspond to sets of edges extended either parallel (for $Z$ logicals) or perpendicularly (for $X$ logicals) to their orientation. There are two sets of each type of logical (one set wrapping around the system in the $\text{\textbf{A}}$ direction and the other in the $\text{\textbf{B}}$ direction) and we can count the number of linearly independent such logicals directly in terms of the properties of $\mathscr{C}_{A,B}$ as we now discuss.

We begin by constructing a basis for the $X$ logical operators. If we restrict our attention to $(i,b)$ vertical edges, then $Z$-checks in Eq.~\eqref{eq:HGP_checks} become equivalent to those of the classical input code $\mathscr{C}_B$. Thus, we can turn any logical of $\mathscr{C}_B$ into an $X$-logical of the HGP. Let $\lambda_B$ denote the support of such a logical. We then have $\mathcal{X}^{\text{\textbf{B}}}_{a,\lambda_B} = \prod_{j \in \lambda'} X_{a,j}$, which is extended in the vertical $\text{\textbf{B}}$ direction, and comprises horizontal $\mathbf{A}$-type edges oriented perpendicular to the direction in which it is extended. Similarly, for any logical $\lambda_A$ of $\mathscr{C}_A$ we can write a logical $\mathcal{X}^{\text{\textbf{A}}}_{\lambda_A,b} = \prod_{i \in \lambda} X_{i,b}$ acting on the vertical edges, and extended in the $\text{\textbf{A}}$ direction. 

The counting of $Z$ logicals works similarly. Restricting to horizontal edges, the $X$-checks become equivalent to the transpose code $\mathscr{C}_A^T$. We can thus construct a $Z$-logical from any logical of this code, or equivalently, from a redundancy of $\mathscr{C}_A$. Let $\rho_A$ denote the support of such a redundancy\footnote{I.e., we have that $\prod_{a \in \rho_A}C_a = +1$, where $C_a$ are checks of the classical code $\mathscr{C}_A$.}. We can turn this into a horizontal logical $Z$ operator of the HGP code, which takes the form $\mathcal{Z}_{\rho_A,j}^{\text{\textbf{A}}} = \prod_{a\in \rho_A} Z_{a,j}$. Similarly, a redundancy $\rho_B$ of $\mathscr{C}_B$ gives rise to a logical of the form $\mathcal{Z}_{i,\rho_B}^{\text{\textbf{B}}} = \prod_{b\in \rho_B} Z_{i,b}$. In the case of $Z$ logicals, the edges are oriented parallel to the direction in which the logical is extended. 

We have thus provided a possible set of $X$ and $Z$ logicals. However, we still need to identify their equivalence classes under the equivalence relation induced by multiplying them with the checks in Eq.~\eqref{eq:HGP_checks}. In other words, we want to know which combinations of logicals are trivial in the sense that they can be obtained as product of checks. Consider the case of $Z$-logicals first. Fix some redundancy $\rho_A$ from $\mathscr{C}_A$ and consider the product $\prod_{a \in \rho_A} B_{a,b} = \prod_{j \in \delta_B(b)} \mathcal{Z}^{\text{\textbf{A}}}_{\rho_A,j}$. In other words, the set of relations between the logicals $\mathcal{Z}^{\text{\textbf{A}}}_{\rho_A,j}$ is induced exactly by the checks of $\mathscr{C}_B$. To count the equivalence classes, we therefore need to find a maximal independent set of $\sigma_j^z$ that cannot occur as products of such checks. Mathematically, this amounts to calculating the dimension of the $\mathbb{Z}_2$ subspace $\text{Im}(\delta_B)^\perp = \text{Ker}(\delta_B^T)$, that is, the space of logicals of $\mathscr{C}_B$, which is given by $k_B$. We have this number of equivalence classes for every independent choice of $\rho_A$, which is counted by $k_A^T$, so that in total we find $k_A^T k_B$ logicals of this type. The same counting is obtained by considering the $X$-logicals $\mathcal{X}^{\text{\textbf{B}}}_{a,\lambda_B}$: deforming these by the $X$-checks in Eq.~\eqref{eq:HGP_checks} amounts to a multiplication by a check of the code $\mathscr{C}_A^T$, giving $k_A^T$ independent classes for each choice of logical $\lambda_B$ from $\mathscr{C}_B$. The analogous counting of logicals acting on $\text{\textbf{A}}$ edges gives $k_A k_B^T$, so that the overall number of logicals is
\begin{equation}\label{eq:k_hypergraph_prod}
    k_\text{q} = k_Ak_B^T + k_A^T k_B.
\end{equation}

This discussion also immediately tells us what the code distances of the quantum code are. In particular, we have
\begin{align}\label{eq:d_hyergraph_prod}
    d_X = \text{min}\{d_A,d_B\}, & & d_Z = \text{min}\{d_A^T,d_B^T\},
\end{align}
where $d_{A,B}$ are the code distances of the classical codes $\mathscr{C}_{A,B}$ and $d_{A,B}^T$ are the code distances of their transpose codes. The overall code distance is the smaller of the two, $ d_\text{q} = \text{min}\{d_X,d_Z\}$. Note that, at best, $d_\text{q} \propto \min\{n_A,n_B\}$, while the overall number of qubits scales as $n_\text{q} \propto n_A n_B$\footnote{More precisely, $n_\text{q} = n_A m_B = m_A n_B$, where $m_{A,B}$ is the number of checks in $\mathscr{C}_{A,B}$. We assume that $m_{A,B}$ scales proportionally with $n_{A,B}$.}. 

A useful way of summarizing the structure of logicals is given by considering the input codes and their transposes as quantum models, as discussed in Sec.~\ref{Sec:Definitions}. It will be convenient to pick a convention where the checks of the transpose code are represented by Pauli $X$, rather than $Z$ matrices, so we would write
\begin{align}
    H(\mathscr{C}_A) = -\sum_{a=1}^{m_A} \prod_{i\in \delta_A(a)} \sigma_i^z & & H(\mathscr{C}_A^T) = -\sum_{i=1}^{n_A} \prod_{a \in \delta_A^T(i)} \tau_a^x.  
\end{align}
We can also write similar Hamiltonians for $\mathscr{C}_B$ and $\mathscr{C}_B^T$. The point of writing the codes in this form is that one can talk about $X$ and $Z$ logicals of the classical codes, as mentioned in Sec.~\ref{Sec:Definitions}. For $\mathscr{C}_A$, the $X$-logicals $\mathcal{X}_{\lambda_A}$ are the logicals of the classical code, flipping between different codewords (ground states of $H(\mathscr{C}_A)$), while the $Z$-logicals $\mathcal{Z}_{\lambda_A}$  are the canonically conjugate variables that can be used to label the different codewords. One can always make an appropriate basis choice where the latter correspond to single-site operators $\sigma_{i_{\lambda_A}}^z$. In the convention we chose, the role of $X$ and $Z$ logicals is reversed in the transpose code: $\mathcal{Z}_{\rho_A}$ labels a logical of $\mathscr{C}_A^T$ (a redundancy $\rho_A$ of $\mathscr{C}_A$), while $\mathcal{X}_{\rho_A}$ labels the ground states of $H(\mathscr{C}_A^T)$ and can be chosen to be in the form $\tau_{a_{\rho_A}}^x$. 

In this language, we can associate a $Z$-logical $\mathcal{Z}^{\text{\textbf{A}}}_{\rho_A,j}$ of the HGP code to a pair $Z$-logicals, $\mathcal{Z}^{\text{\textbf{B}}}_{i,\rho_B}$ corresponding to a logical from $\mathscr{C}_A^T$ and one from $\mathscr{C}_B$. The support of the quantum logical is the Cartesian product of the supports of these two classical logicals. Similarly, the $X$-logicals of the quantum code are products of $X$-logicals of one of the classical codes with the transpose of the other one. This is summarized in Fig.~\ref{fig:HGP_GXC_table}. This also ensures that the quantum logicals inherit their commutation relations from those of the input codes, e.g. $\mathcal{Z}^{\text{\textbf{A}}}_{\rho_A,j_{\lambda_B}}$ and $\mathcal{X}^{\text{\textbf{B}}}_{a_{\rho_A},\lambda_B}$ will anti-commute. 

From this counting of logicals, it follows that one can use the HGP construction to achieve a finite code rate, $k_\text{q} = O(n)$\footnote{This is obtained for example by choosing both $\mathscr{C}_A$ and $\mathscr{C}_B^T$ to have a finite code rate.}; this can be done while also maintaining a non-trivial scaling of the code distance, which is needed to ensure an error threshold as $n\to\infty$. This combination of finite rate and macroscopic distance is not possible in any finite Euclidean dimension, where it is ruled out by the Bravyi-Poulin-Terhal bound~\cite{bravyi2010tradeoffs}. On the other hand, the counting of logicals also shows that the hypergraph product has at most $d_\text{q} \lesssim O(n_\text{q}^{1/2})$, where $n_\text{q} = n_Am_B + m_An_B$ is the number of qubits. This is similar to the toric code and much smaller than the $d_\text{q} = O(n_\text{q})$ scaling that would be required from a good qLDPC code. This upper bound has remained unsurpassed for a long time, until recent breakthrough results~\cite{hastings2021fiber,breuckmann2021balanced,panteleev2022asymptotically}, eventually leading to the construction of good qLDPC codes. The key step in doing so was to go from the simple tensor product of the two input codes to \emph{balanced products}. As we reviewed in Sec.~\ref{sec:products}, modding out the symmetry leads to a reduction in the number of physical bits, which can improve the relative distance $d_\text{q} / n_\text{q}$. When the inputs and their symmetry are chosen approprately, this can be used to obtain the required optimal scaling; we will review the constructions that achieve this in some detail below in Sec.~\ref{subsec:GoodCodes}.

\paragraph{Excitations.} The HGP code inherits some further properties of the input codes, just as in the toric code example. For example, we can consider truncated versions of all of the logical operators above. E.g., consider an operator $\prod_{i \in \tilde{\lambda}} X_{i,b}$, where $\tilde{\lambda} \subset \lambda$ is only a subset of the support of a logical. This will lead to a violation of some of the $Z$-checks of the HGP code. Which checks get triggered is determined simply by $\delta_A^T(\tilde{\lambda})$ and is therefore inherited from the properties of the classical code $\mathscr{C}_A$. In this sense, properties of the excitations in the HGP code derive from those of the classical inputs. 

\subsection{Generalized $X$-cube (GXC) codes}\label{subsec:GXC}

\begin{figure*} 
    \centering
    \includegraphics[trim={1cm 0 0 0},width = 0.95\linewidth]{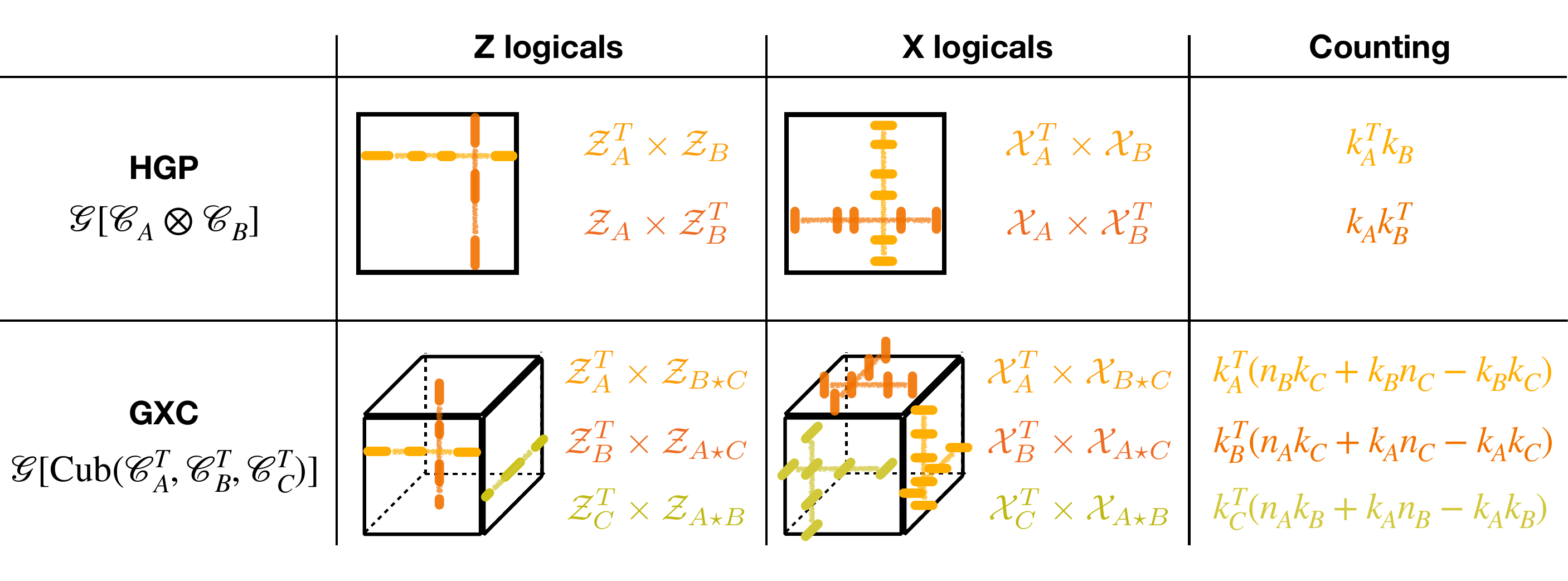}
    \caption{\textbf{Logical operators of hypergraph product and generalized $X$-cube codes.} HGP codes have two sets of $X$ and $Z$ logicals, extended in either of the two directions defined by the underlying product construction (see also Fig.~\ref{fig:TensorCheckProd}). Both types are directly inherited from the logicals of the classical inputs and their transposes, giving the counting on the right. For GXC codes, there are three sets of $Z$ logicals and the corresponding $X$ logicals are inherited from those of the pair-wise check products of the inputs
    }
    \label{fig:HGP_GXC_table}
\end{figure*}

We can perform a similar analysis for codes that are obtained from gauging the classical codes constructed from the three-fold cubic product introduced in Sec.~\ref{subsec:CubicProd}. We refer to this family as \emph{generalized $X$-cube models}, due to their analogy with the $X$-cube model which arises as a special case~\cite{vijay2016fracton}, as we review below. In particular, the generalized $X$-cube model is defined from a triple of classical codes $(\mathscr{C}_A,\mathscr{C}_B,\mathscr{C}_C)$ as\footnote{Using the transposes in the definition of a matter of convention; we chose it to match the usual presentation of the $X$-cube model~\cite{vijay2016fracton}.}
\begin{equation}
    \text{GXC}(\mathscr{C}_A,\mathscr{C}_B,\mathscr{C}_C) = \mathscr{G}[\text{Cub}(\mathscr{C}_A^T,\mathscr{C}_B^T,\mathscr{C}_C^T)].
\end{equation}
To bring to the fore the similarities with the HGP codes discussed in the previous subsection\footnote{See also the next subsection where we connect these two types of codes more directly.}, we will also include an additional Hadamard transformation in the definition, which exchanges $X$ and $Z$ checks compared to our previous convention of the gauging map $\mathscr{G}$\footnote{Thus, it would be more appropriate to call these generalized $Z$-cube codes.}. Our aim in this section is to determine the properties of this code from those of the input codes in a manner similar to the HGP codes above. 

\paragraph{The $X$-cube model.} To gain intuition, we can first consider the usual $X$-cube model, corresponding to the choice $\mathscr{C}_A = \mathscr{C}_B = \mathscr{C}_C = \mathcal{I}_\text{1D}$, with $\mathcal{I}_\text{1D}$ the 1D Ising model. In the convention we have chosen, this has qubits assigned to the edges of a 3D cubic lattice. $Z$-checks act on the $12$ edges along a cube, while each site hosts three different $X$-checks, each acting on the $4$ edges that meet at the site within one of the three orthogonal planes ($A^z$ acts on edges within the $xy$ plane etc.). 

We can form a basis of $Z$ logicals by acting on an entire row of edges parallel to their orientation. This is similar to the case of the toric code in Fig.~\ref{fig:HGP}, and from the perspective of the product construction, we recognize this as being inherited from the redundancy of $\mathcal{I}_\text{1D}$. However, the fracton nature of the $X$-cube model becomes apparent when we consider how this logical can be moved around in the transverse direction. If we denote a logical extended along the $z$ direction as $\mathcal{Z}_{i,j}$ then multiplying it with a set of $Z$-checks turns it into the product $\mathcal{Z}_{i+1,j}\mathcal{Z}_{i,j+1}\mathcal{Z}_{i+1,j+1}$. We recognize this as an inheritance from the 2D plaquette Ising model $\mathcal{I}_\text{1D} \star \mathcal{I}_\text{1D}$: in that code, multiplying a $\sigma_{ij}^z$ operator with one of the plaquette checks turns it into a similar product of three Paulis. Thus, counting the number of independent logicals turns into a calculation of the number of logicals in the plaquette Ising model, which gives $2L-1$ logicals of this type. There are two other sets of $Z$ logicals extended in the other two directions, so that their total number is $k=6L-3$. 

The situation of $X$ logicals is somewhat different, although it obviously has to give the same overall counting. We can form a basis of $X$-logicals by acting on a set of edges extended in a direction \emph{perpendicular} to their orientation; for example, if the logical acts on edges in the $z$ direction and is extended in the $x$ direction, we can denote it schematically  as $\mathcal{X}^z_x$. This is again similar to the toric code example (see Fig.~\ref{fig:HGP}) and originates from the logical of $\mathcal{I}_\text{1D}$. However, the allowed deformations are now different: we can translate the logical in the $z$ direction by multiplying with a set of $X$-checks $A^y$. This feature inherited from the 1D Ising model where flipping a single spin can hop a domain wall over by one site. On the other hand, we cannot move the logical in the $y$ direction; thus, we have $L$ inequivalent logicals of this type. This gives $6L$ $X$-logicals in total (extended in the three possible directions, with two possible orientations in each case). However, not all of these are independent: for example, the product of all $\mathcal{X}_{x}^z$ equals the product of all $\mathcal{X}_{y}^z$. This gives three relations, resulting in the expected counting $k = 6L-3$. 

We can also consider truncating the logical operators and considering the excitations (fractons) thus created. The constraints derived above on the allowed deformations of logicals then turn into mobility constraints of the fractons. One way of understandig it, from the gauging perspective, is that the subsystem symmetries of the ungauged classical code and its Kramers-Wannier dual turn into (global) redundancies of the $X$ and $Z$ checks of the CSS code~\cite{LDPCGauge,vijay2016fracton}, which constrain the patterns in which these checks can be violated. For example, the product of the cubic $Z$-checks on any 2D plane has to equal $+1$, so that any such plane must contain an even number of $Z$-excitations. 

\paragraph{Checks.} We now turn to the case of generic GXC codes. As we will see, the same general structure holds there, \emph{mutatis mutandis}. We start with three input codes $\mathscr{C}_{A,B,C}$ defined by the maps $\delta_{A,B,C}$, which have their bits labeled by $i,j,k$ and their checks $a,b,c$, respectively\footnote{Note that hwere, and in Sec.~\ref{subsec:GXC}, we use $k$ as a bit label, which should not be confused with the number of logical bits. The meaning should be clear from context.}. From these, we can again form the abstract 3D grid discussed in Sec.~\ref{subsec:CubicProd}, which has sites labeled $(i,j,k)$, $\text{\textbf{A}}$-edges $(a,j,k)$ (similarly, $\text{\textbf{B}}$ and $\text{\textbf{C}}$ edges), $\text{\textbf{AB}}$ plaquettes $(a,b,k)$ (similarly, $\text{\textbf{AC}}$ and $\text{\textbf{BC}}$ plaquettes) and cubes $(a,b,c)$\footnote{In other words, this is the 3-dimensional chain complex corresponding to the triple product $\mathscr{C}_A \otimes \mathscr{C}_B \otimes \mathscr{C}_C$.}. On this grid, we place qubits on all the edges, we assign a single $Z$ check to each cube and a triple of $X$-checks to each site as follows:
\begin{align}\label{eq:GXC_checks}
    B_{a,b,c} &= \prod_{\substack{i \in \delta_A(a) \\ j \in \delta_B(b)}} Z_{i,j,c} \prod_{\substack{i \in \delta_A(a) \\ k \in \delta_C(c)}} Z_{i,b,k} \prod_{\substack{j \in \delta_B(b) \\ k \in \delta_C(c)}} Z_{a,j,k}, \nonumber \\
    A^{\text{\textbf{AB}}}_{i,j,k} &= \prod_{a \in \delta_A^T(i)} X_{a,j,k} \prod_{b \in \delta_B^T(j)} X_{i,b,k}, \nonumber \\
    A^{\text{\textbf{AC}}}_{i,j,k} &= \prod_{a \in \delta_A^T(i)} X_{a,j,k} \prod_{c \in \delta_C^T(k)} X_{i,j,c}, \nonumber \\
    A^{\text{\textbf{BC}}}_{i,j,k} &= \prod_{b \in \delta_B^T(j)} X_{i,b,k} \prod_{c \in \delta_C^T(k)} X_{i,j,c}.
\end{align}
Note that restricted to edges of a particular orientation, the $Z$-check has the form of a check product between two of the classical input codes, while the $X$-checks have the form of (the transpose of) a single input code. We will now construct a set of logical operators for this code in terms of the logicals and redundancies of $\mathscr{C}_{A,B,C}$, following the logic of the usual $X$-cube case discussed above.

\paragraph{Logicals.} Let us start by constructing a basis of $X$-type logicals. First, we construct logicals acting entirely on the $\text{\textbf{C}}$-edges $(i,j,c)$. As we noted, restricted to these, the $Z$-checks in Eq.~\eqref{eq:GXC_checks} form the check product $\mathscr{C}_A \star \mathscr{C}_B$, so we merely need to find the logicals of this classical code, which we have already constructed above. For example, one set of logicals is given by 
\begin{equation}\label{eq:GXC_X_logical}
\mathcal{X}_{\lambda,j,c}^{\text{\textbf{C}}} = \prod_{i \in \lambda} X_{i,j,c}, 
\end{equation}
acting on $\text{\textbf{C}}$-oriented edges and extended in the $\text{\textbf{A}}$ direction of the 3D grid. We can also choose logicals extended in the $\text{\textbf{B}}$ direction instead, labeled by some logical $\lambda'$ of the code $\mathscr{C}_B$. We have two more sets of logicals acting on the $\text{\textbf{A}}$- and $\text{\textbf{B}}$-edges, each of which can be extended in either of the two directions perpendicular to its orientation; these are shown in Fig.~\ref{fig:HGP_GXC_table}. 

We can also construct the corresponding basis of $Z$-logicals as follows. Again focusing on $\text{\textbf{C}}$-edges, the $X$-checks in Eq.~\eqref{eq:GXC_checks} reduce to the checks of the code $\mathscr{C}_C^T$. We can therefore construct a $Z$-logical from any logical of this classical code, or, in other words, any redundancy of $\mathscr{C}_C$. For example, picking a redundancy with support $R$, we can write a logical $\mathcal{Z}^{\text{\textbf{C}}}_{i,j,R} = \prod_{c \in R} Z_{i,j,c}$. This gives three sets of $Z$-logicals, extended along the three directions of the grid (see Fig.~\ref{fig:HGP_GXC_table}). 

Having constructed a possible set of logicals, we need to identify their equivalence classes and evaluate the number of logical qubits. This can be done in terms analogous to the discussion of HGP codes above. Starting with $X$-logicals, $\mathscr{X}^{\text{\textbf{C}}}$, we have found that they originate from the logicals of the check product $\mathscr{C}_A \star \mathscr{C}_B$, which we already counted in Eq.~\eqref{eq:CheckProd_k_and_d}. In principle, we can place these logicals along any $\text{\textbf{AB}}$ plane labeled by $c$ but these will not all be independent. Indeed, the product $\prod_{i \in \lambda} A^{\text{\textbf{AC}}}_{i,j,k}$ is equal to $\prod_{c \in \delta_C^T(k)} \mathcal{X}^{\text{\textbf{C}}}_{i,j,c}$. We recognize in this the check of the transpose code $\mathscr{C}_C^T$ and in analogy with the calculation discussed for HGP codes (see the paragraph above Eq.~\eqref{eq:k_hypergraph_prod} in particular), this gives $k_C^T$ distinct choices so we find that the total number of distinct $X$-logicals supported on $\text{\textbf{C}}$-edges is 
\begin{equation}\label{eq:k_Xcube}
    (n_A k_B + k_A n_B - k_A k_B) k_C^T.
\end{equation} 
We find a similar counting for the other two sets of logicals (acting on $\text{\textbf{A}}$ and $\text{\textbf{B}}$-edges, respectively) by appropriately perumting the indiced $A,B,C$. 

We can arrive at the same counting by considering the $Z$-logicals instead. There, we found logicals of the form $\mathcal{Z}^{\text{\textbf{C}}}_{i,j,R}$ where $R$ was a redundancy of $\mathscr{C}_R$. There are $k_C^T$ choices for such redundancies. On the other hand, taking the product $\prod_{c \in R} B_{a,b,c}$ of the $Z$-checks from Eq.~\eqref{eq:GXC_checks}, we find that it equals the product $\prod_{i \in \delta_A(a)} \prod_{j \in \delta_B(b)} \mathcal{Z}^{\text{\textbf{C}}}_{i,j,R}$. This originates from the checks of the code $\mathscr{C}_A \star \mathscr{C}_B$ so that the equivalence classes are in one-to-one correspondence with logicals of the latter. This gives the same overall counting as in Eq.~\eqref{eq:k_Xcube}.

We thus see that the support of quantum logicals corresponds to the Cartesian product of the suppors of a logical from one of the three input codes with the support of a logical from the check product of the other two (see Fig.~\ref{fig:HGP_GXC_table}). This constructive approach to the logicals also provides a formula for the resulting quantum code distance, given by $d_\text{q} = \text{min}(d_Z,d_X)$ where
\begin{align}\label{eq:d_Xcube}
    d_x = \text{min}\{d_A,d_B,d_C\}, & & d_Z = \text{min}\{d_A^T,d_B^T,d_C^T\},
\end{align}
are the $X$- and $Z$-distance, expressed in terms of the code distances of the classical codes $\mathscr{C}_{A,B,C}$ and their transposes. 

\paragraph{Excitations.} Finally, properties of excitations in the GXC code are also inherited from the classical input codes, along with additional constraints on their mobility which arise from the construction itself. For example, let us consider some product of Pauli $X$ operators acting on $\text{\textbf{C}}$-edges and ask what excitations it creates. Based on our discussion so far, this amounts to understanding the nature of excitations in the classical check product code $\mathscr{C}_A \star \mathscr{C}_B$. There are in turn obtained from the two inputs $\mathscr{C}_{A,B}$. For example, consider again a truncated version of the logical operator $\prod_{i \in \tilde{\lambda}} X_{i,j,c}$ where $\tilde{\lambda} \subset \lambda$ is a subset of the support of the logical $\lambda$: this truncated logical now creates excitations on the cubes $(a,b,c)$ with $a \in \delta_A^T(\tilde{\lambda})$ and $b\in\delta_B^T(j)$, i.e., the excitations associated to the code $\mathscr{C}_A$, but each now appearing as a composite of $|\delta_B^T(j)|$ excited cubes. More generally, the ``rectangular'' operator $\prod_{i \in \tilde{\lambda}} \prod_{i \in \tilde{\lambda}'} X_{i,j,c}$ creates excitations in the set $\delta_A^T(\tilde{\lambda}) \times \delta_B^T(\tilde{\lambda}') \times \{c\}$, i.e. at the ``corners'' of the rectangle.

It is also instructive to consider products of Pauli $X$ operators acting on the set of $\text{\textbf{A}}$ edges $(a,j,k)$ and $\text{\textbf{B}}$ edges $(i,b,k)$ within a given $\text{\textbf{AB}}$ plane labeled by a fixed $k$. Restricted to these the $Z$-check $B_{a,b,c}$ in Eq.~\eqref{eq:GXC_checks} becomes $\prod_{i\in\delta_A(a)} Z_{i,b,k} \prod_{j\in\delta_B(b)}Z_{a,j,k}$, in which we recognize the $Z$-check of the HGP code in Eq.~\eqref{eq:HGP_checks}. This gives an alternative perspective on the $X$-logicals, which are indeed the same as logicals of the HGP code within this plane. It also tells us that whatever properties the excitations of the HGP have are also present in the GXC code. We will offer an alternative perspective on this fact in the subsection below. 

We can also consider excitations of the $X$-checks of Eq.~\eqref{eq:GXC_checks}, created by products of Pauli $Z$ operators. For example, a truncated logical of the form $\prod_{c\in\tilde{R}} Z_{i,j,c}$ will excite the $X$-checks $A^{\text{\textbf{AC}}}_{i,j,k}$ and $A^{\text{\textbf{BC}}}_{i,j,k}$ for all $k \in \delta_C(c)$, so that excitations of the GXC code correspond to those of the classical code $\mathscr{C}_C^T$. Another notable feature is that excitations come in pairs and the two labels of the elements in those pairs ($\text{\textbf{AC}}$ and $\text{\textbf{BC}}$ in the example above) depend on the orientation of the logical we truncated. This implies that the logicals cannot be ``bent'' in directions perpendicular to their orientation. For example, if we applied another set of Pauli $Z$ operators on $\text{\textbf{A}}$-edges, these would excite $A^\text{\textbf{AC}}$ and $A^\text{\textbf{AB}}$ checks, which cannot cancel the previously created excitations on any site $(i,j,k)$.

\subsubsection{Coupled layer construction}\label{subsubsec:CoupledLayers}

We can relate to the generalized $X$-cube models of the previous subsection to hypergraph product codes by considering an appropriate generalization of the ``isotropic layer construction'' of the original $X$-cube model discussed in Ref. \onlinecite{vijay2017isotropic}. This gives additional physical insight into the nature of these models; in particular, it gives an interpretation of their logical operators and excitations in terms of the three pairwise hypergraph product codes that one can construct out of the input codes $\mathscr{C}_{A,B,C}$. We will make use of this language when we consider a concrete example below in Sec.~\ref{subsec:NewCodes}.

Let us first review how this construction works in the case of the usual $X$-cube model~\cite{vijay2017isotropic}. One places a toric code on every 2D plane of a 3D cubic lattice, which are initially all decoupled from each other. This means that any given edge hosts two qubits, corresponding to the two planes that intersect at that edge (e.g. an $x$-edge is part of both the $xy$ and $xz$ planes). One then adds a strong local coupling between each pair of such qubits of the form $-t X_e' X_e''$, where the two operators correspond to Paulis acting on the two qubits at the same edge $e$. This coupling commutes with the $X$-checks of the toric codes, but not with the $Z$-checks. In the limit of strong coupling, one gets a constraint $X_e' X_e'' = + 1$ at low energies, yielding a single effective qubit on each edge. The effective Hamiltonian within this low-energy subspace can be found within perturbation theory and it turns out that the lowest order term one can construct out of the toric code plaquette operators that satisfies the constraints is exactly the product of all the $Z$ operators along the $12$ edges of a cube, i.e. the $Z$-check of the $X$-cube model. 

This coupled layer construction helps explain various features of the $X$-cube model in terms of those of the toric code. For example, it sheds light on the similar structure of their logical operators. The $X$-logicals of the toric codes commute with the coupling terms, and are thus also logicals of $x$-cube and they remain confined to their original planes. The plaquette excitations created by truncated logicals (see Fig.~\ref{fig:HGP} again) are now part of two cubes each, and thus correspond to bound states of two fractons that are mobile within the plane. To separate them, one needs to apply multiple such $X$-operators on parallel planes. 

The $Z$-logicals of individual toric codes fail to commute with the $X_e'X_e''$ terms, but one can take the product of two such logicals from two differently oriented planes (which intersect at a line) to get a new $Z$-logical that satisfies the constraint; these become the $Z$-logicals of the $X$-cube model. The fact that they occur at the intersection of two planes explains that they are confined to within a line and cannot be freely moved in the transverse directions. This is related to the fact that when truncated, the excitations they create are \emph{lineons} that can only move within this line. Another way of saying this, is that such a truncated logical violates a pair of checks at each endpoint, originating from the toric codes in the two different planes. For example, a truncated version of the logical along the $z$ direction violates the checks $A^x$ and $A^y$, while that in the $x$ direction violates $A^y$ and $A^z$. If we tried to fuse two such lines together to move the excitations within the $xz$ plane, at their meeting point we would be left with some violations of the $A^y$ terms.

We can now discuss how this construction generalizes to arbitrary GXC codes. We again start with the 3D grid representation, with the same labeling of sites as $(i,j,k)$ etc. Fixing a coordinate, say $k$, defines a 2D $\text{\textbf{AB}}$ plane on this grid, with sites labeled $(i,j)$, horizontal edges $(a,j)$ etc. We can then place the checks of the HGP code $\text{HGP}(\mathscr{C}_A,\mathscr{C}_B)$, defined in Eq.~\eqref{eq:HGP_checks}, on each such plane. Similarly, we can place the codes $\text{HGP}(\mathscr{C}_A,\mathscr{C}_C)$ and $\text{HGP}(\mathscr{C}_B,\mathscr{C}_C)$ on all the $\text{\textbf{AC}}$ and $\text{\textbf{BC}}$ planes, respectively. Just like the simple $X$-cube case, this requires placing \emph{two} qubits on each edge of the grid, each of which will be involved in exactly one HGP code. For example, on the $\text{\textbf{A}}$-edge $(a,j,k)$, one qubit is part of the HGP code on the $\text{\textbf{AB}}$ plane labeled by $k$, while another qubit is part of the HGP code on the $\text{\textbf{AC}}$ plane labeled by $j$. There are three different $X$-checks associated to each site of the grid, for example $A^{\text{\textbf{AB}}}_{i,j,k}$ is the $X$-check of the HGP code on the $\text{\textbf{AB}}$ plane labeled by $k$. On the other hand, each plaquette hosts only a single $Z$-check, corresponding to the plane that the plaquette is a part of. For example, the plaquette $(a,b,k)$ in the $\text{\textbf{AB}}$ plane hosts the check $B^{\text{\textbf{AB}}}_{a,b,k}$.

We can start with the Hamiltonian of this code, as in Eq.~\eqref{eq:H_Gauge}, which is the sum of all the Hamiltonians of all the individual HGP codes with no coupling between them. We can then add a coupling on each edge that couples the two qubits on the same edge. For example, in the aforementioned edge we would have a term $-t X'_{a,j,k} X''_{a,j,k}$, coupling the $x$-components of the two qubits. We now imagine turning the coefficient $t$ to a large value which induces a constraint on the Hilbert space, and we look for the effective Hamiltonian in this low-energy subspace generated by the original HGP terms. Within this subspace, there is now just a \emph{single} qubit on each edge. We can label its two components as $X_{a,j,k} = X'_{a,j,k} = X''_{a,j,k}$ and $Z_{a,j,k} = Z'_{a,j,k} Z''_{a,j,k}$.

The $X$-checks commute with the constraint and therefore remain unchanged, except that now they act on the effective qubits, thus coupling different layers to each other. For example, we have
\begin{equation}
    A^{\text{\textbf{AB}}}_{i,j,k} = \prod_{a \in \delta_A^T(i)} X_{a,j,k}
    \prod_{b \in \delta_B^T(j)} X_{i,b,k},
\end{equation}
which we recognize as the last check in Eq.~\eqref{eq:GXC_checks}. Similarly, the other two $X$-checks, $A^{\text{\textbf{BC}}}$ and $A^{\text{\textbf{AC}}}$, reproduce the remaining $X$-checks in Eq.~\eqref{eq:GXC_checks}.

The $Z$-checks of the HGP codes fail to satisfy the constraint. We thus need to form products of these checks that do. To achieve this, let us pick a cube $(a,b,c)$ of the 3D grid. We can associate to this the following product:
\begin{align}\label{eq:GXC_from_HGP}
    B_{a,b,c} &= \prod_{i\in\delta_A(a)} B^{\text{\textbf{BC}}}_{i,b,c}\prod_{j\in\delta_B(b)} B^{\text{\textbf{AC}}}_{a,j,c} \prod_{k\in\delta_C(c)} B^{\text{\textbf{AB}}}_{a,b,k} \nonumber \\ 
    &= \prod_{i,j} Z_{i,j,c} \prod_{i,k} Z_{i,b,k} \prod_{j,k} Z_{a,j,k}.
\end{align}
In the second line, we recognize the $Z$-check of the generalized $X$-cube model from Eq.~\eqref{eq:GXC_checks}. 

This means that properties of the GXC model can be understood from those of the three underlying HGP codes. For example, $X$ logical operators are inherited directly from the latter, since these again commute with the constraints we used to couple the different layers. For the $Z$-type logicals, we note that these come in pairs; for example, the HGP codes on the $\text{\textbf{AB}}$ planes and those on the $\text{\textbf{AC}}$ planes share a set of $Z$-logicals with support on the $\text{\textbf{A}}$-oriented edges. We can then take these pairs and multiply them together to obtain logicals of the GXC code. 

While the logicals of the GXC code can thus be obtained directly from those of the HGP codes, there are important differences in the excitations and their mobility. The plaquette excitations (i.e., violations of the $Z$-checks, generated by some set of $X$ operators acting on qubits within the plane) maintain whatever properties they had within the plane, but they now correspond to a bound state of multiple violated $Z$-checks of the GXC code. These can be separated out into individual excitations by using the logical operators of the other HGP codes, but these individual excitations will generally have a more restricted mobility, just as in the $X$-cube model. Similarly, while violations of the $X$-checks in the HGP codes directly correspond to violations in the GXC code, they now have three different flavors, corresponding to the three types of checks. These flavors come in pairs, and each pair is associated to a direction ($\text{\textbf{A}}$, $\text{\textbf{B}}$ or $\text{\textbf{C}}$) on the grid and are restricted to move along this direction only. 

\section{Euclidean models}\label{sec:Examples}

\begin{figure*} 
    \centering
    \includegraphics[trim={1cm 0 0 0},width = 1.\linewidth]{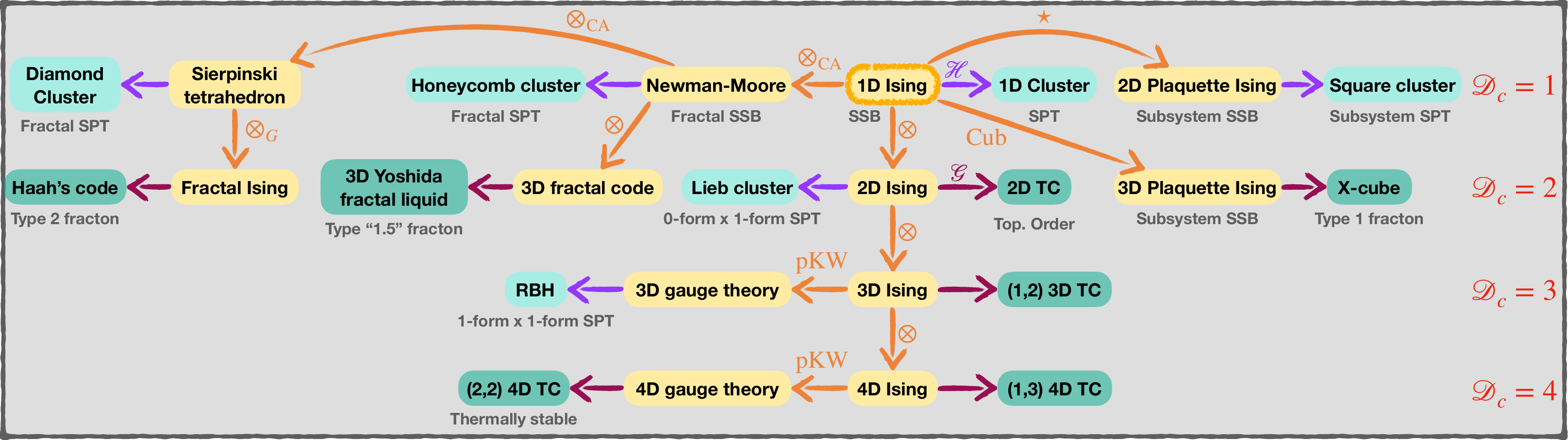}
    \caption{\textbf{The universe from a 1D Ising model.} We show some of the stabilizer models (corresponding to fixed point Hamiltonians associated to gapped phases of matter) that can be constructed out of the 1D Ising model using a combination of different products and the gauging / Higgsing procedures outlined described in Sec.~\ref{sec:products}-\ref{sec:GaugeAndHiggs}. Arrows correspond to the different operations depicted in Fig.~\ref{fig:Factory}. Orange arrows map classical codes to other classical codes; these correspond to phases that break $0$-form symmetries, where the nature of the symmetries is modified along the way. Higgsing (purple arrow) produces cluster states, corresponding to SPT phases with symmetries inherited from the classical codes and their transposes. Products can also increase the code dimensionality $\mathscr{D}_c$ (distinct from the physical dimension); gauging classical codes with $\mathscr{D}_c \geq 2$ (Bordeaux arrows) produces quantum CSS codes which correspond to topological / fracton orders (SSB of higher-form symmetries). 
    }
    \label{fig:MapOfPhases}
\end{figure*}

In this section, we start putting into action the machinery developed in the preceding section, by showing how it can give rise to a large variety of stabilizer models, associated to various phases of matter, already in the context of finite-dimensional lattice models. This works as a kind of ``code lego'', systematically constructing models with desired properties from simple ingredients. We take as our starting point the one-dimensional Ising model, which is arguably the simplest example of both a non-trivial phase and of a classical error correcting code (the repetition code). We then show how a systematic application of the ideas presented above can turn this simple seed into a large tree of different phases of matter, including spontaneous symmetry breaking, SPT, topological and fracton orders; this is summarized in Fig.~\ref{fig:MapOfPhases}. We further illustrate how the idea of modding out spatial symmetries, which lies behind the balanced product construction, can be used to construct new models, which includes a symmetry-enriched topological (SET) phase in two dimensions and a novel fracton phase in 3D. The latter combines the cubic product with the modding out of translation symmetry, resulting in exotic mobility constraints on its fractonic excitations. 

\subsection{The universe from an Ising chain}\label{subsec:IsingUniverse}

Here, we demonstrate how many known phases of matter (represented by the stabilizer models that describe their fixed-point limits) can be obtained by applying the machinery of Fig.~\ref{fig:Factory} to the humble one-dimensional Ising model (which we will also refer to as the ``Ising chain'').

\subsubsection{Ising models and their descendants}

Starting from the 1D Ising chain, we can apply the Higgsing map to produce the 1D cluster state, which is one of the simplest examples of a non-trivial SPT phase~\cite{raussendorf2001one}.

As discussed in Sec.~\ref{sec:products}, taking the tensor product of two 1D Ising chains gives a 2D Ising model. The construction can be iterated further: taking another product produces a $3D$ Ising model and so on, building up the Ising model on hypercubic lattices in arbitrary dimensions. In these cases, the corresponding chain complex is the one that is  associated to the lattice in a natural way, with the vertices, edges, faces, etc. forming the various levels of the chain complex. 

Applying the gauging map to the $D$-dimensional Ising model gives rise to the $D$-dimensional toric code, which has line-like $Z$ logicals with distance $d_Z = L$ and $X$ logicals along co-dimension 1 surfaces with distance $d_X = L^{D-1}$ where $L$ is the linear size of the lattice (i.e., the number of bits in the repetition code used as an input). From a physical perspective, these generate a $(D-1)$-form and a $1$-form symmetry, respectively, which are both spontaneously broken in the toric code phase. We can also apply the Higgsing map to obtain appropriate cluster states from the $D$-dimensional Ising model. These form SPT phases protected by the global $\mathbb{Z}_2$ symmetry of the Ising model and the aforementioned $(D-1)$-form symmetry whose symmetry operators act on closed loops of the lattice~\cite{yoshida2016topological,verresen2022higgs}. 

Given the chain complex representation of the Ising model, we can also consider classical codes that correspond to taking any two subsequent levels of the complex. The simplest non-trivial example of this would be a three-dimensional cubic lattice and placing bits on the edges and checks on the square plaquettes, giving rise to a \emph{classical gauge theory}~\cite{kogut1979introduction}, which we could imagine as being obtained from the Ising model using the kind of ``partial Kramers-Wannier duality'' mentioned in Sec.~\ref{sec:BuildingBlocks}. The classical gauge theory has a small classical code distance, as flipping the 6 edges around a vertex will commute with all the checks. Nevertheless, it can give rise to non-trivial quantum models. While the CSS code obtained from gauging the classical gauge theory is equivalent to the 3D toric code after going to the dual lattice, the cluster model that arises from Higgsing is known as the RBH (Raussendorf-Bravyi-Harrington) model and is an example of an SPT protected by two $1$-form symmetries, exhibiting interesting computational properties as a resource state for measurement-based quantum computation~\cite{raussendorf2005long}. 

Another example of this idea of using two levels in the ``middle'' of the $D$-dimensional chain complex is obtained by going into four dimensions. We can again consider the classical gauge theory with bits on edges and checks on plaquettes. Gauging this now gives the so-called ``$(2,2)$'' version of the toric code model, which is a different phase from the ``$(1,3)$'' toric code that is obtained from gauging the $D$-dimesional Ising model directly. The $(2,2)$ toric code has two $2$-form symmetries (with symmetry operators / logicals supported on 2-dimensional surfaces on the direct and dual lattices, respectively) which it breaks spontaneously. This is related to the fact that in this case, both $X$ and $Z$ checks exhibit local redundancies, such that their violations need to form closed loops. As a consequence, the $(2,2)$ toric code phase is stable even away from zero temperature, unlike its $(1,3)$ version or any of the lower dimensional toric codes~\cite{dennis2002topological,hastings2011topological}.

\subsubsection{Plaquette Ising models and their descendants}

As we already noted in Sec.~\ref{sec:products}, taking the check product of two Ising chains gives rise to the 2D plaquette Ising model, exhibiting line-like subsystem symmetries. This is still a one-dimensional chain complex, so we cannot construct a CSS code out of it, but we can apply the Higgsing map to turn it into the cluster state on a (rotated) 2D square lattice~\cite{LDPCGauge}, which is known to be an SPT protected by these subsystem symmetries~\cite{you2018subsystem,raussendorf2019computationally}. One could take a repeated check product with another Ising model to get the \emph{cubic Ising model}~\cite{you2018subsystem} a 3D model with bits on sites and 8-body checks acting on the corners of a cube which still has line-like subsystem symmetries. Higgsing this would then produce a 3D subsystem SPT~\cite{you2018subsystem}.

As we also observed, the cubic product defined in Sec.~\ref{subsec:SubSysProd} turns three 1D Ising models into the \emph{three-dimensional} plaquette Ising model\footnote{We mention in passing that one could also obtain the same 3D plaquette Ising model by taking the tensor product of three 2D plaquette Ising models and then modding out half of the translations of the resulting six dimensional model.}. As we have also discussed, this model is dual to the $X$-cube fracton model under gauging~\cite{vijay2016fracton}. On the other hand, applying the Higgsing map gives rise to a cluster state on the face-centered cubic lattice, which has both the planar symmetries of the plaquette Ising model and the higher-form-like symmetries associated to the $Z$-logicals of the $X$-cube model~\cite{tantivasadakarn2021long}.

\subsubsection{Models with fractal symmetries}  
 
As discussed in Sec.~\ref{sec:products}, applying the cellular automaton product to two Ising chains gives rise to the Newman-Moore (NM) model, which has fractal symmetries in the shape of Sierpinski triangles~\cite{newman1999glassy}. Higgsing this model gives a cluster state on the honeycomb lattice, which is an SPT protected by these fractal symmetries~\cite{devakul2019fractal}. 

From the NM model, we can build other codes with fractal symmetries using the product machinery. One possibility is to take a repeated CA product with the 1D Ising model. This results in a 3D generalization of the NM model, known as the \emph{Sierpinski tetrahedron} (ST) model~\cite{devakul2019fractal}. Let us change variables and write the polynomial for the NM model as $f = 1 + x + y$. We use Eq.~\eqref{eq:CA_1D} to combine this with an Ising chain described by the polynomial $g=1+z$, to get $1 + (1+x+y)z = 1 + z + xz + yz$, describing a check that acts on a site and its three neighbors on the 3D cubic lattice. Up to a rotation, we can rewrite it as 
\begin{equation}\label{eq:3DNM}
    f_\text{ST} = 1 + x + y + z.
\end{equation}
As the name suggests, this model has fractal symmetries generated by three-dimensional Sierpinski tetrahedra. Higgsing it gives rise to an SPT protected by such symmetries which is equivalent to the cluster state on the diamond lattice~\cite{devakul2019fractal}.

The NM and ST models have no local redundancies and thus have $\mathscr{D}_c = 1$, but we can also use the NM model to build higher dimensional chain complexes. The simplest example is taking the tensor product of the NM model with the 1D Ising chain. This gives rise to an anisotropic 3D model, with a number of NM models along the $x-y$ planes, coupled to each other via Ising interactions in the $z$ direction (see Fig.~\ref{fig:NM_times_Ising}(a)). By gauge duality, this classical code gives rise to Yoshida's 3D anisotropic fractal spin liquid (FSL) model~\cite{yoshida2013exotic,kubica2018ungauging}. This is an example of a hypergraph product code, so one can analyze it along the lines discussed in Sec.~\ref{subsec:HGP}. In particular, it has two types of logicals, one line-like and extended along the $z$ direction and one formed by Sierpinski-triangles in the $xy$ planes. Truncating the former gives rise to a pair of lineon excitations which are free to move in the $z$ direction, while truncating the latter gives rise to three immobile fracton excitations at the three corners of the triangle; these fatures are inherited from the 1D Ising and NM models, respectively. 

Eq.~\eqref{eq:3DNM} also corresponds to one of the two checks appearing in the ``fractal Ising model'' of Ref. \onlinecite{vijay2016fracton}, which is gauge dual to Haah's code~\cite{haah2011local}. We can recover the full classical code by using a combination of products and modding out symmetries as follows. First, take a tensor product of Eq.~\eqref{eq:3DNM} with itself, so that we have two checks
\begin{align}\label{eq:6DNM}
    f_1 &= 1 + x + y + z, \nonumber \\
    f_2 &= 1 + u + v + w.  
\end{align}
This is now a 6 dimensional model with coordinates $x,y,z,u,v,w$ and translation symmetry in all directions. We mod out three of these translations, generated by $T_x^{-1}T_y^{-1} T_u$, $T_x^{-1}T_z^{-1} T_v$ and $T_y^{-1}T_z^{-1} T_w$ to induce the relations
\begin{equation}
    \bar x \bar y u = \bar x \bar z v = \bar y \bar z w = 1.
\end{equation}
Plugging these into Eq.~\eqref{eq:6DNM} we end up with 
\begin{align}
    f_1 &= 1 + x + y + z, \nonumber \\
    f_2 &= 1 + xy + xz + yz,  
\end{align}
which describes the two types of checks in the fractal Ising model of Ref. \onlinecite{vijay2016fracton}. Gauging this gives Haah's code. We note that one could also apply the Higgsing map to the same model to get a cluster state that not only has the ``rigid'' fractal symmetries of the fractal Ising model, but also the ``fractal one-form'' symmetry corresponding to the $Z$ logicals of Haah's code.

\subsection{New codes from balanced products}\label{subsec:NewCodes}

Here, we illustrate the power of combining products with the modding out of spatial symmetries the underlies the balanced product construction by using it to construct two new stabilizer models on two and three dimensional Euclidean lattices, using the 1D Ising and Newman-Moore models as building blocks. In 2D we use the balanced product to construct a symmetry enriched topological (SET) order, which is equivalent to two copes of the toric code with the anyon species permuted by translation symmetry, a feature that arises from the NM model. In 3D we construct a fracton model by combining the cubic product with the modding out of diagonal translations. The resulting code has both line-like and fractal-like logical operators and its excitations exhibit a pattern of mobility constraints that can be understood from the cubic product construction. 

\subsubsection{Balanced product of Ising and Newman-Moore}

We now illustrate the machinery of the balanced product construction by using it to create a new stabilizer code in two spatial dimensions. To do so, we combine the Newman-Moore model with the one-dimensional Ising chain and mod out diagonal translations in the resulting 3D model to map it back to two dimensions. In other words, the quantum code we consider is obtained from the 3D fractal spin liquid model~\cite{yoshida2013exotic} by gluing together sites along a body-diagonal. Interestingly, we find that while the fracton order of the FSL model is gone (as it has to, since there is no two-dimensional fracton stabilizer code~\cite{haah2011local}), a remnant of it remains in the fact that anyons in our 2D model cannot be moved completely freely: translating an anyon by a single site turns it into a different anyon species and instead it needs to be translated three times to get into an equivalent configuration. In other words, we end up with topological order \emph{enriched} by translation symmetry~\cite{essin2013classifying,mesaros2013classification,barkeshli2019symmetry}.

\begin{figure} 
    \centering
    \includegraphics[trim={1cm 0 0 0},width = 1.\linewidth]{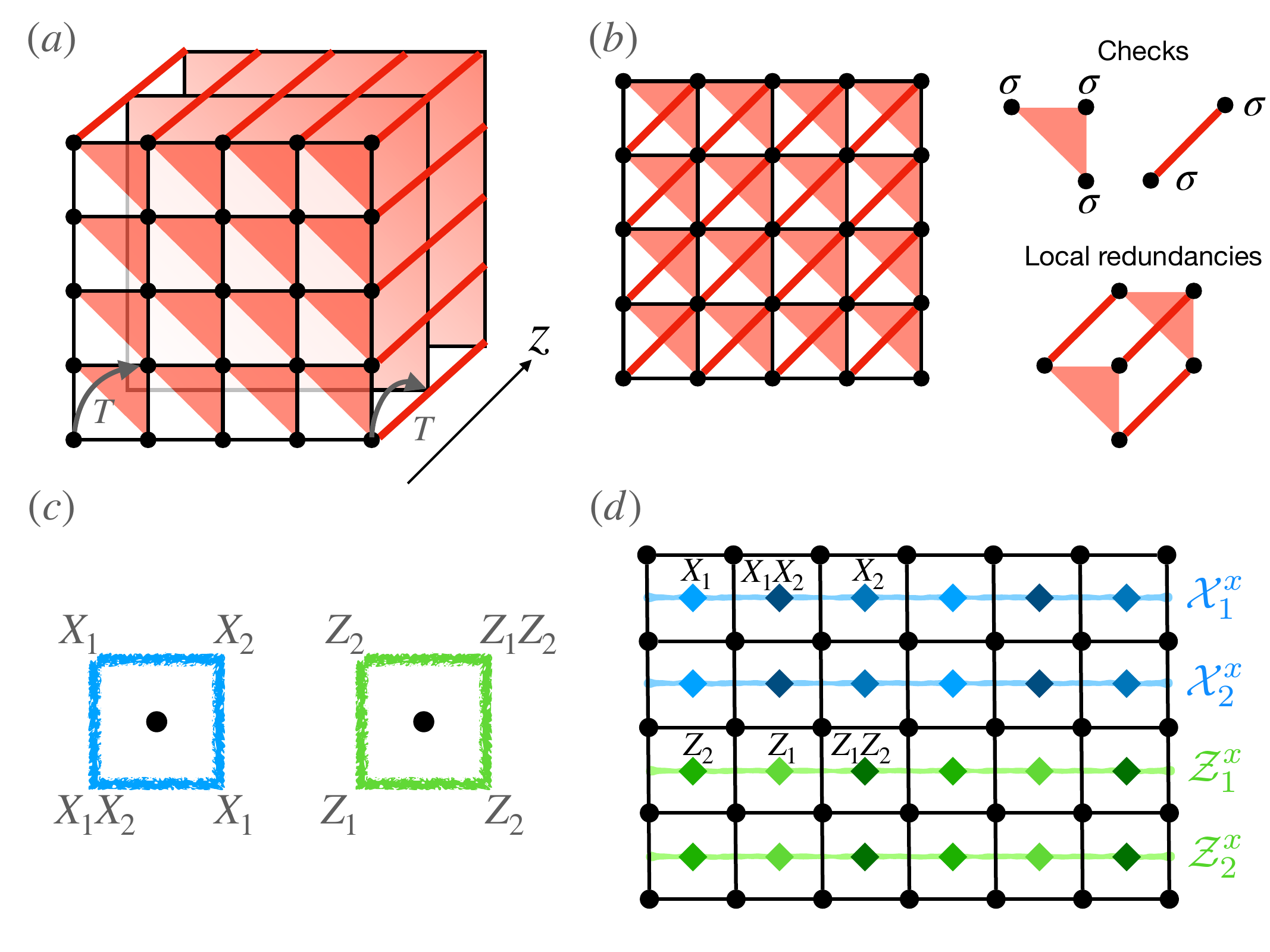}
    \caption{\textbf{Balanced product of Newman-Moore and 1D Ising models.} (a) Tensor product of the two codes, with gray arrows indicating the translation symmetries that are to be modded out. (b) 2D classical code obtained after modding out translations, with two types of checks. A local redundancy is shown in the bottom right, involving five checks around a site. (c) $X$ and $Z$ checks of the CSS code that is gauge dual to the classical code in (b). (d) Two inequivalent $X$ and $Z$ logical operators of the CSS code, stretching along the $x$ direction.. There are two more logicals, stretching in the $y$ direction.
    }
    \label{fig:NM_times_Ising}
\end{figure}

Our starting point is the tensor product of the 1D Ising model on a chain of length $L$ and the Newman-Moore model on a $L \times L$ square lattice, with periodic boundary conditions in all directions. We take the NM model to have checks acting on sites $(i,j)$, $(i+1,j)$ and $(i,j+1)$ on the 2D lattice, labeled by the polynomial $1 + x + y$. We denote the resulting product code as $\mathcal{I}_\text{1D} \otimes \mathscr{C}_\text{NM}$. As discussed in Sec.~\ref{subsec:IsingUniverse}, this gives an anisotropic three-dimensional model, with NM interactions in the $x-y$ plane and Ising coupling along the $z$ direction, shown in Fig.~\ref{fig:NM_times_Ising}(a). 

To turn this model into a balanced product, we mod out translation symmetry $\mathbb{Z}_L$ acting simultaneously on both models. On the Ising chain it acts in the obvious way, as $T_z$, translating each site by one to the right. The NM model has a larger translation symmetry, $\mathbb{Z}_L \times \mathbb{Z}_L$, so we need to choose an appropriate one-dimensional subgroup; we choose the diagonal translations $T_xT_y$\footnote{One can show that other natural choices, such as $T_x$, $T_y$ or $T_xT_y^{-1}$ result in a trivial model.}. Overall, this leads to an identification of all sites related to each other by diagonal translations $T_xT_yT_z$ on the 3D lattice, which in the polynomial representation is written as $xyz = 1$. We can use this to eliminate $z$ and write the resulting two-dimensional classical code $\mathscr{C}_\text{cl}$, generated by two checks with polynomials
\begin{align}\label{eq:NMxIsing_balanced}
    f_1 &= 1 + x + y, \nonumber \\
    f_2 &= 1 + xy, 
\end{align}
i.e., a Newman-Moore model with additional Ising coupling along a diagonal. These come from the Ising interactions in the $z$ direction that have been projected down to the $x-y$ plane, as shown in Fig.~\ref{fig:NM_times_Ising}(b). 

The classical code~\eqref{eq:NMxIsing_balanced} has local redundancies, inherited from the 3D tensor product model. They can be represented by the pair of polynomials $(1+xy, 1 + x + y)$; one such redundancy is shown in Fig.~\ref{fig:NM_times_Ising}(b). We can use these in the gauging procedure to turn the model into a quantum CSS code. For easier representation, we can assign both types of classical checks to the midpoint of the square plaquette that they live on, and their redundancies to sites; therefore, the quantum code has qubits living on the dual (also square) lattice and both $X$- and $Z$-checks assigned to the plaquettes of that lattice. Both checks involve a product of $5$ Pauli operators in an anisotropic fashion (slightly reminiscent of Haah's code~\cite{haah2011local}), as shown in Fig.~\ref{fig:NM_times_Ising}(c).

Let us calculate the properties of this CSS code. The number of qubits equals the number of checks (since there is two per lattice site of each), so we have
\begin{equation}\label{eq:count_GS}
    k_\text{q} = k_\text{cl} + k_\text{cl}^\text{KW} = 2 k_\text{cl},
\end{equation}
where $k_\text{q}$ is the number of logicals in the CSS code, while $k_\text{cl}$ and $k_\text{cl}^\text{KW}$ denote the number of logicals in the ungauged classical code and its Kramers-Wannier dual, respectitely. In the last equality of Eq.~\eqref{eq:count_GS}, we used that the dual classical code $\mathscr{C}_\text{cl}^\text{KW}$ is equivalent to $\mathscr{C}_\text{cl}$, a fact that is easy to check explicitly (and is inherited from the fact that both 1D Ising and NM are invariant under transposition).

The classical GS degeneracy, $k_\text{cl}$ is calculated easily by examining the checks shown in Fig.~\ref{fig:NM_times_Ising}(b). All spins falling on the same diagonal must take equal values, due to the presence of the Ising couplings along these lines. Thus we can assign a single effective variable for each diagonal; these are acted upon the the NM checks as a product of three consecutive spins, $\sigma_{i-1}\sigma_{i}\sigma_{i+1}$. The number of satisfying configurations depends on $L$: if $L$ is a multiple of $3$, then there are $4$ distinct solutions. Apart from the trivial ``all up'' state, these are states of the form $\uparrow\downarrow\downarrow\uparrow\downarrow\downarrow\ldots$; there are 3 of these which are translations of one another. On the other hand, when $L$ is not a multiple of $3$, the latter solutions are not consistent with PBC\footnote{With OBC, the classical model always has $4$ GS; however, in this case Eq.~\eqref{eq:count_GS} is no longer valid.}. 

Thus, from Eq.~\eqref{eq:count_GS} we find that $k_\text{q} = 4$ when $L = 3p$ ($p \in \mathbb{Z}_+$) and $k_\text{q}=1$ otherwise. What are the corresponding quantum logical operators? A possible basis choice for them is shown in Fig.~\ref{fig:NM_times_Ising}(d). They are similar to toric code logicals in that they are made up by products of either $X$ or $Z$ Pauli operators wrapping around the two inequivalent directions of the torus. However, they possess an internal structure, with period 3, similar to the classical GS, which explains their absence when $L \neq 3p$. This gives two independent logicals of each type in both directions (the third one being their product), which we denote by $\mathcal{X}_1^x,\mathcal{X}_2^x,\mathcal{X}_1^y,\mathcal{X}_2^y$ for $X$-logicals, and similarly for $Z$-logicals. We choose a labeling such that for example $\mathcal{X}_{1}^{x}$ anti-commutes with $\mathcal{Z}_{1}^{y}$\footnote{We note that starting from the logicals of the 3D FSL model (constructed in Sec.~\ref{subsec:HGP}) and then modding out the diagonal translations results in a \emph{different} basis of the logicals of the balanced product code. The line-like $X$ and $Z$ logicals of the FSL turn into products of $X_2$ and $Z_1$ along the diagonal. These turn out to be equivalent, up to multiplication by checks, to the product $\mathcal{X}_1^x \mathcal{X}_1^y$ and $\mathcal{Z}_1^x \mathcal{Z}_1^y$. The fractal logicals of the FSL on the other hand are equivalent to the products $\mathcal{X}_1^x \mathcal{X}_2^y$ and $\mathcal{Z}_1^x \mathcal{Z}_2^y$.}. 

Comparing with the fractal spin liquid model we had before modding out translations, we see that the code distance has remained unchanged\footnote{In particular, the basis choice in the previous footnote includes logicals with weight $L$.} despite the fact that we have reduced the total number of qubits by a factor of $L$. In other words, by modding out symmetries, we have boosted the relative distance $d_\text{q}/n_\text{q}$ from $n_\text{q}^{-2/3}$ to $n_\text{q}^{-1/2}$. The same trick underlies the constructions of good qLDPC codes which we will discuss in Sec.~\ref{subsec:GoodCodes}.

Importantly, while the code itself is translation invariant, the basis of logicals we constructed is not. Instead, translating by either $T_x$ or by $T_y^{-1}$ induces the following permutation of logicals:
\begin{align}\label{eq:SET_permutation}
    \mathcal{X}_1^x &\to \mathcal{X}_2^x \to \mathcal{X}_1^x\mathcal{X}_2^x \to \mathcal{X}_1^x, \nonumber \\
    \mathcal{Z}_1^y &\to  \mathcal{Z}_1^y\mathcal{Z}_2^y \to \mathcal{Z}_2^y\to \mathcal{Z}_1^y,
\end{align}
and similarly for the other set of logicals. This means that these translations execute a logical gate of the form $\text{SWAP} \cdot \text{CNOT}$ on this set of logical qubits. Similarly, translating in the opposite direction performs a gate $\text{CNOT} \cdot \text{SWAP}$ which permutes the logical qubits in the opposite order. We note that the ability to execute non-trivial logical gates by translations is advantageous in quantum computing platforms where qubits can be moved around freely, such as in Rydberg atom tweezer arrays~\cite{Bluvstein2022}.

In terms of classifying our model as a phase of matter, as a translationally invariant 2D qubit stabilizer code with $k_\text{q}=4$, it must be equivalent to two copies of the toric code~\cite{haah2021classification}. However, Eq.~\eqref{eq:SET_permutation} implies that translations (which are a symmetry of the model) permute different anyon species. Indeed, anyons are created at the open endpoints of the same strings of Paulis that form logicals and we observe that translating such a string, in either the $x$ or the $y$ direction, leads to a different, inequivalent type. In particular, denoting by $e_{1,2}$ ($m_{1,2})$ the electric (magnetic) excitations of the two toric codes, we have a permutation 
\begin{align}
    e_1 \to e_2 \to e_1 e_2 \to e_1, & & m_1 \to m_1m_2 \to m_2 \to m_1
\end{align}
induced by translations. This implies that the model exhibits \emph{symmetry enriched} topological (SET) order~\cite{essin2013classifying,mesaros2013classification,barkeshli2019symmetry}. This property is an inheritance from the NM model we used to build our code, which was the source of the three-fold periodicity that is present both in the classical ground states and the quantum logicals. Another remnant of the fractonic nature of the FSL can be observed when we try to bend the anyon excitations around a corner: we find that if we naively repeat the same pattern of operators as they appear in Fig.~\ref{fig:NM_times_Ising}(d), this leaves an excitation at the corner; instead, the structure of the operator needs to be modified at the corner to allow for the anyons to bend. 

\subsubsection{3D Balanced cubic product model}\label{subsubsec:BalancedCubicProduct}

We now illustrate the idea of generalized $X$-cube models, introduced in Sec.~\ref{subsec:GXC}, by defining a new fracton model constructed out the 1D Ising models and Newman-Moore models. This is a four dimensional phase which is gauge dual to a classical model that has both planar symmetries and symmetries that are fractal like in two dimensions and extended in a third. Alternatively, it can be understood within the coupled layer description of Sec.~\ref{subsec:GXC}, as a model in which excitations are bound states of toric code anyons and the fracton excitations of the FSL model~\cite{yoshida2013exotic}, which results in exotic mobility constraints. We then mod out the fully diagonal translations of the 4D hypercubic lattice to map this into a three-dimensional model and argue that most if its interesting physical features survive. 

First, consider the cubic product code $\text{Cub}(\mathcal{I}_\text{1D},\mathcal{I}_\text{1D},\mathscr{C}_\text{NM})$, constructed out of two 1D Ising models and a Newman-Moore model, yielding a classical code in four spatial dimensions, whose coordinates we denote $x,y,z,u$. The logicals of this code correspond to planar symmetries in the $xy$ plane and to fractals that take the form of Sierpinski triangles in the $zu$ plane stacked on top of each other in either the $z$ or the $u$ direction. Gauging these symmetries yields a generalized $X$-cube model\footnote{Note that both $\mathcal{I}_\text{1D}$ and $\mathscr{C}_\text{NM}$ are isomorphic to their transposes.} which we can analyze in the general framework developed in Sec.~\ref{subsec:GXC}. Its quantum logicals are either line-like in the $x$ or $y$ direction, or Sierpinski fractals in the $zu$ plane. Truncated versions of these logicals create excitations in groups of $2$ or $3$ that behave like toric code anyons in the $xy$ plane, or like the excitations of the FSL model in either the $xzu$ or $yzu$ volume. These can be further separated to individual fracton excitations by combining fractal and line-like operators (for more details, see App.~\ref{app:CubProd}).

To reduce our 4D model back to three dimensions, we mod out a fully diagonal translations generated by $T_xT_yT_zT_u^{-1}$. In the polynomial language, this corresponds to equating $u = xyz$. Applying this operation to the classical cubic product code yields a model in three spatial dimensions, with three checks per site, corresponding to the following three polynomials:
\begin{align}\label{eq:NMCubic}
    f_1 &= (1+x)(1+y) \nonumber \\
    f_2 &= (1+x)(1+z+xyz) \nonumber \\ 
    f_3 &= (1+y)(1+z+xyz). 
\end{align}
We can further simplify this classical code by redefining the latter two checks as follows\footnote{This will not change the number of quantum logicals we obtain after gauging, but it will affect their concrete form. Nevertheless, they still inherit properties of the 4D GXC code as we shall see.}
\begin{align}\label{eq:NMCubic_redef}
    f_2 &\to f_2 + xzf_1 = 1 + x + z + x^2z = (1+x)(1+z+xz),\nonumber \\ f_3 &\to f_3 + yzf_1 = 1 + y + z + y^2z = (1+y)(1+z+yz).
\end{align}
These are now only 4-body checks (rather than 6-body, as the original $f_{2,3}$ in Eq.~\eqref{eq:NMCubic}) and the act entirely on sites within the $xz$ and $yz$ planes, respectively. In the last equality, we have written them in a way that emphasizes that they correspond to the product of two Newman-Moore-type triangular checks. This allows one to write down the classical logical operators in terms of those of the NM model, as we discuss in App.~\ref{app:CubProd}. The three checks, $f_{1,2,3}$ can be naturally associated to the three types of plaquettes on the cubic lattice (see Fig.~\ref{fig:NMCubic_classical_checks}).

The redundancies also need to be modified appropriately. They read, 
\begin{align}\label{eq:NMCubic_redundancies}
    (1+z+xz)f_1 + (1+y)f_2 &= 0, \nonumber \\
    (1+z+yz)f_1 + (1+x)f_3 &= 0, \nonumber \\
    (x+y)zf_1 +(1+y)f_2 + (1+x)f_3 &= 0.
\end{align}

\begin{figure} 
    \centering
    \includegraphics[trim={1cm 0 0 0},width = 1.\linewidth]{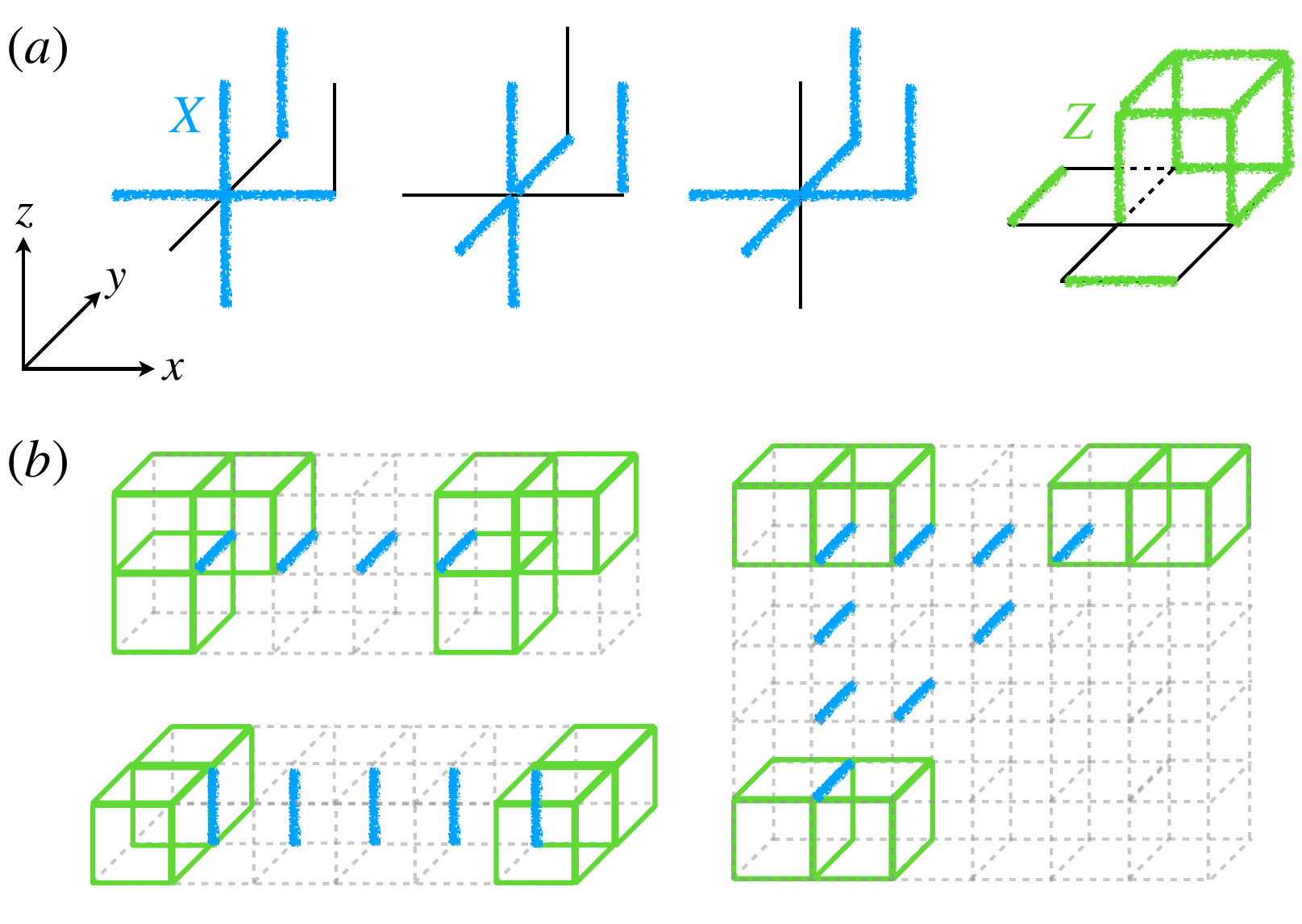}
    \caption{\textbf{3D generalized $X$-cube model constructed from 1D Ising and Newman-Moore models.} The code is obtained by applying the generalized $X$-cube construction followed by a modding out of diagonal translations to produce a 3D model. (a) shows the checks of the resulting quantum code Qubits live on the edges. Each site hosts three 5-body $X$-checks (blue) and cubes host a single 12-body $Z$-check (green). (b) Shows parts of three different types of logical $X$ operators (two line-like and one in the shape of a Sierpinski fractal). We highlight in green the cubes ($Z$-checks) that would be excited if we truncated the logical to the edges shown, which occur in groups of 2 and 3.}
    \label{fig:NM_cubic}
\end{figure}

We are now in a position to write down the corresponding quantum code, obtained from gauging this classical model. In keeping with the conventions established for GXC models in Sec.~\ref{subsec:GXC}, we draw qubits on the edges of the lattice and associate the classical bits with $Z$-checks on cubes, while the $X$-checks correspond to the local redundancies and are assigned to sites. These are shown in Fig.~\ref{fig:NM_cubic}(a). 

The number of logicals of this quantum code is most easily counted by once again using the relation $k_\text{q} = k_\text{cl} + k_\text{cl}^\text{KW}$, where $k_\text{cl}$ is the number of logicals in the classical code we gauged and $k_\text{cl}^\text{KW}$ is the number of logicals of its Kramers-Wannier dual. While the two classical codes in this case are different, it turns out that they share a similar set of logicals, which we count in App.~\ref{app:CubProd}, yielding 
\begin{equation}\label{eq:k_count_3D}
    k_\text{q} = k_\text{cl} + k_\text{cl}^\text{KW} = 2L + 4k_\text{NM}(L),
\end{equation}
where $k_\text{NM}(L)$ is the number of logicals in the NM model on a $L \times L$ lattice\footnote{In terms of the cubic product, this corresponds to $L$ planar symmetries in the $xy$ plane, inherited from the 2D Ising model $\mathcal{I}_\text{1D} \otimes \mathcal{I}_\text{1D}$, and two sets of symmetries inherited from $\mathcal{I}_\text{1D} \otimes \mathscr{C}_\text{NM}$.} . 

The logicals of this code are inherited from those of the 4D GXC model. For example, $X$-logicals are divided into three groups: there are $2L$ line-like logicals composed of either $x$ or $y$ edges within the $xy$ plane, $2k_\text{NM}$ line-like ones composed of $z$ edges extended along the $x$ or $y$ and $2k_\text{NM}$ Newman-Moore fractals either in the $xz$ or $yz$ planes, made out of edges perpendicular to the plane: one representative from each group is indicated in Fig.~\ref{fig:NM_cubic}(b). Also indicated are the excitations (violation of $Z$-checks) that would be created by truncating the logicals: the first and last type create six excitations, either as two triplets at the endpoints of a line-like operator, or two pairs at the three corners of a Sierpinski triangle. These can be separated into six separate excitations by taking products of multiple translated Sierpinski triangles, as we discuss in App.~\ref{app:CubProd}. When truncating the logical acting on $z$-edges, we find two pairs of excitations, that can be separated into four individual fractons at the four corners of a 2D membrane, extended in the $xy$ plane, similar to those that appear in the $X$-cube model. 

Considering $Z$-logicals, there are two families of line-like logicals, extended in either the $x$ or $z$ directions, along with a third set composed of a product of \emph{two} Sierpinski triangles (see App.~\ref{app:CubProd}). The former create lineon excitations that can propagate along the direction of the logicals; one can take a composite of three such lineons to get an excitation that can move freely within the $xy$ plane. We note that a naive truncation of the third type of $Z$ logical creates a large number of excitations, rather than a few isolated ones, although we expect that operators creating such well-separated fractons should also exist. 

\section{Non-Euclidean models}\label{sec:NonEuclidean}

In this section, we discuss review various results in the literature regarding codes and phases on non-Euclidean graphs from the standpoint of the general framework outlined here. We first discuss some simple constructions in recent physics literature and outline how they fit into the product constructions introduced in Sec.~\ref{sec:products} and~\ref{sec:GaugeAndHiggs}. We then go on to give an exposition to the existing constructions of asymptotically good quantum LDPC codes and their relationship to each other. 

\subsection{Topological and fracton phases on graphs}\label{subsec:LaplacianArboreal}

Before delving into the elaborate constructions required to realize good qLDPC codes, we review two simpler examples of codes on generic graphs from recent literature~\cite{gorantla2023gapped,manoj2023arboreal} and how they fit into machinery of product constructions. Both of these papers are motivated by the idea of generalizing the notion of topological order and fracton phases from Euclidean lattices to more generic graphs and while they are not formulated in those terms, the models they construct turn out to be examples of the kinds of codes (hypergraph product and generalized $X$-cube) discussed in Sec.~\ref{sec:GaugeAndHiggs}, illustrating the usefulness of these ideas in approaching the problem of non-Euclidean phases in a systematic manner.  

\subsubsection{Anisotropic Laplacian model} 

Our first example is the set of models introduced in Ref. \onlinecite{gorantla2023gapped}, which serves as the inspiration for our introduction of the (classical) Laplacian model in Sec.~\ref{sec:BuildingBlocks}. The authors of Ref. \onlinecite{gorantla2023gapped} introduce a set of quantum stabilizer models on an underlying geometry which consists of a stack of multiple copies of the same graph $\Gamma$, connected by nearest-neighbor edges between different copies. From the perspective of product constructions, we can identify their models as hypergraph product codes of $\mathscr{L}(\Gamma)$ with the repetition code $\mathcal{I}_\text{1D}$\footnote{Ref. \onlinecite{gorantla2023gapped} also discusses generalizations from qubits to higher dimensional qudit versions of the same model.}. In Fig.~\ref{fig:CodesLocalRed}, we called this model the ``Laplacian gauge theory'' to distinguish it from the corresponding (ungauged) classical code. 

Ref. \onlinecite{gorantla2023gapped} provides general formulae for the ground state degeneracy and the form of logical / symmetry operators of the resulting CSS codes. Due to the relationship between the logicals of hypergraph product codes and those of the underlying classical codes discussed in Sec.~\ref{subsec:HGP}, their results can be re-interpreted in terms of the logicals of the code $\mathscr{L}(\Gamma)$. In particular, they provide a formula for $k$, and a possible generating set of logical operators, in terms of the \emph{Smith decomposition} of the graph Laplacian of $\Gamma$. They also identify lineon excitations, which are free to move between the different copies of $\Gamma$ but not within each layer, which is again a natural consequence of the HGP construction. 

\subsubsection{Arboreal topological order}

Another recent example is set of models constructed in Ref. \onlinecite{manoj2023arboreal}, which considers models insipred by the toric code and $X$-cube Hamiltonians on graphs build out of tree graphs. Translated into our language, the models considered therein can be constructed by applying product constructions to Ising models $\mathcal{I}(\Gamma)$ where $\Gamma$ is either an infinite Bethe lattice or a graph obtained from truncating a Bethe lattice after a finite number of generations with various choices of boundary conditions. The main set of models considered in Ref. \onlinecite{manoj2023arboreal} turn out to be equivlent to HGP codes of these $\mathcal{I}(\Gamma)$ Ising models. The authors also consider a model inspired by the $X$-cube model, which in our language we again identify as the GXC model with the choise $\mathscr{C}_A = \mathscr{C}_B = \mathscr{C}_C = \mathcal{I}(\Gamma)$. In both cases, they analyze how the ground state degeneracy $k_q$ depends on the choice of boundary conditions, as well as properties of their excitations, among other features.

\subsection{Asymptotically good qLDPC codes}\label{subsec:GoodCodes}

Here, we review the constructions, presented in Refs. \onlinecite{panteleev2022asymptotically,dinur2023good,leverrier2022quantum,lin2022good}, which obtain good quantum LDPC codes, achieving the optimal scaling of $k,d \propto n$\footnote{These works build on the initial of breakthrough results~\cite{hastings2021fiber,panteleev2021quantum} that were the first to break the ``$n^{1/2} \text{polylog}(n)$ barrier'' on the code distance.}. As we shall see, all of these are obtained by variants of the balanced product construction~\cite{breuckmann2021balanced}. In this section, we describe the constructions themselves in enough detail to show how the ingredients discussed above enter into them, and describe how the different constructions relate to each other (see Fig.~\ref{fig:goodQLDPC} in particular). In the next section we will discuss the properties that the resulting codes have and how they lead to a good quantum code distance.

\subsubsection{Balanced products of Tanner codes} 

The first construction of a provably good qLDPC code to appear in the literature was that of Ref. \onlinecite{panteleev2022asymptotically} [PK]; a similar construction appears in Ref. \onlinecite{dinur2023good} [DHLV]. In both cases, the authors consider balanced products of two Tanner codes\footnote{Neither of the papers formulate their construction explicitly in these terms. Ref. \onlinecite{panteleev2022asymptotically} formulates their code as a \emph{lifted product}, which, while not explicitly defined as such, turns out to be a special case of a balanced product~\cite{breuckmann2021quantum} Ref. \onlinecite{dinur2023good} starts with an underlying geometric structure called the \emph{left-right (L-R) Cayley complex}, which itself turns out to be a balanced product of two graphs. See also our discussion of quantum Tanner codes below.}. In particular their construction can be described as follows
\begin{itemize}
    \item One starts with a finite discrete group $G$. In anticipation of the product construction, we choose \emph{two} generating sets of $G$, $S_A$ and $S_B$, both of which are symmetric in the sense that if a group element $s$ is in $S_{A/B}$ then so is its inverse $s^{-1}$. 
    \item From $G,S_A,S_B$ we build two Cayley graphs, $\Gamma_{A/B} = \Gamma(G,S_{A/B})$. In $\Gamma_A$ we  edges to correspond to multiplying with a generating element from the left (so that we have edges $(g,ag)$ for $a \in S_A$) and in $\Gamma_B$, multiplying from the right (edges $(g,gb)$ for $g \in S_B$); this will be important when $G$ is non-Abelian, as it is in the cases that yield good codes. In the cases of interest, the generating sets are such that $\Gamma_A$ and $\Gamma_B$ are isomorphic graphs. 
    \item One then builds Tanner codes $\mathscr{C}_A = \mathcal{T}(\Gamma_A,\mathscr{C}_{0,A})$ and $\mathscr{C}_B = \mathcal{T}(\Gamma_B,\mathscr{C}_{0,B})$ on these two graphs. 
    \item In [DHLV], the tensor product $\mathscr{C}_A \otimes \mathscr{C}_B$ is taken. In [PK] one first takes the transpose of one of the input codes, and considers $\mathscr{C}_A \otimes \mathscr{C}_B^T$ instead. 
    \item The resulting product codes are invariant under the action of $G$ acting on the left on $\mathscr{C}_A$ and on the right on $\mathscr{C}_B$. One then mods out the simultaneous (i.e., diagonal) symmetry action to obtain a \emph{balanced product} code.
    \item The gauge dual of the resulting classical codes gives the desired quantum code.
\end{itemize}

In order to obtain good qLDPC codes from this construction, one needs the underlying graphs $\Gamma_{A/B}$ to have good expansion properties; in particular, one can take $G = \text{PSL}(2,\mathbb{F}_q)$, which can yield optimal spectral expanders as mentioned in Sec.~\ref{sec:BuildingBlocks}. The small codes $\mathscr{C}_{0,A/B}$ are then both chosen to be such that $\mathscr{C}_{A/B}$ are good classical codes. Their hypergraph product would then be a code with $k_\text{q} = O(n)$ and $d_\text{q} = O(\sqrt{n})$. Modding out $G$ then reduces the number of qubits $n_\text{q}$, thereby boosting the relative distance $d_\text{q}/n_\text{q}$ from $O(n_\text{q}^{-1/2})$ to $O(1)$. This last step is not automatic, as the distance could also decrease as we mod out $G$. Indeed, to ensure the desired quantum code distance, additional conditions need to be imposed on the small codes. We will describe these below in Sec.~\ref{subsec:LocalMinimality}, when we come to discuss how one proves the good code distance of these quantum codes.  

Apart from the different (but conceptually related) conditions they impose on the small codes, the main difference between [DHLV] and [PK] is the additional transpose taken in the latter. This has the advantage of making the properties of the $X$ and $Z$ logicals more similar to one another\footnote{It is for the same reason that the original definition of HGP codes~\cite{tillich2013quantum} also includes a transpose on one of its classical input codes.}, allowing one to prove good distance for both in one fell swoop. In [DHLV], on the other hand, good $X$ and $Z$-distance need to be proven separately. 

\subsubsection{Quantum Tanner codes} 

\begin{figure} 
    \centering
    \includegraphics[trim={1cm 0 0 0},width = 0.75\linewidth]{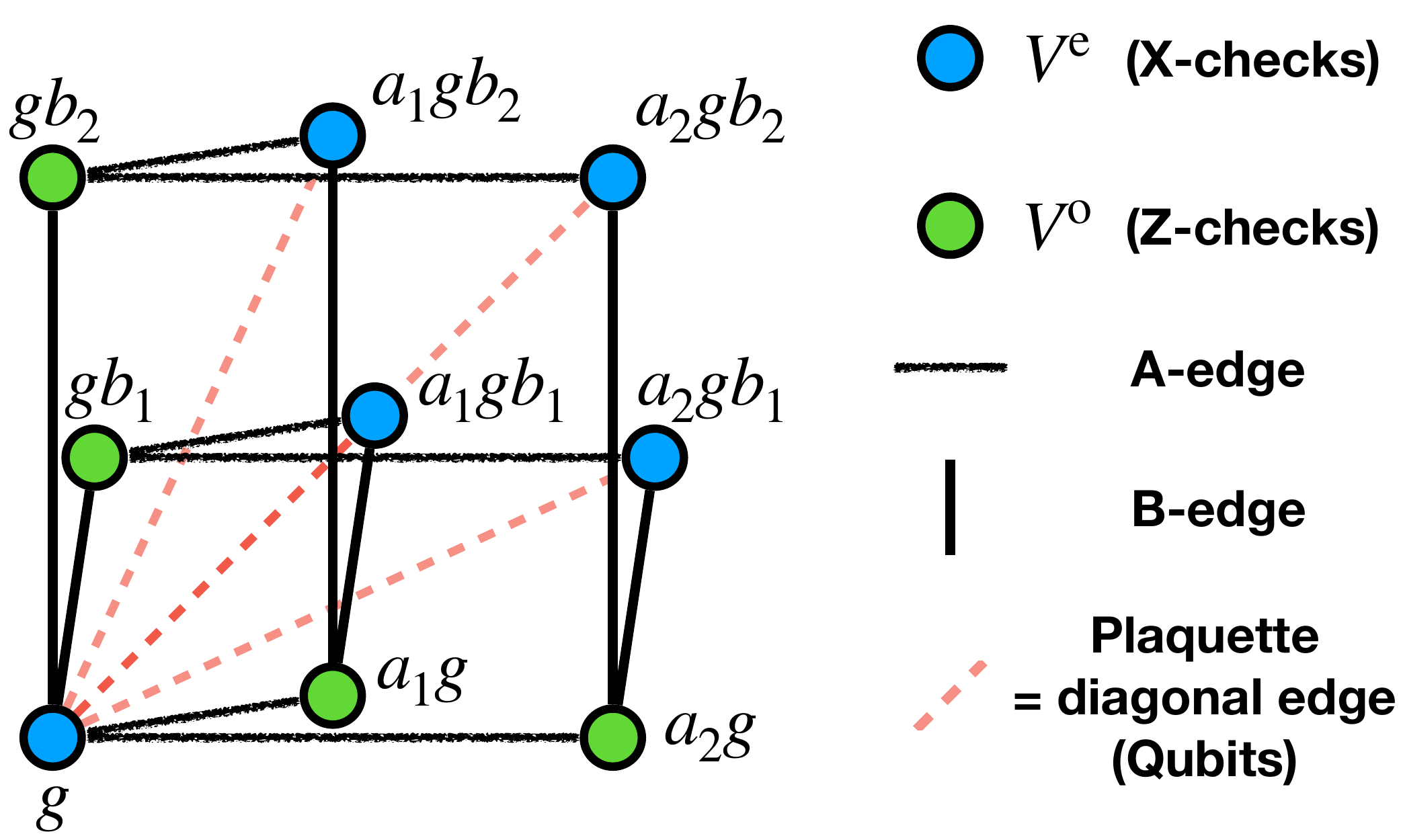}
    \caption{\textbf{Local view of the quantum Tanner code.} Vertices, edges and square plaquettes form the left-right Cayley complex. Vertices are divided into two sublattices, $V^\text{e}$ and $V^\text{o}$, on which $X$ and $Z$ checks are defined. Qubits are associated to plaquettes and the plaquettes adjacent to a given vertex are uniquely labeled by pairs $(a,b)$ so that they can naturally be arranged in a 2D array. Each plaquettes corresponds to an edge (red dashed line) on the modified graph $\Gamma_\square$, connecting vertices of $V^\text{e}$. 
    }
    \label{fig:Book}
\end{figure}

Ref. \onlinecite{leverrier2022quantum} [LZ] introduced another family of codes that includes good qLDPC codes. These are introduced as quantum analogues of the Tanner code construction, wherein and underlying geometrical object is ``dressed'' with small codes. The difference is that the geometrical object in question is itself a $2$-level chain complex, rather than a graph. 

In particular, [LZ] consider a chain complex called the \emph{left-right Cayley complex} (originally introduced in Ref. \onlinecite{dinur2022locally}), denoted $\text{Cay}(G,S_A,S_B)$, which is a generalization of the Cayley graph construction, built out of a discrete group $G$ and two different generating sets $S_{A}$,$S_{B}$. $\text{Cay}(G,S_A,S_B)$ is constructed as follows. 
\begin{itemize}
    \item The vertices correspond to elements of $G$, just like in the Cayley graph.
    \item Edges (which in this case are edges in the usual sense, connecting two vertices) come in two flavors: one adds an edge $(g,ag)$ for every $a \in S_A$ and an edge $(g,gb)$ for every $b \in S_B$.
    \item The resulting graph comes equipped with a notion of square plaquettes, defined by the four corners $(g,ag,gb,agb)$\footnote{In principle, one would also want to ensure that these four corners are indeed four distinct elements of $G$---called the ``total no-conjugacy'' condition in Ref. \onlinecite{dinur2022locally}. However, this condition can be avoided by instead considering four copies of the group and drawing edges between distinct copies, as described in Ref. \onlinecite{leverrier2022quantum}.}. These squares form the faces of the two-dimensional chain complex.
    \item The boundary maps are defined in the obvious way. I.e., the boundary of a plaquette $(g,ag,gb,agb)$ consists of the four edges $(g,ag)$, $(g,gb)$, $(ag,agb)$ and $(gb,agb)$, while the boundary of an edge consists of the two vertices it connects.
\end{itemize}

Given $\text{Cay}(G,S_A,S_B)$, one now wants to define a quantum CSS code on it in a manner resembling the classical Tanner graph construction. To do so, one assigns qubits to the \emph{faces} of the left-right Cayley complex, while $X$ and $Z$ checks will be assigned to the \emph{vertices}. In particular, one considers a bipartition of the vertices as $V = V^\text{e} \cup V^\text{o}$, such that all the edges connect one vertex from $V^\text{e}$ to one vertex from $V^\text{o}$\footnote{Again, one either needs to make sure that the graph corresponding to the lower two levels of $\text{Cay}(G,S_A,S_B)$ is indeed bipartite, or one can replace it with a bipartite graph by taking two copies of the vertices, known as the \emph{bipartite double cover}.}. $X$-checks will be placed on the vertices in $V^\text{e}$ and $Z$-checks on elements of $V^\text{o}$. They will correspond to two appropriately chosen small codes, $\mathscr{C}_{0,X}$ and $\mathscr{C}_{0,Z}$, acting on the \emph{local view}, consisting of all the plaquettes adjacent to a given vertex. 

To see how the checks are defined note that the local view naturally has a product structure associated to it. Due to the construction above, there are $|S_A|\cdot|S_B|$ plaquettes adjacent to any given vertex and they can be uniquely labeled by the pairs $(a,b)$. We can thus arrange the qubits in the local view in a 2D grid, just like the ones we have encountered in our discussion of product constructions in Sec.~\ref{sec:products}. Indeed, $\mathscr{C}_{0,X/Z}$ will both be defined as appropriately defined product codes. In particular, let us choose some small classical code $\mathscr{C}_{0,A}$ ($\mathscr{C}_{0,B}$) acting on an $|S_A|$ ($|S_B|$) number of bits. Out of these two codes, we can now form the codes that give us out desired $X$ and $Z$ checks as follows\footnote{Note that, since we are considering small codes, acting on a finite number of bits here, there is no LDPC restriction on $\mathscr{C}_{0,X/Z}$, which is why we are allowed to use dual codes.}: $\mathscr{C}_{0,X} = (\mathscr{C}_{0,A} \otimes \mathscr{C}_{0,B})^\perp$ and $\mathscr{C}_{0,Z} = (\mathscr{C}_{0,A}^\perp \otimes \mathscr{C}_{0,B}^\perp)^\perp \equiv \mathscr{C}_{0,A} \star \mathscr{C}_{0,B}$. Thanks to the fact that both are defined in terms of the same input codes, the $X$ and $Z$ checks turn out to be mutually commuting\footnote{To see this, consider for example two neighboring vertices connected by an edge labeled by $a \in S_A$. The overlap of the local views of the two vertices corresponds to a row in the $|S_A| \times |S_B|$ grid of qubits, labeled by $a$. By definition of the dual code, checks of $\mathscr{C}_{0,X}$ restricted to this row look like logicals of $\mathscr{C}_{0,B}$ while checks of $\mathscr{C}_{0,Z}$ look like checks of the same code, which commute by definition.}. This completes the definition of the quantum Tanner code.

It might seem surprising that the quantum Tanner code has qubits associated to the $2$-dimensional plaquettes of the underlying chain complex, rather than to edges, as would be more natural based on the general relationship between CSS codes and chain complexes. One can make the construction look more natural from this perspective by drawing it on a different graph $\Gamma_\square$. The vertices of $\Gamma_\square$ will be half of the vertices of the Cayley complex, namely those in $V^\text{e}$, on which we placed our $X$-checks of the quantum Tanner code. Now, every face of the Cayley complex contains exactly two of these vertices and as such, we can associate an edge to every face (see the red lines in Fig.~\ref{fig:goodQLDPC}(f)). Thus, in $\Gamma_\square$, $X$-checks and qubits of the quantum Tanner code indeed correspond to vertices and edges. The other set of vertices of the original complex, $V^\text{o}$, on which we placed the $Z$-checks, now become plaquettes. 

Similar to the case of classical Tanner codes, discussed in Sec.~\ref{sec:BuildingBlocks}, the proof of the goodness of quantum Tanner codes relies on a combination of the expansion properties of the graphs $\Gamma_{A},\Gamma_B,\Gamma_\square$, as well as properties of the classical small codes $\mathscr{C}_{0,A},\mathscr{C}_{0,B}.$ The intuition is again that the expansion of the underlying graphs helps magnify the local properties, encoded in the small codes, into global ones. 

\subsubsection{Relationship between quantum Tanner and balanced product codes}

\begin{figure} 
    \centering
    \includegraphics[trim={1cm 0 0 0},width = 1.\linewidth]{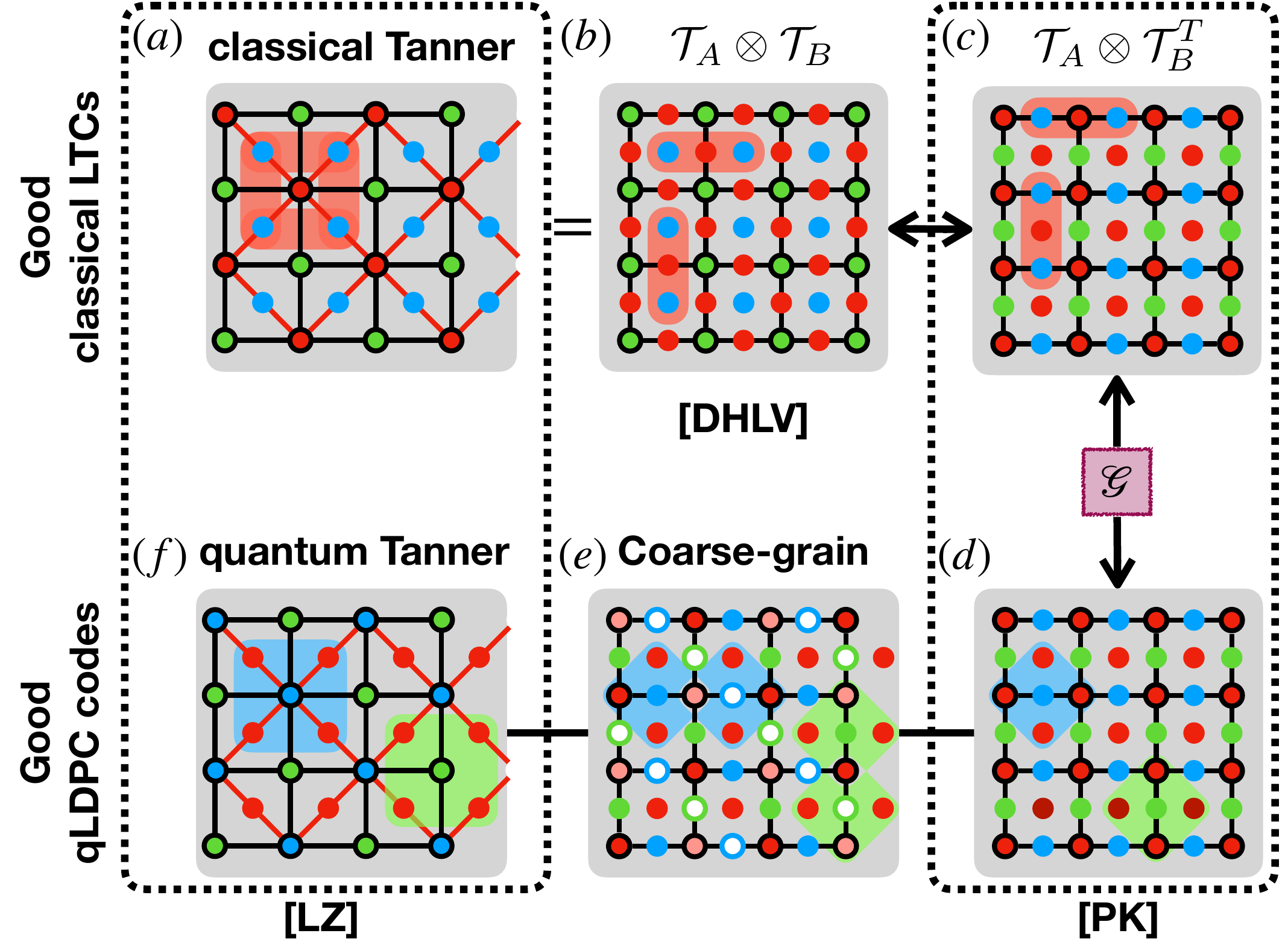}
    \caption{\textbf{Structure of various code constructions and their relationships from Refs. \cite{dinur2023good,panteleev2022asymptotically,leverrier2022quantum}.} We illustrate the structure on an example using two 1D Ising models, envisaged as Tanner codes with bits (checks) on edges (vertices) of a cycle graph. Blue, red and green dots represent bits, checks and local redundancies ($X$-checks, qubits and $Z$-checks) in the classical (quantum) codes. (b) is the classical tensor product of the two Tanner codes which can be equivalently rewritten as the Tanner code (a) on a different graph (the rotated square lattice with red edges), containing half of the vertices. (b) and (c) differ by taking the transpose of one of the inputs; they correspond to the same chain complex but with the roles of the various levels permuted. Gauging the latter gives rise to a version of the toric code (d), with qubits on vertices and plaquettes and $X$-checks ($Z$-checks) on horizontal (vertical ) edges. (e) shows a ``coarse-graining'' of  this toric code by multiplying checks in pairs in such a way that $X$- and $Z$-checks do not overlap on the vertices which can therefore be dropped to obtain another toric code (f) on half as many qubits. This is a version of a quantum Tanner code on the rotated lattice. In the Constructions of Refs.~\cite{leverrier2022quantum,panteleev2022asymptotically,dinur2023good}, the inputs $\mathcal{T}_{A,B}$ are good classical Tanner codes and there is an additional step of modding out symmetries that turns these into balanced products. The resulting classical codes are locally testable codes (LTCs) and the quantum codes are good qLDPC codes. The classical and quantum codes on the left correspond to those appearing Ref. \onlinecite{leverrier2022quantum} [LZ] while those on the right to Ref. \onlinecite{panteleev2022asymptotically} [PK].
    }
    \label{fig:goodQLDPC}
\end{figure}

As should be clear from the definition, the quantum Tanner code construction bears a close similarity to the balanced product codes of [PK] and [DHLV] discussed above: both constructions involve a group $G$ and two of its generating sets, as well as two small codes $\mathscr{C}_{A/B}$. In fact, the two can be related more directly: the quantum Tanner code arises as a ``coarse-grained'' version of the quantum codes of [PK] discussed above, obtained by taking product of its checks and then getting rid of half of the qubits~\cite{LeverrierNote}. We now discuss this set of relationships in some detail.

The first step in connecting the two construction is by noting that the left-right Cayley complex introduced above itself can be understood as the balanced product of the two Cayley graphs, $\Gamma_A$ and $\Gamma_B$, that have appeared in the definitions of the codes of [PK] and [DHLV]. The product $\Gamma_A \otimes \Gamma_B$ has $|G|^2$ vertices, labeled by a pair of group elements $(g,g')$. Horizontal (vertical) edges corresponding to left (right) multiplication on the first (second) group element with generators taken from $S_A$ ($S_B$)\footnote{Here, we consider taking products of graphs, rather than codes. In our definitions, this is equivalent to using $\mathcal{I}(\Gamma)$, the Ising model on graph $\Gamma$, everywhere.}. Modding out the diagonal action of $G$ compresses this product complex back onto a single copy of the group $G$ and we end up exactly with the L-R Cayley complex, with the two types of edges originating from the horizontal and vertical edges of the product. 

Interpreting the L-R Cayley complex as a balanced product of the graphs $\Gamma_{A/B}$ also induces a similar interpretation on the Tanner codes on these two graphs; in particular, their balanced product, which has appeared in our discussion of [DHLV] (and which was introduced previously in Ref. \onlinecite{dinur2022locally}) turns out to be a classical Tanner code (of the kind defined in Sec.~\ref{sec:BuildingBlocks}) on the L-R Cayley complex~\cite{leverrier2022quantum}. More precisely, from the L-R Cayley complex, we can construct the graph $\Gamma_\square$ discussed above, which has edges corresponding to the squares of the L-R Cayley complex. As we discussed, every vertex of $\Gamma_\square$ has degree $|S_A|\times|S_B|$, so that the local view naturally comes endowed with a product structure. We can then define a Tanner code $\mathcal{T}(\Gamma_\square,\mathscr{C}_{0,A} \otimes \mathscr{C}_{0,B})$. Following through the steps of the balanced product, it turns out that the code $\mathcal{T}(\Gamma_A,\mathscr{C}_{0,A}) \otimes_G \mathcal{T}(\Gamma_B,\mathscr{C}_{0,B})$ we encountered before is precisely this Tanner code on $\Gamma_\square$. This gives a useful interpretation into the structure of these balanced products: while modding out the symmetry destroys the tensor product structure \emph{globally}, this structure is still present \emph{locally}, in the vicinity of any vertex.  

Similarly, the quantum codes of [PK] can be interpreted as living on the L-R Cayley complex. Due to the additional transpose in their definition, they have qubits appearing on both the vertices and faces of the complex while their $X$ and $Z$ checks are associated to the two types of edges (``horizontal'' and ``vertical''), respectively. Now, this code can be reduced to the quantum Tanner code as follows. For a vertex $v \in V^\text{e}$, one can consider the horizontal edges meeting at that vertex and take products of the corresponding $X$-checks to get a ``small code'' assigned to $v$. One can do this in a way such that (a) Restricted to the plaquettes in the local view of $v$, the resulting code is equivalent to the small code $\mathscr{C}_{0,X}$ appearing in the definition of the quantum Tanner code and (b) the checks of the new code have no support on the vertices $V^\text{e}$, only on the odd vertices $V^\text{o}$. One can similarly take product of the $Z$-checks on vertical edges around a vertex $v' \in V^\text{o}$ to create a code that mimics $\mathscr{C}_{0,Z}$ and has no support on $V^\text{o}$. Now, since these newly defined $X$ and $Z$ checks have no overlap on the vertex qubits, these can be removed from the code and we can restrict to the plaquette qubits alone. The resulting code is precisely the quantum Tanner code.

The overall structure of this set of relationships an be illustrated already at the level of tensor product codes and their quantum duals (i.e., hypergraph product codes), without the need to mod out any symmetries. We show this in Fig.~\ref{fig:goodQLDPC} for the product of two repetition codes (1D Ising models). To bring out the similarities with the above discussion, we write these as Tanner codes, with bits on the edges of the cycle graph and a single 2-body check on each vertex. The tensor product gives a 2D Ising model with spins on the plaquettes (Fig.~\ref{fig:goodQLDPC}(b)), which can be equivalently rewritten as a Tanner code on a rotated lattice where vertices correspond to half of the original vertices and edges correspond to plaquettes (Fig.~\ref{fig:goodQLDPC}(a)). Vertices of this coarse-grained lattice hosts a small code with 4 bits and 4 checks, corresponding to a tiny product code\footnote{The other set of vertices of the original lattice turn into plaquettes after coarse-graining and we can associate local redundancies between the small codes to them. Note that there is also a single local redundancy within each small product code itself.}. These codes have the structure of the classical codes appearing in Refs. \onlinecite{dinur2022locally,dinur2023good,leverrier2022quantum}. If one instead takes the transpose of one of the repetition codes, one ends up with the analogue of the classical codes in Ref. \onlinecite{panteleev2022asymptotically}, where where bits are assigned to the horizontal edges of the 2D lattice and checks to both vertices and plaquettes, while vertical edges host local redundancies (Fig.~\ref{fig:goodQLDPC}(c)).

We can gauge this last code to get the analogue of the quantum codes in Ref. \onlinecite{panteleev2022asymptotically}, which in this case is simply a ``rotated toric code''~\cite{kovalev2012improved}, with two sets of qubits on vertices/plaquettes and $X$ ($Z$)-checks on horizontal (vertical) edges (Fig.~\ref{fig:goodQLDPC}(d)). Next, one bi-partitions the sites of the 2D lattice into two colors (``even'' and ``odd'') in a checkerboard pattern. The coarse-graining step involves taking the product of two $X$ ($Z$) checks on edges that meet at an even (odd) vertex. This results in a 6-body check that we can naturally associate to that vertex. By taking the product, the support on the vertex itself cancels, so that the 6-body $X$ checks have no support on the even vertices and the $Z$ checks have no support on the odd ones. One can then truncate these checks to a smaller code only on the plaquette qubits, which is drawn in Fig.~\ref{fig:goodQLDPC}(f). In terms of the original lattice, this has qubits on the plaquettes, $X$ checks on half the vertices and $Z$ checks on the other half, mimicking the structure of the quantum Tanner code. It can also be drawn as the usual toric code on the coarse-grained lattice that has already appeared in Fig.~\ref{fig:goodQLDPC}(a) (see red edges in Fig.~\ref{fig:goodQLDPC}(f)). 

Going back to the constructions of [PK], [LZ] and [DHLV], while in this section we focused on quantum codes (panels (d) and (f), as well as the quantum code obtained by gauging (b), not shown in the figure), the corresponding classical codes also have interesting properties, most notably they are \emph{local testable codes} (LTCs), as we will discuss in Sec.~\ref{sec:EnergyBarriers}. Indeed, as we will see, there for all the constructions of good qLDPC codes, the classical codes to which they are gauge dual have this property (while also being good classical LDPC codes). In Sec.~\ref{sec:EnergyBarriers}, we will discuss the relationship of local testability with the good quantum code distance, both of which have to do with the geometry of the underlying two-dimensional chain complex. 

\subsubsection{Balanced products of lossless expanders} 

Finally, let us discuss the codes defined in Ref. \onlinecite{lin2022good} [LH] which differ from the ones above in that they are not constructed in terms of Tanner codes. Instead, they use as ingredients generic classical codes, represented in terms of their Tanner graphs and require that these Tanner graphs have sufficiently good vertex expansion in both directions (i.e., both from bits to checks and from checks to bits) while also having a large symmetry group needed for the balanced product construction.

In particular, [LH] define a property called \emph{2-sided lossless vertex expansion}. We have defined vertex expansion back in Sec.~\ref{sec:BuildingBlocks}, in terms of two parameters $\alpha$ and $\gamma$, the first of which sets how quickly the graph is expanding. \emph{Lossless} expansion refers this parameter being close to maximal. In particular, assume that every bit is adjacent to $w$ checks, and every check involves $w'$ bits. Then the Tanner graph is a $(\gamma,\epsilon)$ 2-sided lossless expander if it has expansion from bits to checks with $\alpha = (1-\epsilon) w$ and from checks to bits with $\alpha = (1-\epsilon) w'$. In particular, the proofs of[LH] require $\epsilon < 1/12$. At the same time, it is required that the codes in question have a large symmetry group $G$ which can be used to construct their balanced products. We note that at the moment, such highly symmetric 2-sides lossless expanders are not known to exist; [LH] shows that if they do, their balanced products provide good qLDPC codes.

\subsection{The geometry of good qLDPC codes}\label{subsec:LocalMinimality}

In the preceding section we described how the various examples of good qLDPC codes in the literature are constructed. We now discuss in some detail the structure of the proofs that these constructions indeed give the desired properties. While the code rate $k_\text{q}/n_\text{q}$ is usually easy to bound, by just counting the number of qubits and checks, and can be made finite even for HGP codes, the linear scaling of the code distance, $d_\text{q} \propto n_\text{q}$ is challenging to show and usually takes up the majority of the papers mentioned. One interesting feature of all these proofs is that they are geometrical in nature: they consider the CSS codes in question from the chain complex perspective. The geometric properties of these chain complexes that give rise to a good quantum code distance then also have implications for the classical codes to which they are gauge dual. In particular, in all the known examples we discussed above, it turns out that in our language both the ungauged classical code and its Kramers-Wannier dual has a property called \emph{local testability}, which we will discuss in the next section. 

The key results in [PK] and [DHLV] concern \emph{locally minimal} lines in the chain complex obtained from the balanced product construction. Given a $2$-complex, defined by the vector spaces $V_{0, 1,2}$ and boundary maps $\delta_{1,2}$, a locally minimal line $c_1 \neq 0$ is a collections of edges (i.e., a vector in $V_1$) such that its length (the number of basis vectors appearing in it) cannot be decreased by adding to it the boundary of any elementary plaquette (basis vector of $V_2$). In particular, one aims to prove what in [DHLV] has been termed the \emph{small-set locally-minimal expansion} property, characterized by a pair of numbers $(\mu,\nu)$. It states that if a locally minimal line $c_1$ contains less than a fraction $\mu$ of all edges of the chain complex ($|c_1| \leq \mu m$, where $m = \text{dim}(V_1)$ is the number of edges), then 
\begin{equation}\label{eq:LocMinExp}
    |\delta_1 c_1| \geq \nu |c_1| ,
\end{equation}
where $\delta_1 c_1$ is the boundary of $c_1$ in the chain complex. 

What locally minimal expansion tells us is that for sufficiently short lines, whose length is smaller than a fraction $\mu$ of the entire system, the size of the boundary is proportional to that of the line itself. One wants to show that this is true for $\mu,\nu = O(1)$. The importance of this property becomes apparent if we consider the case when $c_1$ is a cycle, with zero boundary. Then in order for $c_1$ to be non-trivial, it must be large enough to violate the condition, i.e., $|c_1| \geq \mu m$,. The size of the smallest non-trivial locally minimal cycle is called the \emph{locally minimal distance} of the code, denoted by $d_\text{LM}$. If the chain complex has $(\mu,\nu)$ locally minimal expansion, then $d_\text{LM} \geq \mu m$ is proportional to the total number of edges. This has immediate consequences for the quantum CSS code associated to the chain complex: since logical $Z$ operators are non-trivial cycles, and the smallest such cycles in any equivalence (homology) class is by definition locally minimal, we have that $d_Z \geq d_\text{LM}$. 

One can define a similar notion of co-locally-minimal lines, by requiring that the length of $c_1$ cannot be decreased by the addition of the co-boundary of any vertex (basis element of $V_0$). We can then define $(\mu,\nu)$ small-set co-locally-minimal expansion in analogy with Eq.~\eqref{eq:LocMinExp}, i.e. by requiring that $|\delta_2^T c_1| \geq |c_1|$ for any co-locally minimal lines with $|c_1| \leq \mu m$\footnote{In other words, this is locally-minimal expansion of the dual chain complex}. Similarly, we can define the co-locally-minimal distance $d_\text{LM}^\text{KW}$ as the length of the smallest non-trivial co-locally-minimal co-cycle. Co-locally-minimal expansion then ensures that $d_X \geq d_\text{LM}^\text{KW} \geq \mu m$. Therefore, a sufficient condition for a good quantum distance, $d_\text{q} \propto m$, is to show \emph{both} locally-minimal expansion and co-expansion with some $\mu,\nu = O(1)$ constants. 

As noted above, the construction of [PK], which includes taking the transpose of one of the input Tanner codes, is symmetric between the balanced product chain complex and its dual, so that one can prove properties of locally minimal cycles and co-cycles simultaneously. [PK] show that if the Cayley graphs $\Gamma_{A,B}$ are sufficiently quickly expanding, and the small codes $\mathscr{C}_{0,A/B}$ satisfy appropriate conditions, then the lower bounds $d_\text{LM},d_\text{LM}^\text{KW} \geq \mu m$ for some $\mu = O(1)$ indeed obtain, ensuring a good quantum code distance. 

The situation is more complicated for the codes considered in Ref. \onlinecite{dinur2023good}, where no transpose is taken. In this case, the resulting chain complex is asymmetric; for example, the elementary faces (basis vectors in $V_2$) are only adjacent on $4$ edges, while vertices (basis vectors in $V_0$) can be adjacent on many edges, depending on the degree of the Cayley graphs $\Gamma_{A,B}$, i.e. Indeed, as [DHLV] show, while co-locally-minimal expansion obtains (under appropriate assumptions on $\mathscr{C}_{0,A/B}$), locally-minimal expansion does not. However, they prove a weaker property, which they name \emph{small-set boundary expansion}, characterized by parameters $(\mu,\nu,\lambda)$. It states that if a line $c_1$ contains less than a fraction $\mu$ of all edges, then there exists a surface (set of faces) $c_2$ such that $|c_2| \leq \lambda |c_1|$ and $|\delta_1 c_1| \geq \nu |c_1 + \delta_2 c_2|$. In other words, while $c_1$ might not in itself be locally expanding, we can always deform it by a ``small'' contractible loop to make it so. To see how small set boundary expansion implies a good quantum code distance, consider the case when $c_1$ is a closed loop. In that case, if $c_1$ is smaller than a fraction $\mu$ of all edges, we have that $|c_1 + \delta_2 c_2| \leq 0$, and therefore $c_1 = \delta_2 c_2$ is contractible. Thus, logicals (i.e., non-contractible loops) must contain at least $\mu m$ edges. 

Regarding the codes of [LZ], one can understand their properties as coming from the quantum codes of [PK] of which they are coarse-grained versions, albeit they can be proven more straightforwardly using the definition of quantum Tanner codes directly. In particular, [LZ] show a variant of large locally minimal distance. Consider a closed loop in the chain complex associated to the quantum Tanner code, to which we can associate a product of Pauli $Z$ operators that commute with all $X$ checks. What [LZ] prove is that if the support of this ``Wilson loop'' $\mathcal{W}$ is smaller than some finite fraction of all qubits, then there exists some vertex $v \in V^\text{o}$ such that the support of $\mathcal{W}$ can be strictly decreased by multiplying it with some checks from the small code $\mathscr{C}_{0,Z}$ associated to $v$. This gives a large $d_Z$ distance by the same logic as before, and an analogous argument leads to a large $X$-distance. 

As we mentioned, the proofs require certain properties from the small codes $\mathscr{C}_{0,A/B}$. This includes a lower bound on their $k$ and $d$, in analogy with the requirements needed for the classical Tanner code construction to produce good classical codes. However, in order to ensure the above geometrical properties, one also needs to put additional requirements on the small codes. The details of these differ between [PK],[DHLV] and [LZ] but all three are similar in spirit. Let us discuss the definitions of [DHLV], which is in some sense the conceptually simplest. [DHLV] defines a notion of \emph{robustness} of the small codes as follows. Consider the $2$-complex $\mathscr{C}_{0,A}^\perp \otimes \mathscr{C}_{0,B}^\perp$. Now take $c^1$ to be some set of edges in this complex and $c^0 = \delta c^1$ to be its boundary. We say that the pair of local codes $(\mathscr{C}_{0,A},\mathscr{C}_{0,B})$ is robust with parameter $\kappa$ if $|c^0| \geq \kappa |c^1|$ for any $c^1$. [DHLV] show that such robust local codes exist, and in fact a random choice will suffice. As we can see, robustness bears resemblance to a local version of the locally minimal expansion property. This local property then gets propagated, via the expansion of the underlying graphs, to ensure the locally-minimal expansion of the balanced product code. 

Finally, turning again to the codes of [LH], they show that when their input codes satisfy the the lossless expansion condition, the 2-dimensional chain complex obtained as their balanced product has $(\mu,\nu)$ small-set locally-minimal expansion, as defined in Eq.~\eqref{eq:LocMinExp}, with $\nu = 1/2 - 6 \epsilon$ and some $\mu = O(1)$\footnote{More precisely, rather than the bare size (number of edges / sites) of the line $c_1$ and its boundary, \onlinecite{lin2022good} considers an appropriately weighted version of the size.}. This follows from the checks-to-bits expansion of the underlying classical codes that enter the balanced product. 2-sided expansion is needed to make the construction symmetric and also ensure co-locally-minimal expansion, which is needed to get a bound on the quantum code distance that scales as $d_\text{q} \geq \mu m$. 

\section{Energy barriers and local testability}\label{sec:EnergyBarriers}

In Part I~\cite{LDPCGauge}, we claimed that all known examples of good qLDPC codes originate from gauging classical codes with a particular property, known as \emph{local testability}~\cite{goldreich2005short}. Having reviewed their construction, we are now in a better position to discuss this connection in more detail. 

Let us begin by recalling the definition of a locally testable code (LTC). First, we define \emph{energy barriers} associated to a classical code $\mathscr{C}$, defined by binary matrix $\delta$ as follows. Let $F \leq d/2$ be less than half the code distance of $\mathscr{C}$ and define
\begin{equation}
E_\text{min}(F) \equiv \min_{\Sigma: |\Sigma| = F}(|\delta^\text{T}(\Sigma)|),
\end{equation}
where $\Sigma$ goes over spin configurations with a fixed number $F$ of spins flipped compared to the ``all 0'' state. We will also refer to $F$ as the \emph{Hamming weight} of the configuration $\Sigma$. $E_\text{min}(F)$ is the minimal energy (as measured by the number of checks violated) over all configurations with a given Hamming weight. Locally testable codes are defined by the property that their energy barrier grows as quickly as allowed for an LDPC code, namely
\begin{equation}\label{eq:LTC_def}
    E_\text{min}(F) \geq \kappa F, \qquad 0 \leq F \leq d/2,
\end{equation}
for some $O(1)$ ``soundness'' parameter $\kappa$. Thus, in a locally testable code, no matter how cleverly we arrange our spin flips, they violate a number of checks that grows proportionally with the number of flips, all the way up to half the code distance. We mention that equivalently, we could equivalently define local testability by allowing arbitrary configurations and defining $F$ to be the Hamming distance (number of spin flips) from the closest codeword. 

\subsection{Local testability from locally-minimal expansion}

The construction of good LDPC codes that are also locally testable has been an open problem for a long time and was only resolved with the advent of the same sort of balanced product constructions that gave rise to good qLDPC codes~\cite{dinur2022locally,panteleev2022asymptotically,lin2022c,leverrier2022quantum}. We can understand the connection based on what we said about the properties of these codes in Sec.~\ref{subsec:GoodCodes}. In particular, we saw that the good quantum code distance of these examples followed from a geometric property of the underlying chain complex, namely its small-set locally-minimal expansion, or its weaker version, small-set boundary expansion. The same properties can be used to prove local testability. 

As we saw, locally-mininal expansion ensures a large locally-minimal distance, $d_\text{LM} \geq \mu m$. To see the relationship to local testability, consider an open surface $c_2$ with a boundary $\delta_2 c_2 = c_1$ such that $0 < |c_1| < d_\text{LM}$, so that $c_1$ cannot be locally minimal. Then, by definition, there exists some elementary plaquette $p$ (basis vector of $V_2$) such that $|\delta_2 (c_2 + p)| = |c_1 + \delta_1 p| < |c_1|$. Since the resulting loop continues to be shorter than $d_\text{LM}$, this process can be iterated: we can keep adding elementary plaquettes to $c_2$ until we reach a closed surface and the boundary becomes zero. Since the size of the boundary decreases in each step, this process must terminate in at most $|\delta_2 c_2|$ steps. 

To see the relationship to local testability, consider the classical code $\mathscr{C}_\text{cl}^\text{KW}$, whose bits are associated to the plaquettes of the chain complex and is defined by the matrix $\delta = \delta_2^T$. In terms of this classical code $|\delta_2 c_2|$ is the energy cost of flipping all the spins in $c_2$. Therefore, what we have shown is that this energy cost upper bounds the Hamming distance of the corresponding spin configuration from some nearby codeword (closed surface). This shows that the inequality appearing in the definition of local testability~\eqref{eq:LTC_def} is automatically satisfied for any configuration that obeys $|\delta_2 c_2| < d_\text{LM}$. When $d_\text{LM} \propto m$, the remaining configurations already have an $O(m)$ energy cost, so that they also satisfy Eq.~\eqref{eq:LTC_def}. If we also have a large co-locally minimal distance, $d_\text{LM}^\text{KW} \geq 2$, then this ensures the local testability of the classical code $\mathscr{C}_\text{cl}$, which has bits assigned to vertices of the chain complex. 

As we discussed above, the codes of Ref. \cite{dinur2023good} do not satisfy the definition of small-set locally-minimal expansion but only the weaker condition of small-set boundary-expansion. Nevertheless, this is also sufficient to guarantee local testability. This is shown by the following argument. Consider a trivial loop $c_1 = \delta_2 c_2$ with $|c_1| < \mu m$. We thus have $c_1 + \delta_2 c_2' = \delta_2 (c_2 + c_2') = 0$ for some surface $c_2'$, implying that $c_2 + c_2'$ is a codeword of te classicalal code $\mathscr{C}_\text{cl}^\text{KW}$. By the definition of boundary expansion we then have that $\lambda |c_1| \geq |c_2'|$, where the LHS is the energy associated to spin configuration $c_2$ and the RHS is the Hamming distance of $c_2$ from the codeword $c_2 + c_2'$, as in the definition of local testability~\eqref{eq:LTC_def}. For large loops, we have $|c_1| \geq \mu m \geq \mu \frac{m}{\ell} |c_2|$, where $\ell = \text{dim}(V_2)$ is the number of faces in the chain complex.

\subsection{``Strip'' argument for product codes}\label{subsec:Strip}

\begin{figure} 
    \centering
    \includegraphics[trim={1cm 0 0 0},width = 1.\linewidth]{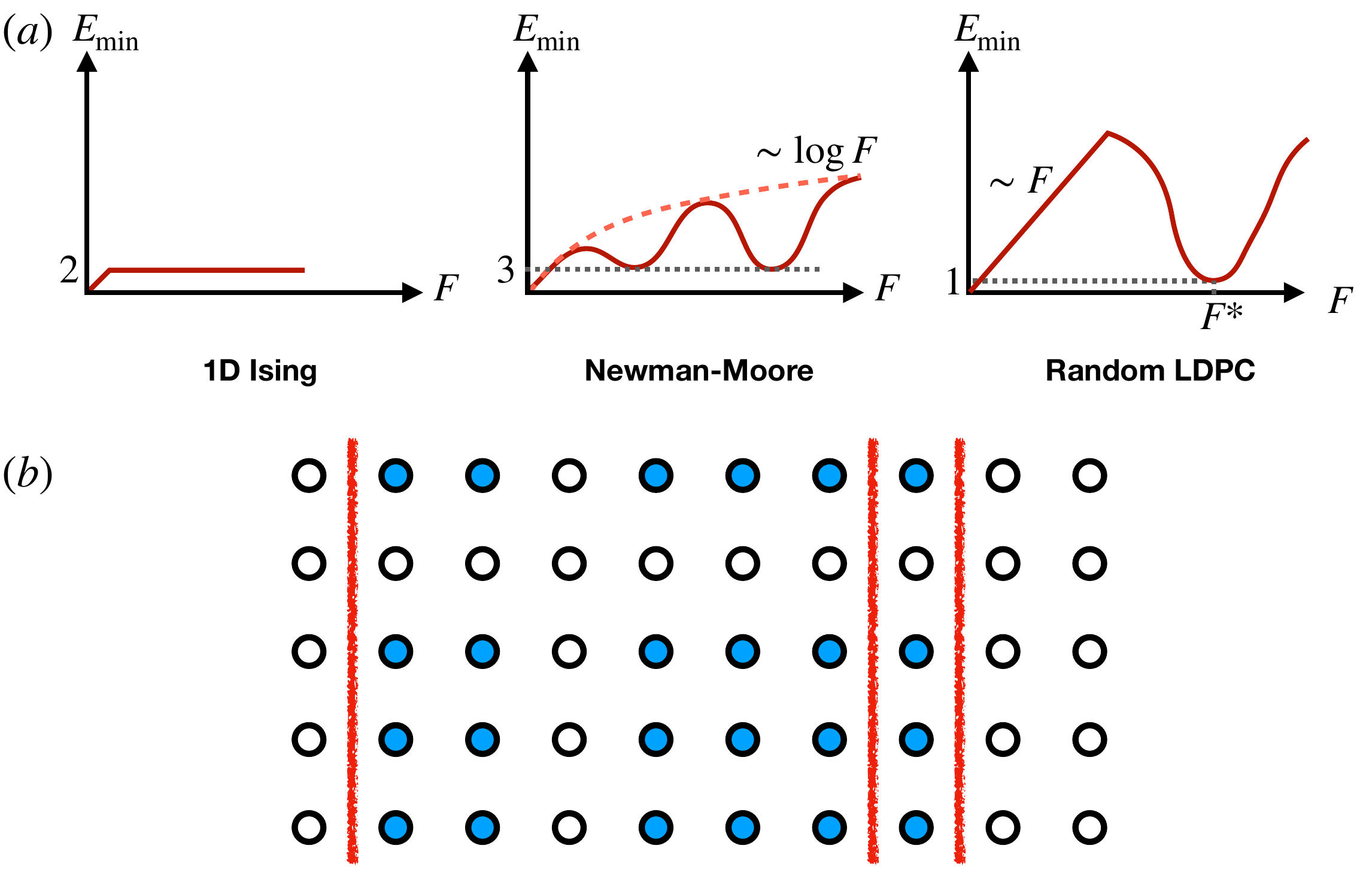}
    \caption{\textbf{Energy barriers and product codes.} (a) Sketch of energy barrier landspace for a few different classical codes without local redundancies, all of which have configurations with many flipped spins ($F$) and low energy costs. (b) A ``strip'' configuration in the tensor product code $\mathscr{C}_A \otimes \mathscr{C}_B$, corresponding to repeating the same spin configuration of $\mathscr{C}_A$ along a logical of $\mathscr{C}_B$. This creates a number of domain walls stretched across the system in the vertical direction (red lines). 
    }
    \label{fig:Strip}
\end{figure}

In the above proofs, local testability and good quantum code distance follow from the same geometric features of the chain complex that is shared between the two codes in question. In Part I~\cite{LDPCGauge} we sketched an argument that aims to more directly connect the two properties. This was motivated by the observation that in the 2D Ising model, the energy barrier $E_\text{min}(F)$ saturates to a value $2L$ which is twice the code distance of the toric code. This followed from the fact that when the number of flipped spins becomes sufficiently large ($F > L^2/4$ in the Ising model), the optimal configuration that minimizes the energy takes the form of a ``strip'' stretching across the system in one direction, whose boundary splits into two disconnected cycles which are both individually non-contractible. This raises the question whether a similar behavior occurs more generally; if it does, then combined with the local testability of the classical code, the good $Z$-distance of its gauge dual would follow. 

For codes that are obtained as tensor or balanced products, we can elaborate more on the conditions under which such a behavior would occur and connect it to the properties of the codes that enter as inputs into these product. First of all, note that the saturation of $E_\text{min}$ in the 2D Ising model is a direct consequence of the fact that the 1D Ising model has a constant energy barrier, consisting of just two domain walls at the endpoints of an arbitrarily large region of flipped spins. We can generalize this idea to other tensor product codes $\mathscr{C} = \mathscr{C}_A \otimes \mathscr{C}_B$ as follows. Let us pick the smallest logical operator of $\mathscr{C}_B$, supported on some subset $\lambda'$ with $|\lambda'| = d_B$\footnote{Without loss of generality, we assume $d_B \leq d_A$ in what follows.} and a configuration of bits in $\mathscr{C}_A$, labeled by a set $\tilde{\lambda}$ of spins that have been flipped compared to the ``all up'' state. We can combine these two into a ``strip'' configuration in $\mathscr{C}$, defined by $\sigma_{ij} = -1$ iff $i \in \tilde{\lambda}$ and $j \in \lambda'$, i.e. the configuration of $\mathscr{C}_A$ is repeated along every column in $\lambda'$. Now let us choose $\tilde{\lambda}$ to be the minimal energy configuration with $|\tilde{\lambda}| = F_A \leq d_A /2$ and denote the corresponding energy with $E_A$. Then, in the tensor product code, the strip configuration has $F = F_A d_B$ flipped spins and energy $E = E_{A} d_B$ (see Fig.~\ref{fig:Strip}(b)). 

Now, the strip thus constructed need not be the minimal energy configuration at a fixed value of $F$. Nevertheless, it puts an upper bound on the energy of the true minimal energy configuration. This has implications for $E_\text{min}(F)$ of the product code, since the true minimal energy is upper bounded by the energy of the strip of the same size. We can, in turn, use this to argue about the code distance of the quantum code obtained from gauging $\mathscr{C}$. 

More concretely, let us consider the case when the two input codes have good distance, $d_{A/B} \propto n_{A/B}$. We will be interested in configurations where the number of flipped spins $F$ is proportional to $n = n_An_B$. To consider strips where this is true, let $F_A = |\tilde{\lambda}|$ satisfy $0 < F_A/d_A \leq 1/2$ as $n_A \to \infty$. Let $E_A^*$ denote the minimal energy obtained among all such configurations and $F_A^*$ the number of spin flips involved in it. We define constants $c_A$, $\alpha_A$ such that $E_A^* = c_A (F_A^*)^{\alpha_A}$ up to terms that are subleading as $n_A \to \infty$. The corresponding strip thus has $F = F_A^* d_B \propto n_A n_B = n$ and $E = E_A^* d_B \propto n_A^{\alpha_A} n_B$. Let us choose $n_A = n_B = \sqrt{n}$; this is then a configuration with $F \propto n$ and $E \propto n^{(1+\alpha_A)/2}$. We note that this shows that $\mathscr{C}$ is not an LTC, unless $\alpha_A = 1$, i.e. $\mathscr{C}_A$ is already an LTC. We could also consider strips in the second direction to conclude that we need $\mathscr{C}_B$ to also be an LTC. 

We will now specialize to the case when $\alpha_A = 0$. More precisely, we want the stricter condition that there exist configurations in $\mathscr{C}_A$ that are $O(n_A)$ Hamming distance from any codeword but have $O(1)$ energy. In Part I~\cite{LDPCGauge} we described such classical codes as having ``point-like excitations''. We thus have $E_A^* = K = O(1)$. To have a non-trivial gauge dual, we assume that $\mathscr{C}_A$ has global redundancies and it is natural to assume that every check is part of some such redundancy\footnote{It this was not the case, then upon gauging, some checks would turn into qubits that are not acted upon by any logical and could be dropped.}; therefore the minimal energy is $K > 1$. If we now construct the corresponding strip configuration, its boundary will consists of $K$ distinct co-cycles, each of which are individually non-contractible and therefore correspond to $X$-logicals of $\mathscr{C}_\text{q}$ (see our discussion of HGP codes in Sec.~\ref{subsec:HGP}). In other words, by choosing the input $\mathscr{C}_A$ to have point-like excitations, the resulting tensor product has precisely the property we required in order to connect the behavior of energy barriers to the quantum code distance of its gauge dual. In particular, we have the following bound:
\begin{equation}
    E_\text{strip} = K d_X \gtrsim E_\text{min}(F = F_A^* d_B \propto n).
\end{equation}
Since $K=O(1)$ by assumption, this establishes a relationship between the code distance $d_X$ and the value of the energy barrier at some Hamming distance $F$ that is proportional to $n$. 

It is interesting to ask what happens to this bound when we generalize from simple tensor product codes to balanced products, which can realize locally testable codes. To go from tensor to balanced product, we need to mod out the symmetry $G$. Assuming the action of $G$ is free, the number of bits becomes $n = n_A n_B / |G|$. If we start out with the above strip configuration, we find that after modding out, the number of flips it contains obeys $F \geq F_A^* d_B / |G| \propto n$, i.e. that after modding out $G$, the strip still contains a finite fraction of all bits. Let us now assume that $\hat{\mathscr{C}}$ is an LTC; we then have 
\begin{equation}
    E_\text{strip} \gtrsim E_\text{min}(F_\text{strip} \propto n) \geq \mu n 
\end{equation}
for some $O(1)$ constant $\mu$. As mentioned, the boundary of the strip (before modding out $G$)  consists of $K$ distinct non-trivial co-cycles. By definition, each of these taken individually becomes an $X$-logical of $\mathscr{C}_\text{q}$, the quantum code dual to $\mathscr{C}$. This would suggest a relationship of the form $d_X \gtrsim n$; i.e., an optimal scaling of the $X$-distance $d_X$ of the balanced product code. However, there are various things that would need to be ascertained to prove this relation, such that the $X$-logicals thus obtained have minimal size within their respective logical sectors, and that there are no significant cancellations between the different co-cycles that make up the boundary of the strip. Finding sufficient conditions for this to obtain, and understanding how they relate to the existing proofs in terms of e.g. the (co-)locally-minimal distance, is an interesting problem for future study.

\section{Conclusions and outlook}

In this paper, we continued the program, began in Part I~\cite{LDPCGauge}, of developing a general theoretical framework for thinking about phases of matter whose fixed points correspond to classical of quantum LDPC stabilizer models, encompassing both Euclidean and non-Euclidean geometries. In particular, we laid out a general machinery, colloquially referred to as the ``code factory'' (Fig.~\ref{fig:Factory}) that begins with some input graphs $\Gamma$, defines classical codes on these and then uses a variety of product constructions and other transformation to turn these into other, increasingly complicated, classical codes with various desired features. These classical codes can then be fed into the gauging and Higgsing dualities, already described in Part I~\cite{LDPCGauge}, to be turned into solvable quantum stabilizer Hamiltonians. To demonstrate the workings of the code factory, we described how it can reproduce a bewildering variety of known phases of matter starting from the simple 1D classical Ising model (Fig.~\ref{fig:MapOfPhases}).

We particularly focused on classical codes with local redundancies, associated with chain complexes of dimension $\mathscr{D}_c = 2$, and their corresponding quantum CSS codes. This included both well-known hypergraph product codes, which we examined from a physical perspective, and a new family of ``generalized $X$-cube codes'' that are constructed out of a triple of cLDPC codes. In both cases, we described how the properties of these classical inputs are manifested in those of the resulting CSS codes. We also showed how combining two constructions with the additional trick of modding out translation symmetries to reduce the spatial dimension (while maintaining the code dimension) gives rise to novel models with interesting features, notable an SET phase and a novel fracton code. Finally, we reviewed the constructions of good qLDPC codes and their relationship to locally testable classical codes from the perspective of product constructions. 

We conclude by mentioning a number of open directions raised by our work. First, as we illustrated in Sec.~\ref{subsec:NewCodes}, variations of the balanced product constructions (i.e., taking products and then modding out spatial symmetries to reduce the number of qubits), can yield interesting models even in the Euclidean case. The examples we considered there used both the 1D Ising and the Newman-Moore model as inputs and as a result, gave rise to quantum codes with distance $d_\text{q} = L$ where $L$ is the linear system size. A natural question is whether using \emph{only} Newman-Moore as input one can construct for example 3D fracton codes with a provably super-linear scaling of $d_\text{q}$, or even approach the optimal bound $d_\text{q} = O(L^{D-1})$ in $D$ spatial dimensions. 

Moving beyond Euclidean models, the variety of exotic phases that can be identified as descendants of the 1D Ising chain (Fig.~\ref{fig:MapOfPhases}) suggests considering what happens if we replace this starting point with some other classical code, e.g. the Ising or Laplacian model on some appropriately chosen graph $\Gamma$ and develop a systematic understanding of the classification of phases that can be obtained from these. The array of such phases would be further enhanced by developing an appropriate non-Euclidean generalization of the cellular automaton product, potentially along the lines of fiber-bundle codes~\cite{hastings2021fiber}. 

Another avenue for future investigations is to develop a more systematic understanding of when  properties of codes are preserved under the operation of  modding out symmetries. This is relevant both to the question of systematically constructing new 3D phases and also to the arguments presented in Sec.~\ref{subsec:Strip} where we argued that one could establish a general relationship between quantum code distances and classical energy barriers for balanced product codes, provided that one can keep track of what happens to non-contractible loops (most notably, the smallest representatives in each class) under this operation. 

The framework presented here gives ways of obtaining many different \emph{gapped} phases of matter from one another and building up complexity gradually along the way. A natural question to ask is whether there exist a similar framework for \emph{gapless} phases. For example, all the phases appearing in Fig.~\ref{fig:MapOfPhases} are based on $\mathbb{Z}_2$ symmetries, inherited from the 1D Ising model we start with. Are there analogues of product constructions for $U$(1) phases, which would exhibit Goldstone modes\footnote{At least in finite dimensions~\cite{laumann2009absence}.}? Gauging of symmetries is well-defined in that case, giving rise to deconfined phases that are feature robustly gapless ``photon'' excitations~\cite{fradkin1979phase} and the corresponding Higgs phases can also be defined~\cite{thorngren2023higgs}. Fractonic versions of $U$(1) gauge theories also exist~\cite{pretko2017generalized,seiberg2020exotic}. Finding appropriate generalizations of product constructions to draw connections between these could significantly enhance our understanding of gapless phases of matter, even in finite Euclidean dimensions, and provide many new examples.

Finally, let us end on a conjectural note that we already touched upon above. Much is understood about the corner of many-body Hilbert space consisting of wavefunctions that can arise as ground states of local Hamiltonians on a lattice, especially if the Hamiltonian in question is gapped, which leads to an exponential decay of correlations~\cite{hastings2006spectral} and is generically expected to impose an area-law on entanglement~\cite{hastings2007area,anshu2022area}. The Hamiltonians we set out to study, gapped but local only on some non-Euclidean graph, provide a larger arena and thus their ground states cover a larger subset of the many-body Hilbert space. How to characterize this subset, given that for example the notion of an area-law loses its meaning on expander graphs, is an exciting open question. We hope that the framework of product constructions and gauge dualities provides a useful way for charting some of the possibilities and getting a handle of the kind of states that can arise in these models. 

{\it Note added.} During completion of this manuscript, Ref. \onlinecite{tan2023fracton} appeared. Although their overall focus is different from ours, there are a couple of points of overlap. In particular, Ref. \onlinecite{tan2023fracton} also identified the models of Ref. \onlinecite{gorantla2023gapped} as hypergraph product codes. The authors also briefly mention (without exposition) that the $X$-cube model can be obtained as a kind of threefold product. 

\begin{acknowledgments}
We thank  Vladimir Calvera, Trithep Devakul, Jeongwan Haah, 
Nicholas O'Dea and especially Anirudh Krishna for enlightening discussions.
T.R. is supported in part by the Stanford Q-Farm Bloch Postdoctoral Fellowship in Quantum Science and Engineering. 
V.K. acknowledges support from the US Department of Energy, Office of Science, Basic Energy Sciences, under Early Career Award No. DE-SC0021111, the Alfred P. Sloan Foundation through a Sloan Research Fellowship and the Packard Foundation through a Packard Fellowship in Science and Engineering.
\end{acknowledgments}

\bibliography{bib}

\appendix

\section{Polynomial formalism}\label{app:Polynom}

We now review the polynomial formalism developed in~\cite{vijay2016fracton} for analyzing translationally invariant codes. We consider a classical code that is invariant under translations in $D$ dimensions. We assume that there are $N$ sites and $M$ checks per unit cell. It is enough to specify these $M$ checks to define the code, since all others are given by their translates.

Consider a single check with support $A$, which specifies the set of spins the check acts on. We can divide this support into $N$ sublattices, one for each site within a unit cell. For a given sublattice, labeled by $I=1,\ldots,N$, we need to specify the set of unit cells that appear in $A$; let us denote this set by $A_I$. The key observation is that this information can be encoded in a polynomial in the following way. Let us fix some unit cell as the origin. Then the location of any unit cell is in one-to-one correspondence with a set of integers $\mathbf{a} = (a_1,\ldots,a_D)$, which the number of translations in each of the $D$ directions that are needed to get from the origin to the given unit cell. We can represent these in the form of a \emph{monomial}, $x_1^{a_1}x_2^{a_2} \ldots x_D^{a_D}$ over some set of dummy variables $x_1,\ldots,x_D$. We can then combine all of these into a polynomial that represents the support of the check within sublattice $I$:
\begin{equation}
    f_I[A] = \sum_{\mathbf{a} \in A_I} x_1^{a_1}x_2^{a_2} \ldots x_D^{a_D}.
\end{equation}
For example, in the $1D$ Ising model, which has a single site and a single check per unit cell, we have $f = 1+x$. In the $2D$ plaquette Ising model, we have $f = 1 +x + y + xy$. 

We thus have $N$ polynomials describing each of our $M$ checks; let as label these by $f_{Ia}$ where $a = 1,\ldots,M$ labels the different checks. These can be combined into an $N \times M$ \emph{stabilizer matrix} $S$ where each matrix element is one of the polynomials $f_{Ia}$. This matrix encodes all the features of the codes (except for things having to do with boundary conditions, which need to be enforced separately). 

Adding polynomials corresponds to taking the product of corresponding checks. Importantly, this means that these are polynomials with binary coefficients, so that $x + x = 2x = 0$ etc, reflecting the fact that the underlying spin-$1/2$ variables square to $+1$. We can also multiply polynomials, which can be interpreted as follows. First consider multiplying a polynomial $f$ describing the support of a check (on a given sublattice) with a single monomial $x_1^{a_1}\ldots x_D^{a_D}$: this corresponds to translating the check by the vector $\mathbf{a}$. More generally, multiplying with an arbitrary polynomial $g$ is equivalent to taking a number of different translated versions of the original check, one for each monomial appearing in $g$, and then multiplying them together. 

Various features of the code have simple descriptions in terms of the matrix $S$ of polynomials. For example, the transpose code corresponds to the matrix $S^\dagger$, which is defined by taking the transpose of $S$ and replacing each variable $x$ with its inverse $x^{-1} \equiv \bar{x}$\footnote{To see why changing $x$ to $\bar{x}$ is needed, consider that if $1+x$ is an Ising check located at the origin of a $1D$ chain, then flipping a bit at the origin will trigger the check at $1$ and the one to its \emph{left}, located at $\bar{x}$.}. Just as columns of $S$ encode the supports of checks that define the code, columns of $S^\dagger$ encode which checks are violated when a particular spin is flipped. Logical operators of the code therefore correspond to a vector $\mathbf{L} = (L_1, L_2, \ldots, L_N)$ satisfying $S^\dagger \mathbf{L} = 0$. Note that when the logical is non-local, the elements $L_{I}$ are not polynomials but formal infinite sums. For example, the logical of the $1D$ Ising model is given by $L(x) = \sum_{a \in \mathbb{Z}} x^a$.

Similarly, redundancies corresponds to a vector $\mathbf{R} = (R_1,\ldots,R_M)$ such that $S \mathbf{B} = 0$. For a global redundancy, $R_a$ are again infinite sums, while for a local redundancy they are polynomials of finite order. There might be a number of distinct local redundancies, say $K$ per unit cell, which can then be combined into a $M \times K$ matrix $R$ such that $S R = 0$. The matrix $R$ itself looks like the stabilizer matrix of some different classical code.  

A CSS quantum code is defined by a pair of stabilizer matrices, $S_X$ and $S_Z$, describing the supports of $X$ and $Z$ checks of the code. Commutativity of checks is equivalent to the condition $S_X^\dagger S_Z = 0$. Given a classical code with stabilizer matrix $S$ and local redundancies $R$, we can always define a corresponding CSS code as $S_X = S^\dagger$ and $S_Z = R$. This is the CSS code that appears in the gauge theory based on the classical code as described in part I. 

\section{Quantum-classical mapping and Kramers-Wannier dualities}\label{app:QuantumToClassical}

\begin{figure}[t!]
    \centering
    \includegraphics[trim={1cm 0 0 0},width = 1.\linewidth]{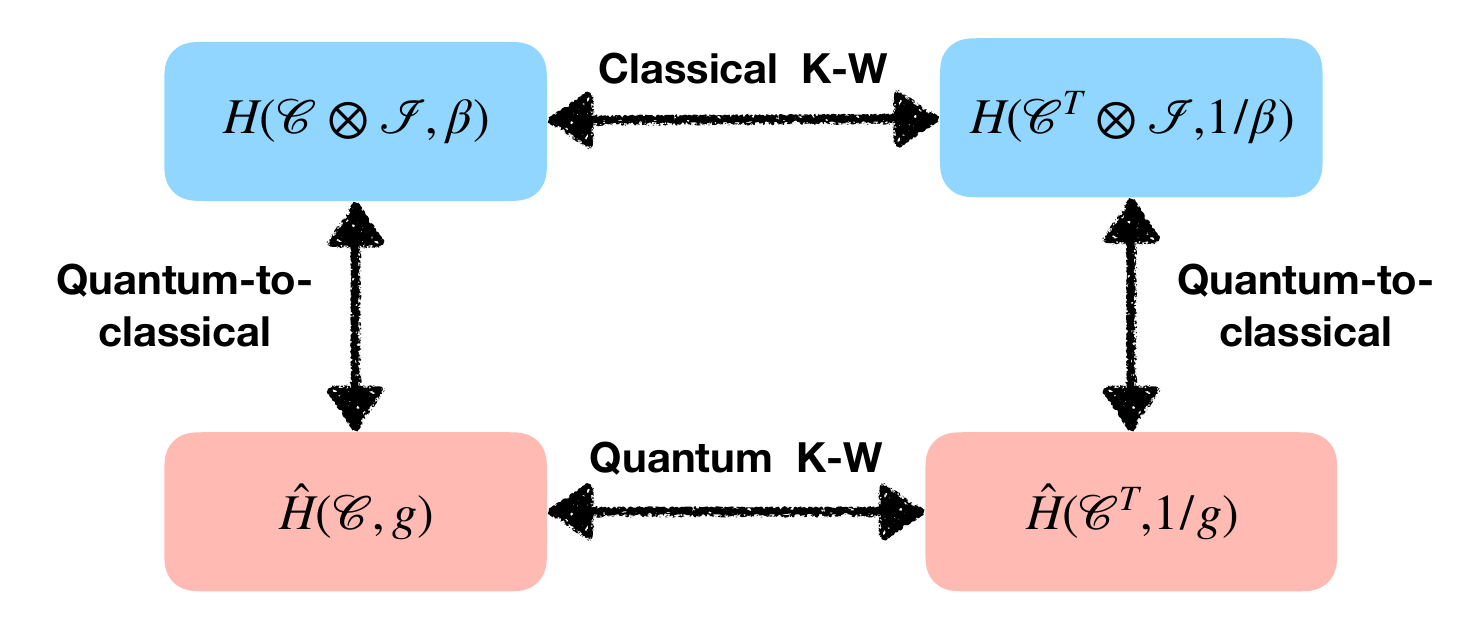}
    \caption{\textbf{Quantum-classical mappings and Kramers-Wannier dualities.} The quantum-classical duality maps the quantum Hamiltonian $\hat{H}(\mathscr{C},g)$, corresponding to classical code $\mathscr{C}$ in a transverse field $g$, to the partition function of a classical code that has the structure of the tensor product $\mathscr{C} \otimes \mathcal{I}_\text{1D}$, with $\mathcal{I}_\text{1D}$ the 1D Ising model. The quantum KW duality maps $\hat{H}(\mathscr{C},g)$ to $\hat{H}(\mathscr{C}^T,1/g)$, changing the code for its transpose and low to high field. The classical KW duality maps a partition function of $\mathscr{C} \otimes \mathcal{I}_\text{1D}$ at temperature $\beta$ to that of $\mathscr{C}^T \otimes \mathcal{I}_\text{1D}$ at a dual temperature $\beta_*$. The two operations (taking the classical dual and taking the KW dual) commute.
    }
    \label{fig:KWDualities}
\end{figure}

In this paper and its predecessor~\cite{LDPCGauge}, we encountered two different notions of Kramers-Wannier duality. One we discussed in Sec.~\ref{sec:BuildingBlocks} is analogous to that of the 2D classical Ising model, mapping between two classical codes associated to dual chain complexes. The other one is a mapping between quantum Hamiltonians, which relates the ``quantized'' version of a code $\hat{H}(\mathscr{C})$ (see Eq.~\eqref{eq:QuantizedHam}), to that of its transpose $\hat{H}(\mathscr{C}^T)$, similar to the KW duality of the 1D quantum transverse field Ising chain. In the Ising case, the two models, and their respective dualities are related to each other through the quantum-classical correspondence of partition functions. Here we briefly outline how this works out more generally. 

Let us start with some classical code $\mathscr{C}$ embedded into a quantum system as in Eq.~\eqref{eq:QuantizedHam}. We can then make the model ``quantum'' by adding a transverse field to get $\hat{H}(\mathscr{C},g) = -\sum_a \hat{C}_a - g \sum_i \sigma_i^x$, with $\hat{C}_a = \prod_{i\in \delta(a)} \sigma_i^z$. We will assume that the code $\mathscr{C}$ has no local redundancies. The quantum Kramers-Wannier duality (see Part I~\cite{LDPCGauge} for details) maps $\hat{C}_a$ into $\tau_a^z$ and $\hat{\sigma}_i^x$ into a check of the transpose code $\mathscr{C}^\text{T}$ of the form $\prod_{a \in \delta^T(i)} \tau_a^x$. After a Hadamard transformation and an overall rescaling, we find that the gauging map relates $\hat{H}(\mathscr{C},g)$ to $\hat{H}(\mathscr{C}^\text{T},1/g)$ and vice versa. This is the appropriate version of (quantum) Kramers-Wannier duality, which relates the ordered (small $g$) phase of one code to the disordered (high $g$) phase of its transpose. Of special interest are cases where the code and its transpose are isomorphic (related by a symmetry of $\mathscr{C}$), $\mathscr{C} \cong \mathscr{C}^\text{T}$, in which case the duality relates different points in the \emph{same} phase diagram\footnote{This is fairly general for translation invariant codes with one check per unit cell. For example, it applies to the 2D plaquette Ising and Newman-Moore models.}; assuming a single phase transition in this case, it has to occur at $g=1$ and is self-dual under the quantum KW duality. 

Consider now a different classical code $\mathscr{C}$, which does have local redundancies. We can therefore represent it as a $\mathscr{D}_c = 2$ chain complex and we can apply the classical Kramers-Wannier duality of Sec.~\ref{sec:BuildingBlocks} (taking the dual of the chain complex) to map it to a different code $\mathscr{C}^\text{KW}$. Considering the code as a model of statistical mechanics, we can associate to it a partition function 
\begin{equation}
    \mathscr{Z}_\text{cl}(\mathscr{C},\beta) = \sum_{\{\sigma_i = \pm 1\}} e^{-\beta \sum_{a=1}^{m} \prod_{i\in\delta_1(a)} \sigma_i},
\end{equation}
where $\delta_1$ is the boundary map relating checks to bits in the chain complex. We can relate this to the partition function of the dual code, $\mathscr{Z}_\text{cl}(\mathscr{C}^\text{KW},\beta)$ through an appropriate Kramers-Wannier duality. In particular, one can define high- and low-temperature expansions of the classical partition function. The former represents $\mathscr{Z}_\text{cl}(\mathscr{C},\beta)$ as a sum over all cycles (closed loops) of the chain complex corresponding to $\mathscr{C}$. The latter is an expansion in terms of domain wall configurations of $\mathscr{C}$, which form contractible closed loops in the \emph{dual} chain complex. Since one expansion is in terms of cycles, and the other in terms of cocycles, this implies a relation
\begin{equation}\label{eq:ClassicalKW}
    \mathscr{Z}_\text{cl}(\mathscr{C},\beta) \, ``="  \, \mathscr{Z}_\text{cl}(\mathscr{C}^\text{KW},\beta_*),
\end{equation}
where the two temperatures are related to each other through $\tanh{\beta_*} = e^{-2\beta}$, so that small $\beta$ corresponds to large $\beta_*$ and vice versa. 

In Eq.~\eqref{eq:ClassicalKW}, we put the equality in inverted commas because it only holds up to corrections arising from the fact that one side (the high temperature expansion) involves a sum over \emph{all} cycles while the other only over \emph{contractible} ones. While this difference, having to do with ``boundary conditions''~\cite{LDPCGauge} is indeed negligible in the Ising model, or finite-dimensional Euclidean examples more generally, it can be significant in other contexts~\cite{placke2023random}. A similar issue arises in the quantum KW duality: to recover the full Hilbert space of the $\sigma$ variables (bits in $\mathscr{C}$) in the dual theory of $\tau$ variables (bits of $\mathscr{C}^T$), one needs to sum over all choices of ``boundary conditions'' in the latter (see Part I~\cite{LDPCGauge} for details); again, in non-Euclidean models, this could lead to important differences between the two theories, such as a mismatch between their critical fields $g_c$.

The two types of KW duality, quantum and classical, can be related via a quantum-to-classical mapping as follows. Given $\hat{H}(\mathscr{C},g)$, we can relate its \emph{quantum} partition function, $\mathscr{Z}_\text{q}(\mathscr{C},g,\beta) \equiv \text{tr}\left(e^{-\beta \hat{H}(\mathscr{C},g)}\right)$, to that of an appropriate classical model. Going through the usual steps~\cite{sachdev1999quantum} of splitting the Hamiltonian into two parts, Trotterizing, and inserting basis states in the $\sigma_i^z$ basis, we get $N$ copies of the system (with $\Delta\tau \equiv \beta/N$ being the Trotter step); within each copy, the terms $\hat{C}_a$ take their expectation values, turning into checks of the classical code $C_{a,t} = \prod_{i \in \delta(a)}\sigma_{i,t}$ with $t=1,\ldots N$ labeling the copies. Between the copies, we end up with matrix elements of the transverse field, which can be written in the usual way as $\braket{\sigma_{i,t}|e^{g\hat{\sigma_i^x}}|\sigma_{i,t+1}} \propto e^{K \sigma_{i,t}\sigma_{i,t+1}}$ with $e^{-2K} = \tanh{g}$; that is, the copies are coupled by the usual Ising-terms. 

Combining these terms, we end up with the partition function of a classical statistical mechanics model, that closely resembles the product code $\mathscr{C} \otimes \mathcal{I}_\text{1D}$, the only difference being that the checks of the two input codes come with different coupling constants, $\Delta \tau$ within the copies and $K$ between them. To the extent that this anisotropy can be neglected, we end up with the partition function $\mathscr{Z}_\text{cl}(\mathscr{C} \otimes \mathcal{I}_\text{1D},\beta)$; in this case, we can use the correspondence to relate critical properties at the quantum phase transition of $\hat{H}(\mathscr{C},g_c)$ to those of the classical model $\mathscr{C} \otimes \mathcal{I}_\text{1D}$ at its critical temperature $\beta_c$.

We thus have three sets of dualities: classical KW, quantum KW and quantum-to-classical. The relationships between these are illustrated in Fig.~\ref{fig:KWDualities}. The key observation is that, using the properties of the tensor product and the fact that the 1D Ising model is isomorphic to its transpose, we have that $(\mathscr{C} \otimes \mathcal{I}_\text{1D})^\text{KW} \cong \mathscr{C}^T \otimes \mathcal{I}_\text{1D}$. We thus find that the order of the two mappings commute: we can either take the quantum KW dual of $\hat{H}(\mathscr{C},g)$ to get $\hat{H}(\mathscr{C}^\text{T},1/g)$ and then map it to a classical partition function via the quantum-classical mapping or we can use the quantum-classical mapping on the original $\hat{H}(\mathscr{C},g)$ and then apply the classical KW duality: either way, we end up with the same classical model. If $\mathscr{C} = \mathscr{C}^\text{T}$, then the classical KW duality maps the low- and high-T phases of $\mathscr{C} \otimes \mathcal{I}_\text{1D}$ into each other. Self-duality is achieved when $e^{-2K} = \tanh{\Delta\tau}$; this corresponds to $g=1$, indicating that $g$ plays the role of temperature in the classical phase diagram. 

\section{More details on the generalized $X$-cube code}\label{app:CubProd}

Here we provide some more details on the code introduced in Sec.~\ref{subsubsec:BalancedCubicProduct}, which was obtained by applying the cubic product construction to two Ising chains and the Newman-Moore model and the modding out diagonal translations to reduce the resulting 4D model back into 3 dimensions. We first describe some features of the 4D GXC model, obtained by gauging the cubic product. Then we describe how the fatures of this 4D model are reflected in the 3D model obtained after modding out translations.

\subsection{4D GXC model}\label{subsec:4DGXC}

First, we consider the ``generalized $X$-cube'' model $\text{GXC}(\mathcal{I}_\text{1D},\mathcal{I}_\text{1D},\mathscr{C}_\text{NM})$, where $\mathcal{I}_\text{1D}$ is the 1D Ising model and $\mathscr{C}_\text{NM}$ is the Newman-Moore model. This is a CSS code in four spatial dimensions, which we can analyze in the framework of Sec.~\ref{subsec:GXC} to uncover its logicals and excitations which inherit their properties from the classical input codes.

Let us denote the four coordinates of the 4D hypercubic lattice as $x,y,z,u$. The CSS code is defined on three sets of qubits, assigned to $x$-edges, $y-edges$ and $zu$-plaquettes, respectively. As such, the support of each check will be described by a set of three polynomials. There is a single $Z$-check (assigned to the hypercubes) and three $X$-checks (assigned to the sites). We can summarize these in the following stabilizer matrices
\begin{align}
    S_Z &= \begin{pmatrix}
        (1 + \bar{y}) (1+\bar{z}+\bar{u}) \\
        (1 + \bar{x}) (1+\bar{z}+\bar{u}) \\
         (1 + \bar{x}) (1 + \bar{y})
    \end{pmatrix}, \\ 
    S_X &= \begin{pmatrix}
        1+x & 1+x & 0 \\
        1+ y & 0 & 1+y\\
        0 & 1+z+u & 1+z+u
    \end{pmatrix}.
\end{align}

The logicals of this code will be built out of the logicals of the classical input code. $\mathcal{I}_\text{1D}$ has a single such logical that flips all spins and is represented by the polynomial $1 + x + \ldots + x^{L-1}$. The NM model has a number of logicals $k_\text{NM}(L)$ that changes erratically with the system size $L$, with the largest value $k_\text{NM}(L) = L-1$ when $L$ is of the form $2^p-1$ for an integer $p$. In general, we will denote by $f_l$, $l=1,\ldots,k_\text{NM}(L)$ a set of polynomials corresponding the some basis set of these logicals. For example, in the aforementioned case of $L=2^p-1$, one possible choice is given by $f_1(z,u) = (1 + z + u)^L + 1$\footnote{Note that we have taken periodic boundary conditions so that $x^{L} = 1$ and similarly for the $y,z,u$ coordinates.}. In terms of these, we can write a basis set of $X$-logicals as
\begin{align}\label{eq:4DXLogicals}
     \mathcal{X}_1 &= \begin{pmatrix}
       1+y+\ldots+y^{L-1} \\ 0 \\ 0
    \end{pmatrix},  \mathcal{X}_2 = \begin{pmatrix}
        f_l(z,u) \\ 0 \\ 0
    \end{pmatrix}, \nonumber \\ 
     \mathcal{X}_3 &= \begin{pmatrix}
        0 \\ 1+x+\ldots+x^{L-1} \\ 0
    \end{pmatrix},
     \mathcal{X}_4 = \begin{pmatrix}
        0 \\ f_l(z,u) \\ 0
    \end{pmatrix}, \nonumber \\ 
    \mathcal{X}_5 &= \begin{pmatrix}
        0 \\ 0 \\ 1+x+\ldots+x^{L-1}
    \end{pmatrix},  \mathcal{X}_6 = \begin{pmatrix}
        0 \\ 0 \\ 1+y+\ldots+y^{L-1}
    \end{pmatrix},
\end{align}
along with their translates. Here, $\mathcal{X}_1,\mathcal{X}_2$ are acting on $x$-edges, $\mathcal{X}_3,\mathcal{X}_4$ on $y$-edges and $\mathcal{X}_5,\mathcal{X}_6$ on the $zu$-plaquettes. $\mathcal{X}_1$ and $\mathcal{X}_6$ are extended along the $y$ directions and $\mathcal{X}_3,\mathcal{X}_5$ along the $x$ direction, while $\mathcal{X}_2,\mathcal{X}_4$ form fractal patterns in the $zu$-plane, inherited from the NM model. The corresponding $Z$-logicals take the form
\begin{align}\label{eq:4DZLogicals}
    \mathcal{Z}_1 &= \begin{pmatrix}
        1 + \bar x + \ldots + \bar x^{L-1} \\ 0 \\ 0
    \end{pmatrix} \nonumber \\ \mathcal{Z}_2 &= \begin{pmatrix}
        0 \\ 1 + \bar y + \ldots + \bar y^{L-1} \\ 0
    \end{pmatrix}, \nonumber \\ 
    \mathcal{Z}_3 &= \begin{pmatrix}
        0 \\ 0 \\ f_l(\bar{z},\bar{u})
    \end{pmatrix},
\end{align}
acting on $x$-edges, $y$-edges, and $zu$-plaquettes, respectively. $\mathcal{Z}_{1,2}$ are linear, while $\mathcal{Z}_3$ is a Sierpinski fractal in the $zu$-plane. 

We can also consider truncated logicals and the excitations they create. For example, we can truncate line-like logicals at a finite distance $1 + x + \ldots x^r$; in the Ising model, this would create a pair of excitations at the locations of its endpoints, as can be seen from the equation $(1+x)(1+\ldots+x^r) = 1 + x^{r+1}$. For the NM model, we can consider a Sierpinski triangle whose side length is $2^q$ for some integer $q$, corresponding to the polynomial $f_{S}^{(q})(u,z) = (1+z+u)^{2^q - 1}$. In the NM model, this creates a triple of excitations, one at each corner of the triangle, as can be seen from $(1+z+u) (1+z+u)^{2^q - 1} = 1 + z^{2^q} + u^{2^q}$. We can thus replace the appropriate polynomials in Eqs.~\eqref{eq:4DXLogicals} and \eqref{eq:4DZLogicals} with their truncated versions to consider truncated logicals and the point-like excitations they create in the 4D quantum code.

For the $X$-logicals of Eq.~\eqref{eq:4DXLogicals}, their naive truncation creates groups of excited $Z$-checks clumped together. For example, the truncated version of $\mathcal{X}_1$, with support $1 + y + \ldots + y^{r}$ creates excitations corresponding to the locations  $(1+y^{r+1})(1+z+u)$, i.e. a \emph{triple} of excitations at both endpoints. Similarly, we we replace the the logical $f_l(u,z)$ in $\mathcal{X}_2$ with a finite Sierpinski triangle $f_{S}^{(q})(u,z)$, we find excitations at the locations $(1 + y) (1 + z^{2^q} + u^{2^q})$, i.e. a pair of excitations at each corner. However, one can combine these two operators to create individual, well-separated excitations. I.e., the product of $X$ operators with support $(1 + y + \ldots + y^{r}) (1+z+u)^{2^q - 1}$, i.e. a stack of Sierpinski triangles, on the $x$-edges will create six $Z$-excitations at the locations $(1+y^{r+1})(1 + z^{2^q} + u^{2^q})$, all separated from one another. Similarly, there re 2D membrane operators in the $xy$-plane, acting on the $uz$-plaquettes, create four isolated excitations at their corners.  

Truncated versionf of the $Z$-logicals $\mathcal{Z}_{1,2}$ create lineon excitations that are free to move within the $x$ and $y$ directions, respectively. A triple of these excitations can be combined to obtain a composite particle that can move within the entire $xy$ plane. In particular, the product of Pauli $Z$ matrices with support 
\begin{equation}
    \begin{pmatrix}
        (1 + x + \ldots + x^r) (1 + z^{2^q} + u^{2^q}) \\ 
        (1 + y + \ldots + y^{r'}) (1 + z^{2^q} + u^{2^q}) \\       (1+z+u)^{2^q - 1} 
    \end{pmatrix},
\end{equation}
which corresponds to three ``bent'' line operators in the $xy$ plane, glued together by a Sierpinski triangle in the $uz$ plane, creates excitations only at the endpoints in the vicinity of $x^r$ and $y^{r'}$.  

\subsection{3D Classical codes}

\begin{figure}[t!]
    \centering
    \includegraphics[trim={1cm 0 0 0},width = 0.7\linewidth]{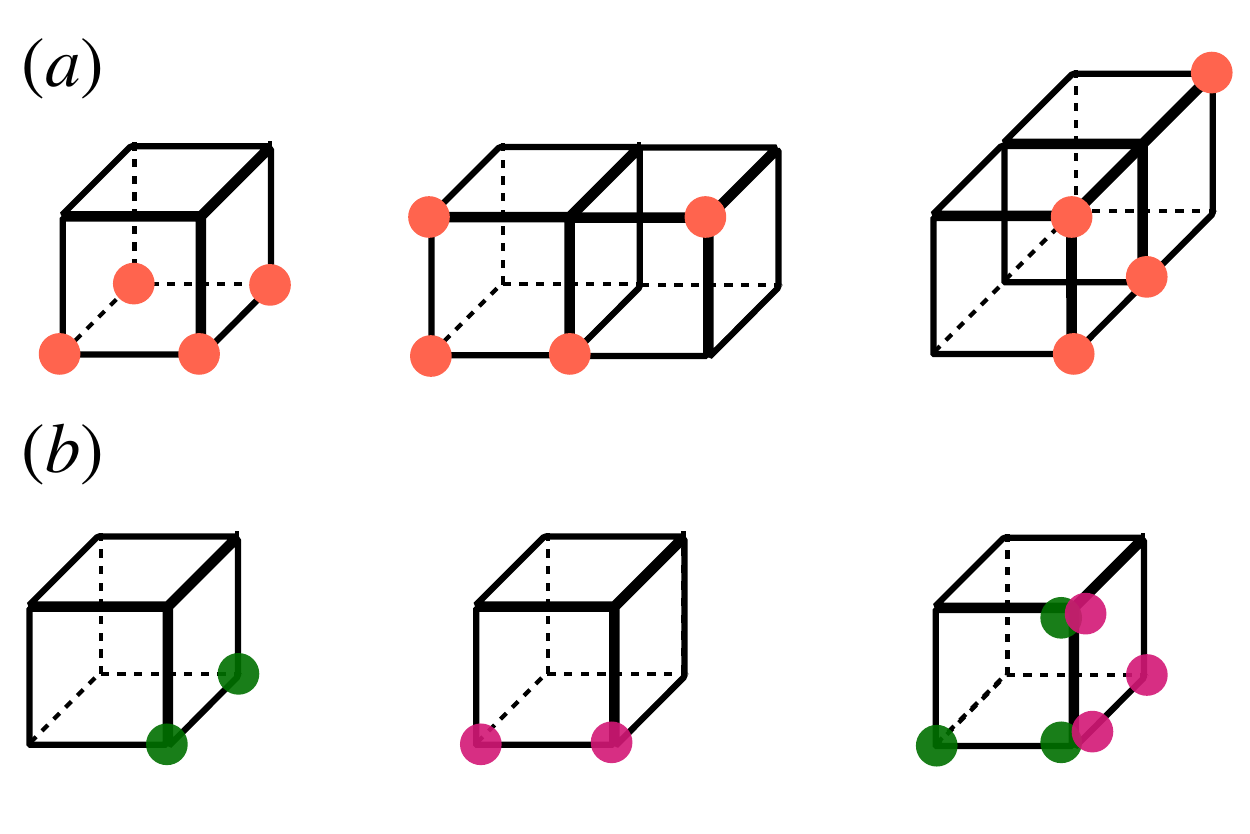}
    \caption{Checks of the classical code from Sec.~\ref{subsubsec:BalancedCubicProduct} and its Kramers-Wannier dual. 
    }
    \label{fig:NMCubic_classical_checks}
\end{figure}

We now turn to the 3D model obtained from the above 4D code after modding out fully diagonal translations. First, we consider the corresponding classical code $\mathscr{C}_\text{cl}$ and its Kramers-Wannier dual, $\mathscr{C}_\text{cl}^\text{KW}$ and count the number of their logical operators. This gives some intuition about the quantum code, since the latter is obtained from gauging the symmetries corrsponding to these classical logicals. Moreover, the counting also directly yields the number of quantum logicals through the relation $k_\text{q} = k_\text{cl} + k_\text{cl}^\text{KW}$. 

The classical code in question was defined in Eqs.~\eqref{eq:NMCubic},\eqref{eq:NMCubic_redef}. The checks of this code and of its Kramers-Wannier dual are shown in Fig.~\ref{fig:NMCubic_classical_checks}(a) and (b) respectively. The former we have already examined to some extent in the main text: it consists of bits on a cubic lattice and one 4-body check for each plaquette. The ones on the $xy$ plane are simple the four bits along a plaquette, while the other two extend to two nearby plaquettes and are equal to the product of two Newman-Moore checks, shifted in either the $x$ or $y$ direction. 

The dual classical code has \emph{two} bits on the cubes of the 3D lattice, corresponding to the first two redundancies in Eq.~\eqref{eq:NMCubic_redundancies} (in principle, we should include a third bit for the third redundancy, but since this is not linearly independent of the first two, we choose to drop it). For simplicity, in Fig.~\ref{fig:NMCubic_classical_checks} we draw the checks of this code on the dual lattice, so that bits once again live on the sites and we denote the two species of bits by two colors (green and purple). There are three checks per unit cell, corresponding to the three orientations of an edge; along two orientations ($x$ and $y$) we find that the checks are simply Ising interactions, one for one species of bits and one for the other. The third check, associated to horizontal edges, is 6-body and has the form of the product of two Newman-Moore checks, oriented along the $xz$ and $yz$ planes, acting on the two different species. 

Let us first calculate $k_\text{cl}$. Looking at Fig.~\ref{fig:NMCubic_classical_checks}(a), it is clear that all checks share a planar symmetry, consisting of flipping every bit along an $xy$ plane. These give $L$ logical operators. Another set of logicals can be constructed from the observation that the checks in the $xz$ plane are the product of two Newman-Moore checks. As such, they are invariant under any of the Newman-Moore logicals applied within the plane. The planar-fractal operator does not commute with the other two checks. However, we can make it commute by extending it in the third direction and repeating it along \emph{every} $xz$ plane; since the remaining two checks act on an even number of bits on every $y$-directed row of edges, they will commute with this extended fractal, which gives $k_\text{NM}(L)$. A similar set of logicals is given by placing a NM logical along the $yz$ plane and repeating it along the $x$ direction. It is easy to check that all these logicals are linearly independent\footnote{To see this, observe that no logical operator of the NM model contains a full row of spins in its support.} This gives the counting of $k_\text{cl} = L +  2k_\text{NM}(L)$, which can also be confirmed numerically. 

Turning to the dual code (Fig.~\ref{fig:NMCubic_classical_checks}(b)), we find that, despite having a very different set of checks, it has an equivalent set of logical operators. To see this, first enforce the conditions corresponding to the first two (Ising-like) checks: these force one species of bits to be fully aligned along the $y$ direction, resulting in $L^2$ effective bits. The last check defines a NM model on these effective bits; therefore we have a set of logicals that correspond to NM logicals in the $xz$ plane extended in the $y$ direction, acting on the first species of bits alone. Similarly, we have a set of NM logicals in the $yz$ plane, extended along $x$, acting on the second species. Finally, all the checks are invariant under flipping the spins along the $xy$ plane in \emph{both} species. We thus again end up with $k_\text{cl}^\text{KW} - L + 2 k_\text{NM}(L)$. 

\subsection{3D quantum code}

We can now turn to the 3D CSS codes whose checks are define in Fig.~\ref{fig:NM_cubic}(a) and perform a similar analysis to the one we carried out for the 4D Generalized X-cube model in Sec.~\ref{subsec:4DGXC}. 

We once again have three qubits per units cell, associated to the three types of edges $x$, $y$ and $z$. We begin by writing the stabilizer matrices for the $Z$ check as
\begin{align}
    S_Z &= \begin{pmatrix}
        (1 + \bar{y}) (1+\bar{z}+\bar{z}\bar{y}) \\
        (1 + \bar{x}) (1+\bar{z}+\bar{z}\bar{x}) \\
         (1 + \bar{x}) (1 + \bar{y})
    \end{pmatrix}, \\ 
    S_X &= \begin{pmatrix}
        1+x & 0 & 1+x \\
        0 & 1+y & 1+y\\
        1+z+yz & 1+z+xz & (x+y)z
    \end{pmatrix}.
\end{align}

The terms of the form $1+z+yz$ correspond to Newman-Moore checks, so that the quantum code inherits NM logicals, which we again label as  $f_l$. In particular, The $X$ logicals are
\begin{align}\label{eq:3DXLogicals}
     \mathcal{X}_1 &= \begin{pmatrix}
       1+y+\ldots+y^{L-1} \\ 0 \\ 0
    \end{pmatrix},  \mathcal{X}_2 = \begin{pmatrix}
        f_l(y,z) \\ 0 \\ 0
    \end{pmatrix}, \nonumber \\ 
     \mathcal{X}_3 &= \begin{pmatrix}
        0 \\ 1+x+\ldots+x^{L-1} \\ 0
    \end{pmatrix},
     \mathcal{X}_4 = \begin{pmatrix}
        0 \\ f_l(x,z) \\ 0
    \end{pmatrix}, \nonumber \\ 
    \mathcal{X}_5 &= \begin{pmatrix}
        0 \\ 0 \\ 1+x+\ldots+x^{L-1}
    \end{pmatrix},  \mathcal{X}_6 = \begin{pmatrix}
        0 \\ 0 \\ 1+y+\ldots+y^{L-1}
    \end{pmatrix},
\end{align}
which we recognize as having the same structure as the 4D code in Eq.~\eqref{eq:4DXLogicals}. There are $L$ distinct classes of $\mathcal{X}_1$ ($\mathcal{X}_3$) logicals, corresponding to translating them in the $x$ ($y$) direction, and there are $k_\text{NM}(L)$ classes of the other four types of $X$ logicals\footnote{For $\mathcal{X}_2$ and $\mathcal{X}_4$ this comes from the possible choices for the NM logical $f_l$. For $\mathcal{X}_5$ and $\mathcal{X}_6$, one obtains this by noting that A product of a line of $X$-checks corresponds to the product of three such logicals, arrangedi the shape of a NM logical.}, in agreement with the counting in Eq.~\eqref{eq:k_count_3D}.

The corresponding $Z$-logicals read
\begin{align}\label{eq:3DZLogicals}
    \mathcal{Z}_1 &= \begin{pmatrix}
        1 + \bar x + \ldots + \bar x^{L-1} \\ 0 \\ 0
    \end{pmatrix} \nonumber \\ \mathcal{Z}_2 &= \begin{pmatrix}
        0 \\ 1 + \bar y + \ldots + \bar y^{L-1} \\ 0
    \end{pmatrix}, \nonumber \\ 
    \mathcal{Z}_3 &= \begin{pmatrix}
        0 \\ 0 \\ f_l(\bar{x},\bar{z}) f_{l'}(\bar{y},\bar{z})
    \end{pmatrix},
\end{align}
$\mathcal{Z}_1$ and $\mathcal{Z}_2$ correspond to $L+k_\text{NM}(L)$ equivalence classes each, corresponding to the number of logicals in the 2D classical code with checks $(1+x)(1+z+xz)$ in the $xz$ plane, which correspond to horizontal lines and NM fractals (closely related to the logicals of the 3D classical codes we encountered in the previous subsection). Most interesting is the last set of logicals, $\mathcal{Z}_3$, which consist of combinations of Sierpinski fractals in the $xz$ and $yz$ planes; there are $2k_\text{NM}(L)$ classes of these, corresponding to the two independent choices of NM logical. There is an alternative basis for these logicals, given by \begin{align}
\mathcal{Z}_3' &= \begin{pmatrix}
        0 \\ 0 \\ (1 + x\bar y + \ldots + (x\bar y)^{L-1}) f_l(\bar{x},\bar{z}) 
    \end{pmatrix}, \nonumber \\    
    \mathcal{Z}_3'' &= \begin{pmatrix}
        0 \\ 0 \\ (1 + x\bar y + \ldots + (x\bar y)^{L-1}) f_{l'}(\bar{y},\bar{z})
    \end{pmatrix}. \nonumber 
\end{align}
These correspond to picking a Sierpinski fractal in either the $xz$ or $yz$ plane and then extending it in the third direction, but with an additional translation in each step.

The excitations created at the endpoints of truncated versions of $X$-logicals again come in groups of 2 and 3, in exactly the same way as they did in the 4D code. However, the operators that separate these into individual excitations now look different, since they have been compressed down into 3 dimensions. For example, consider the operator corresponding to the set of polynomials
\begin{equation}
    \begin{pmatrix}
        (1 + y  + \ldots + y^r) (1 + z +yz)^{2^q-1} \\ 0 \\ 0
    \end{pmatrix}.
\end{equation}
This has the form of a product of many Sierpinski triangles in the $yz$ plane, translated along the $y$ direction and it creates six separated excitations at locations corresponding to $(1+y^{r+1})(1 + z^{2^q} + (yz)^{2^q})$. There are also rectangular operators, with support $(1+x+\ldots  x^r) (1 + y + \ldots y^{r'})$ acting on the $z$-edges that create four separated excitations at their corners.

The truncated versions of the logicals $\mathcal{Z}_1$ and $\mathcal{Z}_2$ create lineon excitations that can move in the $x$ and $y$ directions, respectively. We can again combine three such lines to form bound states that can move within the $xy$ plane, e.g. by considering the $Z$-operator with support
\begin{equation}
    \begin{pmatrix}
        (1 + x + \ldots + x^r) (1+z+yz) \\ 
        (1 + y + \ldots + y^{r'}) (1 + z + xz) \\       
        1
    \end{pmatrix},
\end{equation}
which creates no excitations at the origin, but only in the vicinity of $x^r$ and $y^{r'}$. 

Finally, let us mention that we do not know how to write down a product of Pauli $Z$ operators acting on $z$-edges alone that creates a finite number of well-separated excitations, although this should be possible, since the code $\mathscr{C}_X$ does not have any local redundancies once we drop the third $X$ check in Fig.~\ref{fig:NM_cubic} (which is trivially a product of the first two) from the set of generators. To see the issue, consider a truncated version of the logical $\mathcal{Z}_3'$, which takes the form
\begin{equation}
    \begin{pmatrix}
        0 \\ 0 \\ (1 + x\bar{y} + \ldots (x\bar y)^r) (1 + \bar x + \bar x \bar z)^{2^{q}-1}.
    \end{pmatrix}.
\end{equation}
This operators creates excitations along two entire Sierpinski triangles (located at $1$ and $(x\bar y)^{r+1}$) and along the three one-dimensional lines connecting the corners of these triangles.

\end{document}